\documentclass[genTeX,type1rest,usecmfonts]{nrc1}   %% {nrc one}

\usepackage{graphicx}
\begin{document}

\title{Atomic X-ray Spectroscopy of Accreting Black Holes}

\author{D.A. Liedahl}

\address{Lawrence Livermore
National Laboratory, 7000 East Ave, L-473, Livermore, CA 94550, U.S.A.}

\author{D.F. Torres}

\address{Lawrence Livermore
National Laboratory, 7000 East Ave, L-413, Livermore, CA 94550, U.S.A.}

\shortauthor{Liedahl and Torres}

\maketitle

\begin{abstract}
Current astrophysical research suggests that
the most persistently luminous objects in the Universe are powered by the
flow of matter through accretion disks onto black holes. Accretion disk systems are
observed to emit copious radiation  across the electromagnetic spectrum, each
energy band providing access to rather distinct regimes of physical conditions and
geometric scale. X-ray emission probes the innermost regions of the accretion disk,
where relativistic effects prevail. While this has been known for decades,
it also has been acknowledged that inferring physical conditions in the relativistic regime
from the behavior of the X-ray continuum is problematic and not satisfactorily constraining.
With the discovery in the 1990s of iron X-ray lines bearing signatures of relativistic distortion
came the hope that such emission would more firmly constrain models of disk accretion
near black holes, as well as provide observational criteria by which to test general
relativity in the strong field limit. Here we provide an introduction to this phenomenon. 
While the presentation is intended to be primarily tutorial in nature, we aim also to 
acquaint the reader with trends in current research. To achieve these ends, we present 
the basic applications of general relativity that pertain to X-ray spectroscopic 
observations  of black hole accretion disk systems, focusing on the Schwarzschild 
and Kerr solutions to the Einstein field equations. To this we add treatments of 
the fundamental concepts associated with the theoretical and modeling aspects of 
accretion disks, as well as relevant topics from observational and theoretical X-ray spectroscopy.
\end{abstract}

\section{Introduction}

The publication of Einstein's General Theory of Relativity (GR) in
1915 was the culmination of several years of effort that began as early
as 1907, when Einstein began to address
the incompatibility between the Special Theory of Relativity
and Newtonian gravitation. The nature of the inconsistency is clear.
For example, Newton's formulation of gravitation is embodied in the Poisson
equation for the gravitational potential, $\nabla^2 \phi =4\pi G \rho$, where
the source term $\rho$ is the mass density, and it is implied that the field responds instantaneously
to changes in $\rho$, thereby violating special relativity. 
To Einstein, this inconsistency foreshadowed a complete revision of gravitation theory.

For a relativistic theory of gravity, the characteristic length scale
for physical effects near a mass $M$, on dimensional grounds, is $GM/c^2$, 
which is known as the {\it gravitational radius}. In assessing the magnitude of 
GR effects in the vicinity of a mass $M$, at a distance $r$ from the mass, 
one compares $r$ with this characteristic length scale. Therefore, we expect 
only small perturbations to Newtonian physics whenever $GM/rc^2 \ll 1$. 
This is just the case for the first proposed tests of GR, and in fact, for all 
possible terrestrial and solar system experiments. Most notable were the 
problem of the precession of the perihelion of Mercury and the prediction 
of the bending of starlight as it grazed the Sun. In both cases,
$GM/rc^2 \ll 1$, and the magnitude of the effects are indeed subtle. 

The advance of Mercury's perihelion is 574
arcseconds per century, only 43 of which are left unexplained by Newton's $1/r^2$ law
after accounting for perturbations of Mercury's orbit caused by the other planets.
In 1914, using the nearly completed GR theory, Einstein found that the rate of advance of the
perihelion $\Delta \omega \approx (3GM/c^2r_{\rm orbit})\, \omega_{\rm orbit}$, which works out
to 43 arcseconds per century. Einstein was thus able to account for
a discrepancy that had been known for sixty years, and was delighted, being moved to exclaim: 
\begin{quote}
For a few days, I was beside myself with joyous excitement. 
\end{quote}

The quantitative confirmation of the deviation of a ray of Sun-grazing starlight
brought worldwide acclaim to Einstein (an authoritative discussion can be found in \cite{pais}). 
In 1911, again before the final version of GR was in hand, Einstein predicted that 
the deflection should amount to $2GM/R_{\odot}c^2=0.87$ arcsec, where
$R_{\odot}$ is the radius of the Sun. This result could not have been especially satisfactory
to Einstein, since it can be derived from Newtonian mechanics, if one assumes that
light is corpuscular, composed of particles with a mass equivalent $E/c^2$. 
The result is referred to as the  ``Newton value,'' and is one half of the 
revised value derived by Einstein in 1915 (1.74 arcsec). In 1919, the result 
of a measurement of the deflection that confirmed Einstein's prediction  was 
announced to the Prussian Academy, after which Einstein's reputation soared, both 
inside and outside academic circles, and the notions of warped space 
and time -- warped spacetime -- found their way into mainstream physics.  

The revolutionary implications of GR forced dramatic revisions of the 
concepts of space and time, not the least of which was the abandonment 
of the action-at-a-distance model of gravity. By contrast, the applications 
of GR in the early 1900s could be regarded as subtle corrections to Newtonian theory. 
However, discoveries of the 1960s --- the discovery of compact X-ray binary sources
\cite{giacconi}, the recognition of the nature of quasars \cite{schmidt}, and the
discovery of radio pulsars \cite{hewish} --- provided examples of environments for which general
relativity constituted substantially more than a gentle correction to Newtonian physics.  

The succession of rapid-fire astronomical discoveries of the 1960s 
marks the birth of relativistic astrophysics. Perhaps the most remarkable
idea to present itself during that era was that black holes could exist in nature. 
In \S2, a brief history of the black hole concept is presented. Here, we remark only that,
in 1916, only a few months after Einstein published 
his final version of GR, Karl Schwarzschild discovered
the solution to the field equations corresponding to the exterior of
a spherically symmetric mass distribution. This solution
is the starting point for the study of black holes. Prior to the 1960s, however, 
references in the scientific literature to what eventually came to be called black 
holes were sporadic and met with skepticism. 

Taken alone, GR does not prescribe a minimum black hole mass. However, those 
objects currently identified as black holes are believed to have formed from 
the catastrophic collapse of massive stars, and, combining GR with the equations 
of stellar structure, a minimum mass of about 3$M_{\odot}$ is required
\cite{chapline} \cite{ruffini}. According to GR, a classical theory,
matter undergoing gravitational collapse to form a black hole is crushed 
inexorably until it ends up in a point, known as the {\it singularity}. The 
singularity is enshrouded by a  mathematical surface known as the {\it event horizon}, 
within which matter and light are trapped, destined to merge with the singularity. 
The event horizon has a radius $2GM/c^2$ (about 3 km per solar mass), if the black hole is not
spinning. The event horizon can be as small as $GM/c^2$, in the case of a black hole spinning at its
maximum rate.  A black hole is a simple object, in the sense that spacetime outside its event horizon
can be described entirely by its mass, spin, and charge.  Therefore, black holes
belong to a three-parameter family -- absurdly simple compared to, say, main-sequence stars. And yet,
black holes so violate one's physical intuition that even Einstein doubted that they could 
exist in nature. In the words of Igor Novikov \cite{novikov}:
\begin{quote}
Of all conceptions of the human mind perhaps the most fantastic is the black hole.
Black holes are neither bodies nor radiation. They are clots of gravity.
\end{quote}

Outside the event horizon, the behavior of matter is dictated, in part, by the spacetime geometry.
If radiating matter of sufficient quantity exists near the event horizon of a black hole, such that
it is observable at Earth, then some of the effects of warped spacetime are observable, and we gain
experimental access to physical environments for which $GM/rc^2$ is not negligibly small. 
This regime is known as the {\it strong field limit}, 
where relativistic effects are of fundamental importance.  Nature has obliged us, 
by providing (at least) two classes of black hole systems -- black hole X-ray binaries
and active galactic nuclei (AGN) -- in which prodigious quantities of matter are flowing 
``onto'' the black hole, resulting in the release of copious electromagnetic radiation. 
In fact, detections of X-ray line radiation from regions where $GM/rc^2 \approx 1$ have been claimed.

The flow of material referred to above is called {\it accretion},
defined as the capture of matter by an object's gravitational field, 
where it is presumed that the fate of the captured material is coalescence 
with the gravitating body, or, in the black hole case, passage through the event horizon.
The physics of accretion has been an active area of research for over three decades.
For the purposes of this paper, the importance of accretion is tied primarily to the role it
plays in the conversion of gravitational potential energy into radiation.
Owing to the prevalence of angular momentum in the cosmos, accretion often involves
a disk. Accretion disks provide the means by which to dissipate angular momentum,
allowing accretion, which is accompanied by the release of energy.
The extraction of energy is especially efficient if the inner edge of the accretion disk extends
to small radii, so that the ratio $M/R_{\rm inner}$ is large. This is the case for
accretion onto compact objects -- white dwarfs and, especially, neutron stars and black holes. 
The luminosities of the accreting compact objects that populate the Milky Way Galaxy range
as high as $\sim 10^5$ $L_{\odot}$\footnote{The symbol $\odot$ refers to the standard
solar value of the subscripted quantity. Throughout the paper, we refer to the mass, radius,
and bolometric luminosity of the Sun -- $M_{\odot}=1.99 \times 10^{33}$ g; 
$R_{\odot}=6.96 \times 10^{10}$ cm;  $L_{\odot}=3.9 \times 10^{33}$ erg s$^{-1}$
\cite{allen}.}, while AGN can exceed that by seven, eight, even nine orders of 
magnitude \cite{fabian97}. 

Accretion disks present a number of theoretical challenges. 
The complexity of the problem, if one hopes eventually
to work  from first principles, is suggested by S. Shapiro and S. Teukolsky \cite{shapiro}: 
\begin{quote}
In the general accretion case, one must solve the time-dependent, multi-dimensional, 
relativistic, magneto-hydrodynamic equations with coupled radiative transfer!''
\end{quote} 
Given the context of the earlier parts of this section, the above quote begs the question
as to whether or not the strong field regime of GR can be tested through studies of accreting black holes.
Considering that accretion onto black holes is one of the dominant light-producing mechanisms 
in the Universe, developing reliable working models of disk accretion is, from the 
astrophysics standpoint, likely to be considered a higher priority than testing GR, 
although the latter is certainly a long-term goal. Moreover, it is probably fair 
to say that, {\it at present}, discerning possible inadequacies in strong field GR, 
a ``clean'' theory, using ``messy''  accretion disk models is problematic. In the meantime, recognizing
that GR has stood up to every test to which it has been subjected, the theory in full is generally 
taken as being correct, and incorporated into the array of theoretical tools 
used to study black hole accretion. 

Accretion-powered objects are bright X-ray sources \cite{giacconi} \cite{elvis}. 
The observed variability of the X-ray flux \cite{oda} \cite{vanderklis} \cite{kuneida}, 
rapid compared to  variability at longer wavelengths,
implies that X rays are produced preferentially in the inner regions of the disk.
\footnote{The time scale for appreciable variations in the observed flux is associated
with the light-crossing time of the source, so that the linear extent of the source
is given by $R \sim ct_{\rm cross}$.}
The X-ray continuum spectra vary from source to source, and can vary in time as well,
but for our purposes it is adequate to think of accreting black holes as exhibiting power-law
continua that extend up to hard X-ray energies ($\sim$ hundreds of keV) \cite{mushotzky} \cite{white84}. 
If an optically thick accretion disk surrounds the compact object, and if the
X-ray continuum flux irradiates the disk, then spectral signatures of the interaction
of the X-ray continuum and the disk are expected to imprint themselves
on the overall spectrum \cite{basko} \cite{guilbert} \cite{lightman}. This imprinting is
known as ``reflection.''
Among the expected reflection features are a blend of X-ray lines from iron lying in the
6.4-7.0 keV spectral range. These lines are produced as the result of $1s$ photoionization
of multi-electron iron ions. It was also shown that, if the iron lines are generated
very near a black hole, relativistic effects will skew the line, i.e., relativistic
Doppler effects would broaden the line to several $\times$ 10,000 km s$^{-1}$, and
the gravitational redshift would produce observed emission down to several hundred to
a few thousand eV below the rest energy (see Fig. 1), depending on the radial extent
of the disk and its inclination with respect to
the plane of the sky \cite{fabian89}. Precedence for the possibility of observing
relativistic effects in emission lines was set by observations of the elliptical
galaxy Arp 102B in the visible band, where the shapes of  
H$\alpha$ and H$\beta$ emission lines were analyzed 
in the context of a relativistic accretion disk model \cite{chen89}. Broad X-ray 
iron line emission was detected in the Galactic black hole X-ray sources Cyg X-1 
\cite{barr} and 4U1543-47 \cite{woerd}, possibly implying relativistic broadening, 
although the poor spectral resolution available at that time
made definitive conclusions problematic. 
Alternative mechanisms for producing broad lines were shown to be unlikely 
\cite{kallmanwhite}. The breakthrough observation was that of the Seyfert 1 
galaxy MCG--6-30-15, using the {\it ASCA} observatory, which exhibited evidence 
for highly-broadened and skewed iron line emission \cite{tanaka} \cite{nandra97}.

\begin{figure}[t]
\centering
\includegraphics[width=14cm,height=6cm]{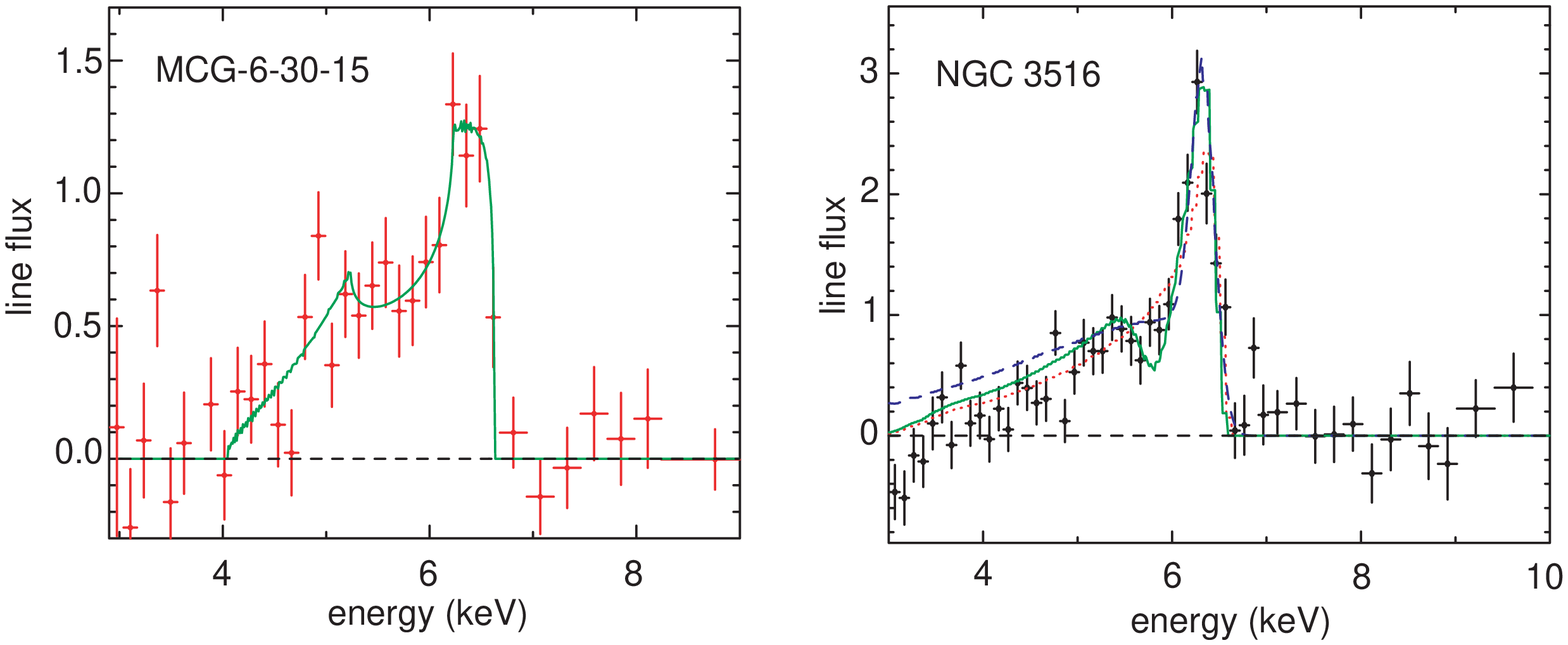}
\caption{Broad iron K$\alpha$ emission from two Seyfert 1 galaxies:
MCG--6-30-15 ({\it left panel}) \cite{tanaka} and NGC 3516 ({\it right panel})
\cite{nandra99} as recorded with the {\it ASCA} Solid-State Imaging
Spectrometer ({\it crosses}) following subtraction of a continuum model. 
The line flux is scaled in multiples of $10^{-4}$ photons cm$^{-2}$ s$^{-1}$ 
keV$^{-1}$. Line profiles consist of a narrow emission component
at 6.4 keV, the approximate rest energy of near-neutral iron ions,
and a broad redshifted wing, believed to result from relativistic
effects within a few gravitational radii of a supermassive black hole.
 Figures courtesy of K. Nandra (see \cite{nandra01}). }
\end{figure}

The discovery of skewed iron line emission has motivated a great deal of theoretical
research, and has intensified observational searches for more and better X-ray 
data of black hole systems. By the current paradigm of black hole accretion disks
\cite{mushotzky93}, the study of iron line emission from the inner orbits of
the accretion disk surrounding black holes provides the most direct view of the
region within a few gravitational radii of the central engine. 
Our intention is to provide a primer covering the three topics that
unite this phenomenon: (1) black holes; (2) accretion theory and modeling; and (3) 
X-ray spectroscopy and spectral modeling. Having here presented a quick
synopsis, in the following sections each is elaborated in turn. It is fair to note
that the relativistic disk interpretation is not unanimously accepted. Not surprisingly,
with far-reaching consequences at stake, alternative models for producing the broadened
line have been and are being investigated (e.g., \cite{elvis00}, \cite{titarchuk}). However,
a comparison of the successes and failures of the various models would take us too
far afield, and we focus here on the relativistic disk interpretation.  

We also omit from detailed discussion the subjects of aperiodic and
quasi-periodic variability in the X-ray emission from black hole systems, 
subjects that are extremely rich from a phenomenological standpoint, as evidenced
by data collected with the {\it Rossi X-ray Timing Explorer (RXTE)}. 
In black hole X-ray binary systems, the observed  time scales for variability are 
of the same order as the dynamical time scales expected for accretion flows near 
the black hole. Therefore, the temporal behavior of the X-ray emission
provides another powerful view of the relativistic dynamics associated with the inner
accretion disk. Although studies of the timing properties of black hole systems have, 
for the most part, constituted
a somewhat ``orthogonal'' approach relative to spectroscopic analysis,
there is a great deal of potential in analyzing the X-ray behavior
of accreting black holes in the spectral and temporal domains when
measured simultaneously. For reviews, see \cite{vanderklis} 
\cite{klis} \cite{mcclintock03}.

We have endeavored to provide the reader with a sufficient number of references 
to begin an exploration of the literature, but we make no claim of unbiased completeness. 
The collection of papers cited is a reflection of our own experience with the subject 
matter. We note that excellent reviews of this topic have recently appeared \cite{fabian00}
\cite{reynolds03}. We take a more tutorial approach in this paper, and hope that readers will find
our paper to be complementary, although there is some unavoidable overlap.  

The organization of the paper is as follows: In \S2 we briefly sketch the historical
development of the black hole concept, from Michell's dark star to Schwarzschild's
solution of Einstein's field equations to the golden age of theoretical black hole research.
Properties of the two size classes of black holes that are convincingly established -- stellar-sized
and supermassive -- are outlined in \S3. Also in that section we comment on the more
speculative intermediate-mass black holes, and the microquasar phenomenon.
A few of the core concepts associated with accretion power are treated in \S4,
including the $\alpha$-disk model and the concept of accretion efficiency.
In our treatment of general relativity, we forego a treatment of the theory proper,
and instead focus on a few of the applications relevant to line spectroscopy of 
accreting black holes, including the implications of both the Schwarzschild (\S5) 
and Kerr (\S6) solutions. These include the gravitational redshift, the nature of
circular orbits, and light motion. Vertical structure calculations of accretion disk
atmospheres and the basic atomic physics processes that give rise to iron K spectra 
are the subjects of \S7. Topics here include X-ray photoionization codes,
the X-ray fluorescence mechanism, the fluorescence yield, resonant Auger
destruction, and X-ray reflection. We conclude in \S8.

\section{History of the Black Hole Concept}

Given the association of black holes with Einstein's GR, it might be
natural to conclude that the notion of objects from which light
cannot escape was engendered by 20th Century physics.
However, allusions to the idea were already around  hundreds of years earlier. In 1676, Olaus
Roemer, noting  variations in the period of Jupiter's moon Io, discovered that the speed of light
is finite. In 1728, James Bradley, through observations of stellar aberration,
produced further confirmation and a more accurate value of the speed of light, 295,000
km s$^{-1}$. The concept of escape velocity, which, for a spherical mass
$M$ of radius $R$, is $v_{\rm esc}=(2GM/R)^{1/2}$, was present in Newton's time. In
1783, the English geologist John Michell combined these previous
pieces of information, and realized that it would be theoretically
possible for gravity to be so overwhelmingly strong that nothing
-- not even light -- could escape. Michell's statements were made in the
context of a corpuscular theory of light, the  in-vogue approach
at the time: 
\begin{quote}
Suppose the particles of light to be attracted
in the same manner as all other bodies...[then]
there should exist in nature bodies from which
light could not arrive at us.
\end{quote}
Michell went further, and proposed that from the motion of companion stars\footnote{
Michell is credited with the idea of binary stars.}
\begin{quote}
... we might still perhaps infer the
existence of the central objects with some degree of probability.
\end{quote}
Michell's objects were dubbed {\it dark stars}. Pierre Simon Laplace
published similar comments in his 1795 {\it Exposition du Systeme
du Monde}, and added that 
\begin{quote}
... it is therefore possible that the
greatest luminous bodies in the Universe are on this account
invisible.
\end{quote} 
Late 19th Century experiments fueled the
rise of quantum mechanics and the wave interpretation of light,
and the ideas of dark stars were soon forgotten.

The mathematical discovery that would lead to the 
modern concept of the black hole was introduced in 1916, as
the first solution to Einstein's field equations.
At the outbreak of World War I, in August 1914, Karl Schwarzschild
volunteered for military service. He served in Belgium as
commander of a weather station, in an artillery unit in France
calculating missile trajectories, and in Russia. During this time he
wrote two papers on Einstein's relativity theory (and one on
Planck's quantum theory, explaining  the splitting of the spectral
lines of hydrogen by an electric field -- the Stark effect). 
Einstein was later to say of Schwarzschild's work:
\begin{quote}I had not expected that one could formulate the exact solution of
the problem in such a simple way.
\end{quote}
Regrettably, while in Russia, Schwarzschild contracted a fatal disease
and went home to die at the age of 42, never believing in the
physical nature of what his solution implied for dark stars.
In fact, a back-of-the-envelope calculation shows that for an object to 
exist within a sphere of radius $2GM/c^2$, its mean
density must satisfy $\bar{\rho}>3c^6/32\pi G^3 M^2$, or $\bar{\rho} > 10^{16}$ g cm$^{-3}$ 
for a stellar-mass object, which must have seemed absurd at that time.

With the end of the First World War, further astronomical tests of
general relativity led to popular and scientific interest in Einstein's ideas. 
Most notable was an expedition led by
Arthur Eddington, the goal of which was to obtain a measurement of the bending of 
starlight by the Sun during a solar eclipse.
Meanwhile, research on stellar structure and evolution thrived during these years.
In 1930, S. Chandrasekhar computed the first white dwarf models,
taking into account special relativistic effects in the degenerate electron
equation of state, based on the Fermi-Dirac statistics introduced in 1926, and on
the ideas of W. Fowler regarding the equilibrium between electron
degeneracy pressure and gravity. Chandrasekhar discovered that no white dwarf
could sustain a mass larger than 1.4$M_\odot$. It was apparent
that if the electron degeneracy cannot withstand gravity, further
collapse is inevitable. 

In 1934, W. Baade and F. Zwicky predicted that this further
collapse would strip the atoms of their electrons, packing the
nuclei together, while forming a neutron star. These stars were
expected to be about 10 km in diameter, but with densities on
the order of a billion tons per cubic inch. The
neutron had been discovered by Chadwick only two years earlier.
In this same work, Baade and Zwicky coined the term
{\it supernova} and introduced its concept. In acknowledgement of one of the most
prescient papers ever written, even the comic strip of the {\it Los
Angeles Times} would comment on the issue. Zwicky would later say of this insert:
\begin{quote} This, in all modesty, I claim to be one of the most concise
predictions ever made in science. More than thirty years were to pass 
before the statement was proved to be true in every respect.
\end{quote}

In 1939, Oppenheimer and Volkoff \cite{volkoff}, and, a few months later, Oppenheimer and
Snyder \cite{snyder} realized that this inevitable stellar
collapse implies a stellar evolutionary scenario for the formation
of black holes. In the first paper, the first detailed
calculations of the structure of neutron stars were performed, establishing the
foundation of the general relativistic theory of stellar
structure. The second paper focused on the collapse of a homogeneous
sphere of a pressure-free fluid using general relativity.
Oppenheimer and Snyder's star was precisely spherical,
non-spinning, non-radiating, uniform in density, and with no
internal pressure. It was this set of assumptions, perhaps, that
gave rise to skepticism regarding the applicability of the
results to real imploding stars. The most impressive prediction in
the Oppenheimer and Snyder work was the fact that an external
observer -- one far from the star -- would see the implosion come
to a halt at the event horizon, whereas one riding with the falling
material would witness the whole process to the end, finding herself
drawn into a point of infinite density, notwithstanding the meager
hopes of surviving the trip until contact with the singularity. They concluded:

\begin{quote}
When all thermonuclear sources of energy are exhausted, a
sufficiently heavy star will collapse. Unless [something can
somehow] reduce the star's mass to the order of that of the sun,
this contraction will continue indefinitely.
\end{quote}

These discoveries in the field of stellar evolution directed 
attention to the Schwarzschild model of the spacetime 
exterior to a star. But yet again, a new world war 
would soon divert the efforts of most scientists towards military
and nuclear research. Work on black hole physics recommenced in the mid-1950s, when
D. Finkelstein discovered an alternative coordinate system for the Schwarzschild
geometry that helped to clarify the Oppenheimer and Snyder results
as being caused by time dilation in a gravitational field. J. A.
Wheeler, first an opponent of the black hole concept, began to
accept it, and tried to construct a quantum mechanical view of the
inner singularities. Finally, Kerr discovered the spinning black
hole solution to the Einstein field equations in 1963 \cite{kerr}. 

The term ``black hole'' was introduced in 1967 by Wheeler. 
In a 2003 interview with P. Davies, he explains his choice: 
\begin{quote}
The occasion was a meeting
in the fall of 1967 at the Institute of Space Studies in New York
to consider this marvelous work of Jocelyn Bell and Anthony
Hewish on the pulsars. What could be the cause of these absolutely
regular pulses from some object out in space, and one obvious
possibility was vibration of a white dwarf star, another was the
rotation of a neutron star. But then I thought that to keep one's
mind open, to look at all the possibilities one ought really to
look at the gravitationally completely collapsed object. Well, the
very words sounded so foggy, so ethereal, so far from touchability
that nobody resonated to that as a possibility to be investigated.
So in desperation I adopted the words Black Hole. Well, here at least was a name.
\end{quote}

The golden age of theoretical research in black hole physics started in
1964, and was to last for at least a decade.Ê In this decade,
computer codes used in hydrogen bomb research were
adapted to study stellar collapse, topological methods and
thermodynamical ideas were introduced into the study of black holes,
the ``no hair theorem'' was demonstrated (a black hole has no
characteristics indicative of the star from which it came, and
that only three parameters -- mass, angular momentum, and charge --
are needed to describe it), and the cosmic censorship conjecture
was stated (there are no naked singularities, but rather,
they are dressed with horizons), among many other results. 
After 1975, whereas important theoretical progress
was still being made, the field slowly started to be dominated by
an astronomical search for black holes in the Universe, at all
scales. By that time, there were ample astrophysical reasons to expect
objects described by solutions to Einstein's field equations to become part 
of the ontology of the Universe.

\section{Accretion Power}

Accretion is defined as the capture of matter by an object's gravitational field,
where it is presumed that the fate of the captured material is coalescence with
the gravitating body, the results of which are
an increase in the object's mass, as well as the conversion of gravitational potential energy
into other forms. The introduction of accretion flows into astrophysics is often traced to 
the 1939 paper by Hoyle and Lyttleton \cite{hoyle}. Interestingly, 
its title is ``The Effect of Interstellar Matter on Climatic Variations,'' which
examined the possibility that variable accretion of interstellar clouds by the Sun 
could be linked to the Ice Epochs of Earth's deep geological past. The notion of accreting
black holes lay more than two decades in the future. The astronomical discoveries of
the 1960s provided impetus for the rapid development of accretion theory,
currently an active area of astrophysical research.
We provide here a brief sketch of the theory as it pertains to our subject matter.
A collection of seminal papers on the topic can be found in \cite{treves}.
For a comprehensive textbook introduction to the physics of accretion, see \cite{frank}.
An up-to-date review article, with a pedagogical approach, can be found in \cite{blaes}.

\subsection{Basic Concepts}

To impart the fundamentals of accretion physics, we start with the Newtonian view of gravitation
and a neutron star onto which matter falls.
A test mass $m$ that free falls from rest at ``infinity'' and comes to rest on the surface
of a star of mass $M$ and radius $R$, loses an energy $GMm/R$, which, for the
purposes of this discussion, we assume to be converted into radiation.
Therefore, given a more-or-less steady infall of matter, the energy generation rate --- the accretion
luminosity --- is $L_{\rm acc}=GM\dot{M}_{\rm acc}/R$, where $\dot{M}_{\rm acc}$ is the mass
accretion rate, the dot denoting differentiation with respect to time. It is
sometimes convenient to express $L_{\rm acc}$ in a special relativistic context,
treating the energy release per unit time as being equivalent to
a conversion of rest mass energy into radiation per
unit time. Thus $L_{\rm acc}=\eta \dot{M}_{\rm acc}c^2$, where $\eta$ is the dimensionless {\it accretion
efficiency}. For the case of a spherical star, as discussed here, $\eta=GM/Rc^2$. 
The appearance of relativistic correctness implied by the adornment of 
these last two expressions with factors of $c^2$ is entirely artificial, since the
derivation of the accretion luminosity was obtained with purely Newtonian physics.
Later, however, we will see that expressing $L_{\rm acc}$ in terms of $\eta$ follows 
quite naturally in the general relativistic context, and we continue to use $\eta$ to
relate $L_{\rm acc}$ to $\dot{M}_{\rm acc}$.

Matter in an accretion flow is subject not just to gravitational forces, but
to radiation forces, as well. The radiation produced near the surface of the
accreting object exerts a pressure on the accretion flow. For radiation emerging from
a gravitating source in a frequency interval $d\nu$, the force on a particle at distance $r$, in 
terms of an interaction cross-section $\sigma(\nu)$, is
\begin{equation}
F_{\rm rad}=\int_0^{\infty} d\nu ~ \frac{L_{\nu}}{4\pi r^2 h\nu} \, \frac{h\nu}{c} \, \sigma(\nu),
\end{equation}
where $L_{\nu}/4\pi r^2 h\nu$ is the areal rate at which
photons in the frequency range $[\nu, \, \nu+d\nu]$ arrive at $r$, and $h\nu/c$ is the photon momentum.
Thomson scattering sets the baseline for this integral, so that, ignoring
relativistic corrections to the Thomson cross-section $\sigma_T$, the net force,
including gravity, is
\begin{equation}
F=\frac{\sigma_TL}{4\pi cr^2}-\frac{GMm_p}{r^2}
\end{equation}
assuming for simplicity a fully ionized hydrogen plasma ($m_p$ is the proton mass).
For accretion to proceed, we must have $F<0$, which imposes an upper limit
to the accretion luminosity, known as the {\it Eddington luminosity}, which is given by
\begin{equation}
\label{eq:ledd}
L_E=\frac{4\pi Gm_pcM}{\sigma_T}= 1.3 \times 10^{38}~\frac{M}{M_{\odot}}~~~{\rm erg~~s^{-1}}.
\end{equation}
By the same reasoning, there is an upper limit to the mass accretion rate. 
From the relation defining the proportionality between $L$ and $\dot{M}$, $L=\eta \dot{M}c^2$, 
we define the {\it Eddington accretion rate},
\begin{equation}
\label{eq:medd}
\dot{M}_E=\frac{4\pi Gm_pM}{\eta c\sigma_T}=
1.4 \times 10^{18}~\biggl(\frac{\eta}{0.1}\biggr)^{-1} \, \frac{M}{M_{\odot}}~~~{\rm g~~s^{-1}}.
\end{equation}
If a black hole accretes matter at the Eddington rate, then its growth rate
$dM_{\rm BH}/dt \propto M_{\rm BH}$, and the mass grows exponentially with time, with
an $e$-folding time $\tau_{\rm growth} =\eta c \sigma_T/4\pi Gm_p$ $\sim 5 \times
10^7$ yr, if $\eta=0.1$.

Perhaps the simplest model of mass transfer in an X-ray binary system, to the extent that
it is straightforward to obtain order-of-magnitude estimates, is that pertaining to a
{\it high-mass X-ray binary} (HMXB). A more detailed description can be found in \cite{davidsonost}.
In HMXBs, a high-mass (O or B) star, radiates a bright UV continuum that transfers 
momentum to the stellar atmosphere through  absorption in resonance lines, thus driving
a mass outflow, or a stellar wind. In close proximity to the OB star is a neutron star,
which captures part of the wind. In order to obtain numerical values of various quantities,
we assume here that the mass of the neutron star is 1.4 $M_{\odot}$, and that its radius is
$10^6$ cm, so that $\eta=0.21$. A simple estimate of the rate at 
which mass is captured by the neutron star can be
found by finding the distance $r_{\rm acc}$ at which 
the wind kinetic energy per unit mass, determined by the wind
velocity $v_{\rm w}$, is equal to the gravitational 
potential associated with the neutron star.
This gives the {\it accretion radius}, $r_{\rm acc}=2GM/v_{\rm w}^2$.
Since the neutron star accretes at the rate $\dot{M}_{\rm acc} = \pi r_{\rm acc}^2 \rho_{\rm w} v_{\rm w}$,
we can determine the accretion luminosity once we know the wind mass density $\rho_w$ 
at the position of the neutron star. This can be estimated by a mass continuity equation, assuming
that the wind is spherically symmetric with respect to the OB star: 
$\dot{M}_{\rm w}=4\pi a^2 \rho_{\rm w} v_{\rm
w}$, where $\dot{M}_{\rm w}$ is the total mass loss rate of the star, and $a$ is the separation between the centers
of mass of the components of the binary. Putting all this together, we find
\begin{equation}
\label{eq:accrate}
\dot{M}_{\rm acc} \sim \biggl( \frac{GM}{a v_{\rm {\rm w}}^2} \biggr)^2 
\, \dot{M}_{\rm w}=2 \times 10^{16}~
a_{12}^{-2} \, (v_{\rm w})_8^{-4} \, (\dot{M}_{\rm w})_{-6}~~~{\rm g ~~s^{-1}}.
\end{equation}
To obtain the numerical estimate, we have written the various quantities in terms of multiples of
typical values for those quantitities: $a_{12}$ is the binary separation expressed as
$a/(10^{12}$ cm); $(v_{\rm w})_8$ is the wind velocity expressed as $v/(10^8 ~\rm{ cm ~s^{-1}})$; 
$(\dot{M}_{\rm w})_{-6}$ is the mass loss rate expressed as 
$\dot{M_{\rm w}}/(10^{-6}M_{\odot}~ {\rm yr^{-1}})$.

In obtaining an estimate of the mass accretion rate in a typical HMXB, there is no reason 
we could not have substituted a black hole for the neutron star. But let us now 
proceed to determine the resulting luminosity and radiation temperature.
Using $\eta=0.21$, derived above for the 
neutron star case, the rate of mass accretion found in Eq. (\ref{eq:accrate}) results in a luminosity 
$L_{\rm acc} \sim 10^{36}$ erg s$^{-1}$, or approximately $10^3$ $L_{\odot}$. 
An order-of-magnitude estimate (actually a lower bound) of the radiation temperature 
associated with the accretion luminosity can be obtained
by assuming that the radiant energy takes the form of a blackbody. If the
energy is released uniformly over a spherical surface of radius $R$, which we assume
to coincide with the neutron star surface, then
$L_{\rm acc}=4\pi R^2 \sigma T_{\rm bb}^4$, where $\sigma$ is the Stefan-Boltzmann constant.
Thus we find $T_{\rm bb}=(GM\dot{M}_{\rm acc}/4\pi\sigma  R^3)^{1/4}$, or, again in terms
of typical system parameters,
\begin{equation}
T_{\rm bb}=3.2 \times 10^7 ~(v_{\rm w})_8^{-1}  \,a_{12}^{-1/2} \,(\dot{M}_{\rm w})_{-6}^{1/4}~~~{\rm K},
\end{equation}
which implies that the emission peaks in the X-ray band.

The simple model described above is consistent with observations of typical 
HMXBs. Is this model, on the other hand, relevant to black hole systems? 
For example, is it valid to simply evaluate $\eta$ for a black hole by 
inserting, say, $R=2GM/c^2$ into
$\eta=GM/Rc^2$, i.e., $\eta_{\rm BH}=0.5$? Clearly, 
the answer must be ``no,'' since the calculation of $\eta$ for a
neutron star is based upon the assumption that the matter 
comes to rest on a material surface. In the cold fluid
approximation, as used above, we might expect all of the 
energy release  in black hole accretion to occur below the
event horizon, leaving $\eta=0$. In a more realistic treatment, 
one assumes a finite gas temperature, and includes
thermal energy balance in the calculations. Some fraction 
of the accretion energy is radiated away before crossing
the event horizon. Such a scenario applies to accretion 
of the interstellar medium by an isolated black hole.
Detailed calculations \cite{shapiro73}  show, however, that 
for an ambient ISM density and
temperature of 1 cm$^{-3}$ and $10^4$ K, respectively, the efficiency $\eta \sim 10^{-10}$, and the
expected luminosity is expected to be only $\sim 10^{21}$ erg s$^{-1}$ per solar mass of the accretor, 
far too dim to be observed (cf., \cite{shvartsman} \cite{meszaros}). The inefficiency of spherical
accretion by a black hole is borne out by the realization  that, in spite of the large number of
black holes postulated to populate the Milky Way Galaxy, only the  merest fraction (perhaps 1 in
$10^7$) has been identified.\footnote {By ``identified,'' we mean ``inferred.''  The designation
{\it black hole candidate} is applied when alternative explanations do not suffice.}

In acknowledging that radial accretion onto a black hole is unlikely to produce a light source
sufficiently bright that it is observable at Earth, we turn instead to accretion via a disk. We will
see that disk accretion can be an extraordinarily efficient means by which to extract energy from
matter near a black hole.

\subsection{Accretion Disks}

Angular momentum favors the formation of disks around accreting masses.
The surfaces of constant gravitational potential between two stars in
a binary system are such that a saddle point forms on the line of centers
(the imaginary line connecting the centers of mass) between the two stars.
The equipotential surface that intersects this saddle point is called the {\it Roche lobe}.
If matter from one star reaches this saddle point, known as the inner Lagrangian
point, or L1, then it can accrete onto the other star. Since there is relative
motion between the stars, however, the overflowing matter cannot strike the 
accreting star directly, but rather goes into an orbit \cite{lubow} dictated 
by its specific angular momentum. If sufficiently large quantities of matter
spill through L1, however, these orbits cannot persist. Instead, collisions
result in loss of energy and circularization of the orbits. Viscosity spreads
the disk, transporting angular momentum outward and matter inward. This model
describes the mass transfer mechanism in X-ray binaries, a pair of stars in close orbit,
one of which is a gravitationally collapsed object --- a neutron star or a black hole.
The mass donor star is known as the companion. The X-ray binary phase of a
Roche lobe overflow system is initiated when the companion expands to fill its Roche lobe, a
consequence of  normal stellar evolution \cite{verbunt}. The accretion of matter, if the
overflow rate is sufficiently large, can power a highly luminous X-ray source. In this case,
however, as contrasted to the HMXB case, the disk itself radiates much of the luminosity,
irrespective of the presence of a material surface at the gravitating center. 

Disk accretion by Roche lobe overflow was first invoked 
to explain cataclysmic variable systems \cite{kraft},
in which a normal star transfers mass to a white dwarf 
($R_{\rm wd} \sim 10^9$ cm) in a close binary system.
The fundamental theoretical aspects of accretion disks
in X-ray binaries were worked out by Pringle and Rees \cite{pringle}
and by Shakura and Sunyaev \cite{ss73}. The latter authors derived
an analytic model of disk flow, where the viscously dissipated energy is assumed to be
locally radiated as blackbody emission, with temperature given by
\begin{equation}
\label{eq:tdisk}
T_{\rm bb}(r)=\biggl[\frac{3GM\dot{M}}{8\pi \sigma r^3}~
(1-\sqrt{R_*/r}) \biggr]^{1/4},
\end{equation}
where $R_*$ is the radius of the accretor, or, in the black hole case,
$R_*$ is the innermost stable circular orbit of disk material (see \S6).
Equation (\ref{eq:tdisk}) is similar to the expression derived for HMXBs,
showing that, for similar parameters, X-ray-emitting temperatures are expected.
The energy dissipated per unit area is
\begin{equation}
\label{eq:ldisk}
Q(r)=\frac{3GM\dot{M}}{8\pi r^3}~(1-\sqrt{R_*/r}),
\end{equation}
from which Eq. (\ref{eq:tdisk}) follows. The disk luminosity $L_{\rm disk}$ can be
found by integrating $Q(r)$ over both disk faces, from $R_*$ to some
outer radius $R_{\rm outer}$. A good approximation for $L_{\rm disk}$ is
found by letting  $R_{\rm outer} \rightarrow \infty$, yielding
the result $L_{\rm disk}=GM\dot{M}/2R_*$ \cite{frank}. If the accretor has
a solid surface, the remainder of the energy is dissipated as it comes to rest on the surface.
If the accretor is a black hole, then the remainder simply disappears behind the event horizon.
In this latter case, if the black hole does not spin, the inner disk radius is
taken to be $6GM/c^2$ (see \S5), which gives $L_{\rm acc}=\dot{M}c^2/12$. i.e., $\eta \approx 0.083$. 
For a black hole spinning at its theoretical maximum rate, we set $R_*=GM/c^2$, which gives
$\eta=0.5$. In \S5 and \S6, we derive relativistically correct values for the accretion efficiencies
of non-spinning and spinning black holes, respectively.

In terms of accretion disk dynamics, the essential breakthrough introduced by Shakura
and Sunyaev \cite{ss73} was the $\alpha$-prescription, a parameterization of
the vertically-averaged stress that assumes it can be written as $\alpha P$, where $P$ is the
vertically-averaged total pressure in the gas (gas pressure plus radiation
pressure), and where $\alpha$ is a constant. The
$\alpha$-prescription permits the solution of the coupled set of eight equations that determine
the disk structure. Although the $\alpha$-prescription sidesteps a good bit of  
complex physics, it turns out that many results are fairly insensitive to the 
value of $\alpha$.  For example, the disk central temperature, i.e., $T(r,z=0)$,
is proportional to $\alpha^{-1/5}$, whereas, Eqs. (\ref{eq:tdisk}) and (\ref{eq:ldisk})
show that some results of interest have no dependence at all on $\alpha$. 
For this reason, the $\alpha$-disk has seen wide use.
Indeed, the optically thick $\alpha$-disk model has been accepted with only a few 
modifications since its inception in 1973 (\cite{abramowicz88}
\cite{chen93}; \cite{luo}). Among these modifications are the development of various branches of
self-consistent accretion flow solutions such as advection-dominated accretion flows
\cite{ichi} \cite{narayan95} that successfully describe the spectral behavior of
black hole XRBs \cite{lewin} \cite{tanaka96}, advection-dominated inflow/outflow 
systems \cite{blandford99}, and convection-dominated accretion flows \cite{quataert}.
Comparisons of the $\alpha$-disk model with observations of X-ray binaries and  AGN are presented in
\cite{frank} and \cite{malkan} \cite{krolik98}, respectively. Modern accretion flow 
models are discussed in \cite{chen2}.

In spite of the successes and ease of use of the $\alpha$-disk model, a
physical understanding of $\alpha$ (or something like it) has been an important goal
for accretion disk modelers, since it determines the essential mechanism that
drives the transport of angular momentum. 
The viscous mechanism in accretion disks allows accretion to occur by
transporting angular momentum outward, matter inwards, and by
dissipating gravitational energy into heat inside the disk. The viscosity
mechanism has been identified in theory as a magneto-rotational
instability arising from the entanglement of magnetic fields caused by
differential rotation of gas in Keplerian orbits \cite{balbus}. 
In MHD models, this mechanism provides the energy
dissipation and angular momentum transfer needed to naturally 
produce mass accretion with self-sustained magnetic fields $B$ that
are much smaller than the equipartition level ($B^2/8\pi \ll \rho v^2$),
where $\rho$ is the gas mass density and $v$ is the thermal velocity.
Numerical magnetohydrodynamic (MHD) models \cite{gammie} show that
$\alpha$ is not constant, but that it ranges from $\sim 10^{-3}$--$10^{-1}$. 

While the X-ray continuum in accreting neutron star systems is interpreted, in part, as thermal emission from
the disk, the presence of (apparently) non-thermal continuum radiation in these sources, and AGN, 
as well, is not naturally explained with the $\alpha$-disk formalism.  In fact, the high-energy emission is
especially problematic in AGN, if we consider Eq. (\ref{eq:tdisk}). Anticipating a result that
is derived in \S6, set $R_*=6GM/c^2$, which corrresponds to the inner disk edge in a non-spinning
black hole, and express the mass accretion rate as a fraction $f_E$ of the Eddington
accretion rate, $\dot{M}=f_E \dot{M}_E$.  Finally, if we express
$r$ as a multiple of the gravitational radius
$GM/c^2$, so that $x=c^2r/GM$, then Eq. (\ref{eq:tdisk}) can be written as
\begin{equation}
\label{eq:tdisk2}
T_{\rm bb}(x)=\biggl[\frac{3m_p c^5 f_E}{2\eta \sigma \sigma_T GM}~
\frac{(1-\sqrt{6/x})}{x^3} \biggr]^{1/4},
\end{equation}
The key result to take away from this expression
is the scaling $T_{\rm bb}(x) \propto M^{-1/4}$. Thus if we accept the $\alpha$-disk model
for stellar-sized accreting black holes ($M_{\rm BH} \sim 10 M_{\odot}$), 
we are faced with a glaring inconsistency if
we wish also to apply it to AGN ($M_{\rm BH} \sim 10^7 M_{\odot}$); all other things being equal, 
$T_{\rm bb}^{\rm agn}/T_{\rm bb}^{\rm xrb} = (M_{\rm BH}^{\rm xrb}/M_{\rm BH}^{\rm agn})^{1/4}$,
which means that a $10^7$ K blackbody for an X-ray binary is scaled down to about $3 \times 10^5$ K
for an AGN disk. The AGN emission is thus expected to peak in the UV band, with very little
luminosity appearing in the hard X-ray region. While the UV peak, the ``big
blue bump,'' is a well known component of AGN spectra \cite{shields}
\cite{sargent}, the $\alpha$-disk model says nothing
about the X-ray flux that characterizes AGN \cite{elvis94}, and refinement is required in order to
match disk models to observations. For example, a more realistic accounting of
radiation transport effects in the disk atmosphere partially ameliorates
this problem \cite{czernyelvis} \cite{white89} \cite{ross92}.

The favored explanation of the X-ray luminosity of accreting black holes
posits that the blackbody component originates in a cold, optically thick accretion disk,
whereas the hard X-ray power-law component is produced in an optically thin
hot corona by thermal {\it Comptonization} of disk photons \cite{maraschi} \cite{poutanen}.
Comptonization refers to the deformation of a radiation field as it interacts
through Compton scattering with an electron distribution, where a self-consistent
solution determines both the spectral shape and the electron temperature
\cite{kompaneets} \cite{rybicki} \cite{peebles}.
Constructing theoretical models of continuum production is complicated by the 
need to include $e^+$- $e^-$ pair production in a manner that is self-consistent 
with the radiation field \cite{kogan} \cite{zdziarski} \cite{svennson} \cite{ghisellini}. 

In disk accreting systems where a hard X-ray source is present, the
disk is exposed to this radiation and will be heated by it. In fact, 
radiative heating can exceed internal viscous heating in some regions of the
disk. The temperature structure of the disk can thus be controlled by the X-ray field,
photoionizing the gas, suppressing convection,  and increasing the scale
height of the disk \cite{mario01}. Photoionization in the disk is balanced by
radiative and dielectronic recombination \cite{liedahl1}, and possibly three-body recombination
\cite{bautista} and charge transfer recombination \cite{kingdon}.
These recombination processes produce discrete line emission, which constitute
spectral components distinct from the fluorescence component of the spectrum, potentially
providing corollary information relating to the disk structure.
A quantity that is commonly used to describe emission lines that are superimposed
on a continuum is the equivalent width. Since we make several references to the equivalent
width in later sections, we define it here for convenience.
The experimentally measured line equivalent width is defined by
\begin{equation}
\label{eq:ew}
W_{\epsilon}=\int_0^{\infty} d\epsilon ~
 \frac{(F_{\epsilon})_{\rm line}}{(F_{\epsilon})_{\rm cont}}.
\end{equation} 
The equivalent width is usually quoted in units of eV or keV. 
Theoretical predictions of $W_{\epsilon}$ follow from calculations of accretion disk models,
which can then be compared to observations. For example, early model calculations of the
expected iron K$\alpha$ equivalent width, based on the response of an X-ray irradiated
slab, showed that 90-150 eV was attainable \cite{george91}. 

The production of line emission, regardless of the formation mechanism, 
and the transfer of these lines through the overlying atmosphere requires, 
in principle, a detailed calculation
of the opacity distribution along their lines of flight. The spatial distribution
of the line emissivity, as well as a proper accounting of the probability of escape must, 
therefore, account for the vertical structure of the disk. This includes, in 
the limit that the vertical structure is
in steady-state, the density, temperature, and charge state distribution. Thus each annulus of the disk
can be thought of as a stellar atmosphere, where the equations of energy
flow must be solved self-consistently with the ionization equations, possibly constrained
by hydrostatic equilibrium. The explicit vertical structure of the disk, however, does
not follow directly from the $\alpha$-disk model. For example, the
surface density -- the line integral of the density in the vertical direction -- appears
as one of the variables in the disk equations. We return to the subject of the 
vertical structure of accretion disks as it bears on X-ray fluorescence in \S7.

\section{Astrophysical Black Holes}

In this section, we present a brief survey of black holes from an astronomical point of view,
touching on the gross characteristics of black holes in X-ray binaries and AGN.

\subsection{Black Hole X-ray Binaries}

Currently about 250 X-ray binaries are known in our galaxy \cite{liu1} \cite{liu2},
possibly representing an underlying population of more than 1000 objects. 
X-ray binaries can be separated into two populations: the {\it low-mass X-ray 
binaries} (LMXBs), where ``low-mass'' refers to the companion star to the compact 
object, an older population, concentrated near the Galactic bulge; and the 
HMXBs (see \S3.1), younger systems, concentrated in the spiral
arms. Mass transfer from the companion to the compact object is
found to typically differ for these two classes as well. In
general, LMXBs transfer mass to  their companions through Roche lobe overflow (Fig. 2). 
Mass transfer in HMXBs is generally mediated by stellar winds.

A subset of X-ray binaries, twenty or so, are identified as black holes paired with a star
that serves as a mass donor for accretion \cite{lewin} \cite{charles}.  The most famous 
example of this class is Cygnus X-1, identified in 1972 as a black hole candidate \cite{bolton}.
A more common subclass of X-ray binaries are those for which the compact object is a neutron star,
rather than a black hole. When the compact object in an X-ray binary system is shown to be
more massive than about $3M_{\odot}$, the compact object is a good black hole candidate. For
a small group of X-ray binaries, the mass measurement has been performed
with high precision. This group of X-ray sources is usually
referred to as {\it dynamically confirmed black holes}. 
\footnote{The relevant experimentally inferred quantity is the mass function,
defined by $f(M)~\equiv~P_{\rm orb}K_{2}^{3}/2\pi G$
=$M_{1} \sin^3i/(1+q)^{2}$ where $P_{\rm orb}$ is the
orbital period and  $K_{2}$ is the semi--amplitude of the velocity curve
of the secondary, $M_{1}$ is the black hole mass, $i$ is the
orbital inclination angle, and $q~\equiv~M_{2}/M_{1}$, where
$M_{2}$ is the mass of the secondary, is the mass ratio. The
equation for $f(M)$ implies that the value of the mass function is
the absolute minimum mass of the compact primary. A secure
value of $f(M)$ may be sufficient to show that the mass of the
compact X--ray source is at least 3$M_\odot$. } A
continually updated list of such objects can be found, thanks to J.
Orosz\footnote{http://mintaka.sdsu.edu/faculty/orosz/web/} (see
\cite{mcclintock03} for a more detailed discussion).
Among the dynamically confirmed black hole X-ray binaries, there are several recurrent X-ray
novae, all of them LMXBs, and only three
persistent sources -- Cyg X-1, LMC X-1, and LMC X-3 -- the latter of which
are HMXBs. These special binaries span a large range in the parameter space
of basic properties; for example, XTE J1118+480 has $P_{\rm orb} = 0.17$ days and a
binary separation of $\approx 2.8 R_\odot$, whereas GRS 1915+105 has $P_{\rm orb} = 33.5$ days and
a binary separation $\approx95 R_\odot$. 

Clearly, if a less strict standard of evidence is accepted, many
more black hole candidates can be identified (see, e.g., \cite{mcclintock03} 
for a recent discussion). Still, given the small number of black hole candidates, 
it may be surprising that, from stellar evolutionary considerations, together with the 
incidence of the  X-ray binaries that are believed to contain black holes, it is thought that 
the Milky Way Galaxy contains about $\sim$300 million stellar-mass black holes 
\cite{heuvel} \cite{brown} \cite{timmes} \cite{agol}.  This number implies, 
assuming $\sim10 M_\odot$ per black hole \cite{mcclintock03},  that about 
4\% of the total baryonic mass (i.e., stars plus gas) of the Galaxy is in the form of black holes. 

The X-ray spectra of X-ray binaries are dominated by continuum
radiation. The X-ray luminosities of persistent
XRBs are intrinsically variable, and range anywhere from (typically) $L_x \sim 10^{36}
~\rm{erg~s}^{-1}$ to as high as the Eddington luminosity
of $\rm{few} \times 10^{38} ~\rm{erg~s}^{-1}$.  In black hole X-ray novae, the
range in X-ray luminosities is more dynamic, with quiescent luminosities
as low as $L_x \sim 10^{30} ~\rm{erg~s}^{-1}$ \cite{garcia}. 
Based upon spectral observations in the soft X-ray band (say, 1--10 keV),
two main categories of emission states have been distinguished, depending 
mostly on the slope of the
power-law ($F_E \propto E^{-\alpha}$) describing the continuum emission of the system. Here, the
differential photon count rate is given in units of photons 
cm$^{-2}$ s$^{-1}$ keV$^{-1}$.
The so-called low state (relatively low photon count rate) 
features non-thermal X-ray flux, typically with
slopes near 1.6 -- 1.7. Because of the hardness (i.e., relatively high ratio
of hard X-ray flux to soft X-ray flux) of such indices,
this state is also referred to as the low/hard state. The high state,
by contrast, is characterized by intense quasi-thermal flux. In this
state, most of the radiated energy is concentrated in a blackbody
component, while the power-law contribution becomes softer, with
power-law indices typically larger than 2.
The usual interpretation is that the blackbody
component originates in a cold, optically thick accretion disk,
whereas the power-law component is produced in an optically thin
hot corona by thermal Comptonization of disk photons \cite{poutanen} 
\cite{dove} \cite{esin97} \cite{esin98}. 

The first broad Fe K$\alpha$ line observed was reported in the
spectrum of Cyg X-1, based upon {\it EXOSAT} data \cite{barr}. 
Recently, the {\it Chandra} X-ray observatory was used to observe Cyg X-1 with
the High Energy Transmission Grating Spectrometer in an intermediate X-ray
state \cite{jmiller02}. A narrow Fe line was detected at
$E=6.415\pm 0.007$ keV with an equivalent width (see Eq. \ref{eq:ew}) of
$W_{\epsilon}=16^{+3}_{-2}$ eV, along with a broad line at $E=5.82\pm 0.07$
keV with $W_{\epsilon}=140^{+70}_{-40}$ eV. A smeared photoelectric edge was also detected
at $7.3\pm 0.2$ keV. These results are interpreted
in terms of an accretion disk with irradiation of the inner disk
producing the broad Fe K$\alpha$ emission line and edge, and
irradiation of the outer disk producing the narrow line. The broad
line is thought to be shaped by Doppler and gravitational effects and, to a
lesser extent, by Compton reflection (see \S7).

\begin{figure}[t]
\centering
\includegraphics[width=12cm,height=9cm]{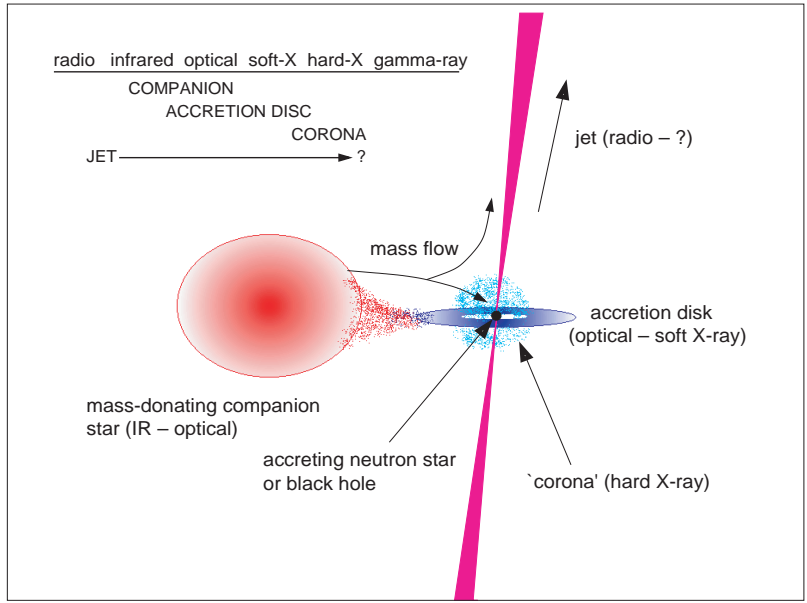}
\caption{Current basic understanding of the physical components and sites of 
emission in an X-ray binary system. In this example, the companion loses mass to 
the compact object through Roche lobe overflow. The accreting matter forms a disk, which
can be thought of as a series of annular rings in quasi-Keplerian orbits, whose radii
decrease under the action of stresses between neighboring annuli.  Radiant energy is released,
giving rise to X-ray emission at small radii. Hard X-ray emission impinges on the
optically thick disk, which ``reflects'' a fraction of the incident flux, with spectral
imprinting in the form of lines, absorption, and deformed continuum. A fraction of the accretion
energy is redirected into kinetic energy, giving rise to the jets. A hot corona sandwiches the disk
at small radii, which can affect the reflection spectrum through Compton scattering.
From Fender and Maccarone \cite{fender03}. }
\end{figure}

Early studies of relativistically smeared Fe K$\alpha$ lines from
X-ray binaries relied on proportional counter detectors with relatively
poor spectral resolution (e.g., the
{\it Ginga}'s LAC and the {\it RXTE}'s PCA, have an energy
resolution of $\approx~1.2$~keV at Fe K$\alpha$). The response
matrices of the detectors are uncertain at the 1--2 \% level,
while the Fe line profile is typically only 1--5 \% above the
X-ray continuum \cite{mcclintock03}. Therefore, interpretations of
results from these instruments must be approached with caution.
{\it BeppoSAX}, with a resolution of $\approx~0.6~$keV at 6.4 keV,
has also been used to observe several black hole candidates.  The
observed iron emission generally appear to be rather symmetric, and may be
more a product of Compton scattering than relativistic broadening.
However, in the cases of  GRS~1915+105 \cite{martocchia}
and V4641~Sgr \cite{jmiller02a}, the iron lines are skewed, possibly
implying relativistic smearing. In addition,  an observation using the
{\it XMM} EPIC--MOS1 detector led to the report of a
broad, skewed Fe K$\alpha$ emission in XTE~J1650--500, perhaps
suggesting the presence of a rapidly spinning black hole \cite{jmiller02b}.

\subsection{Active Galactic Nuclei}

Approximately one in a hundred galaxies shows evidence for
energy output that appears to be unrelated to normal stellar processes -- rapid bulk
motions, a bright, non-stellar radiation continuum, rapid aperiodic variability,
and high-luminosity  emission lines originating near the galactic nucleus. 
Members of this subset of galaxies are known as {\it active galaxies}, and the nuclei
are referred to as AGN \cite{krolikbook}.
The classification scheme for active galaxies is somewhat daunting.
We list here the various classes as they have been designated in the
literature \cite{woltjer}: (1) radio galaxies; (2) radio quasars; (3)
BL Lac objects; (4) optically violent variables; (5) radio quiet quasars; 
(6) Seyfert 1 galaxies; (7) Seyfert 2 galaxies; (8) low-ionization nuclear
emission-line regions (LINERs); (9) nuclear H II regions; (10) starburst
galaxies; and (11) strong IRAS galaxies. Not all among this group
are active by virtue of black hole accretion. For example, starburst
galaxies are sites of anomalously large star formation rates, probably
the result of a gravitational encounter with another galaxy, leading
to an enhanced IR luminosity. We do not delve here into the phenomenology
that gives rise to this complex taxonomy. In what follows, we reserve the term
AGN for active galaxies believed to harbor a supermassive black hole.

Within the unification scheme \cite{antonucci}, the
underlying model for all classes of AGN is intrinsically similar.
At the very center of the galaxy sits a {\it supermassive black hole}
($\sim10^6$ -- 10$^{10}\,M_{\odot}$), which accretes
galactic matter through an accretion disk.
Broad emission lines are produced in clouds orbiting above the
disc at high velocity (the Broad Line Region \cite{netzer}), and this
central region is surrounded by an extended, dusty, molecular
torus. A hot electron corona sandwiches the inner regions of the disk, probably
playing a dominant role in generating continuum X-ray emission. 
Two-sided jets of relativistic particles emanate
perpendicular to the plane of the accretion disc, the generation
of which is still not fully understood. 

As pointed out in \cite{mushotzky02}, X-ray surveys are an
excellent means by which to locate AGN. Surveys using the
{\it Chandra} and {\it XMM-Newton} observatories find
a remarkable $\sim10^3$ X-ray sources per square degree, the vast
majority of which are undoubtedly AGN \cite{mushotzky00} \cite{hasinger}.
To date, however, the most secure detections of
supermassive black holes (accompanied by a ``rule out'' of
alternative models, such as the existence of dense clusters of
stars or exotic particles) come from stellar proper motion in the
Galactic center and the H$_2$O megamaser of the nearby
Seyfert 2 galaxy NGC 4258 \cite{miyoshi} (see \cite{ferrarese} and
references therein for further discussion). Optical stellar and gas
dynamical studies, generally using the Hubble Space Telescope, have
revealed a large concentration of nuclear mass in several tens of
candidates, believed to be black holes. It is true, however, that
these methods currently lack sufficient angular resolution to
probe the spacetime at distances on the scale of the black hole horizon (e.g., see
\cite{magorri}). 

It is worth noting that nuclear activity is not a prerequisite
for the presence of a black hole. In fact, most of the galaxies probed
for supermassive black holes are not really active, but dormant
quasars, relatively close to Earth, and for which the Keplerian signatures
could be more easily discovered because of the higher spatial resolution \cite{kormendy}.
The most famous example is provided by our own Galaxy \cite{melia}.
After charting the kinematics of stars swirling around the
central regions of the Milky Way \cite{schodel} \cite{ghez1} \cite{ghez2}, it has been
found that the total mass of the region enclosed within a radius of
$2 \times 10^{15}$ cm is approximately $ 3.7 \times 10^6$
$M_\odot$, far more compact than what is possible for a stable
distribution of individual objects; total gravitational collapse is required,
according to GR. Indeed, the only remaining alternative candidate, other 
than a black hole, to describe the behavior of the innermost stellar orbits in the center
of the Milky Way Galaxy comes from particle physics, and is known as a boson
star \cite{diego00}. Mass estimates for central supermassive  black holes in about twenty
nearby galaxies are also available (see review in \cite{richstone}). Therefore, if we assume
that the Milky Way Galaxy is not unique in this respect, it is possible that a large fraction of 
galaxies, active or not, harbor supermassive black holes near their dynamical centers.

Currently, the Seyfert 1 galaxies MCG--6-30-15 \cite{wilms} \cite{vaughan},
Mrk 766 \cite{mason}, and NGC 3516 \cite{nandra99} provide the most robust 
detections of the relativistically broadened iron  K$\alpha$ line. 
In addition to the preceding papers, a concise set of case studies 
of these three objects is provided in \cite{reynolds01}.

\subsection{Microquasars and Jets}

Some X-ray binaries are sources of jets, high-velocity streams of oppositely
directed particles originating near the compact object. 
These sources are called {\it microquasars}
\cite{mirabel1} \cite{mirabel2}, exploiting an analogy discussed in
more detail below. Figure 2, from \cite{fender03},
is a sketch of an X-ray binary, presenting the major physical
components and sites of emission in such systems. The basic
underlying idea behind the analogy between quasars and
microquasars is that the physics in all black hole systems is
essentially  governed by  scaling laws, whose order parameter is
the black hole mass. For instance, the scales of length and time
of black hole related phenomena are proportional to the mass of
the black hole. For a given critical accretion rate, the
bolometric luminosity and length of relativistic jets are also
proportional to the mass of the black hole. For a black hole of
mass $M$ the density and mean temperature in the accretion flow
scale with $M^{-1}$ and $M^{-1/4}$, respectively.  The maximum
magnetic field at a given radius in a radiation dominated
accretion disk scales with $M^{-1/2}$, which implies that in the
vicinity of stellar-mass black holes the magnetic fields may be
10$^4$ times stronger than that found near supermassive black
holes \cite{sams}.

One of the most impressive phenomena occurring in both
quasars and microquasars is the ejection of blobs of plasma at
apparently superluminal speed \cite{rees66}. Very Long Baseline Interferometry
at radio wavelengths allows position measurements down to milli-arcseconds,
thereby permitting detections of small changes in position on the sky.
The apparent velocity is obtained by multiplying the observed
proper motion by the distance to the source. Such superluminal sources were
discovered as radio-galaxies and quasars, where the central black
hole is supposed to have millions of solar masses. It was found
that this very same phenomenon could also
occur in Galactic sources, after the report on GRS 1915+105 by
Mirabel and Rodriguez \cite{mirabel1}.

\subsection{Intermediate Mass Black Holes}

Although consensus has not yet been reached,
there are two types of data suggesting the existence of
intermediate mass black holes, with masses between $\sim 20 M_\odot$ and
several thousand $M_\odot$ \cite{cmiller04}. First, there are numerous X-ray point sources,
dubbed {\it ultraluminous X-ray sources}, that are not associated
with AGN, and that have fluxes far beyond the
Eddington limit of a stellar size black hole system. Second,
several globular clusters show clear evidence for an excess of
dark mass in their cores, which appears to be a single object. The
number of such intermediate objects, as well as their actual existence, is
yet under debate, and strongly depends on the mechanism that forms
them. For both intermediate and supermassive black holes, the
formation processes are not as well understood as they are for their
stellar counterparts.

\section{Black Hole Accretion in the Schwarzschild Metric}

In this section we present in some detail a few of the results of general
relativity that pertain to black hole accretion. As mentioned earlier,
there are only three vacuum solutions for black holes. The simplest -- the
Schwarzschild metric -- describes the geometry of spacetime outside the
event horizon of a black hole with zero charge and angular momentum;
mass alone describes the geometry. A clear and concise derivation of the
Schwarzschild solution can be found in, for example, \cite{weinberg}. 
Before proceeding to the Schwarzschild solution, we introduce the unit 
conventions that have been adopted for dealing economically with 
calculations involving relativity.

\subsection{Geometrized Units}

In the domain of special relativity, space and time are linked by the invariant
spacetime interval.
By convention, we can rewrite the familiar expression for the proper time interval in
flat spacetime $d\tau^2=c^2 \, dt^2 -d{\bf x}^2$ as $d\tau^2= dt^2 -d{\bf x}^2$ by expressing
time in cm; 1 cm of time is, in conventional units, the time required for
light to travel 1 cm, or $t_{\rm conv}=t_{\rm cm}/(3 \times 10^{10} ~{\rm cm ~s^{-1}})$.

In general relativity, the dimension of length is given to other quantities as well.
For example, as mentioned in \S1, the characteristic length scale in black hole physics
is the gravitational radius\footnote{ As an easy-to-remember conversion, the gravitational 
radius of an object is $1.477$ km per solar mass.} $R_g=GM/c^2$. Thus one 
often encounters dimensionless terms of the form $r/(GM/c^2)$. In  
adopting geometrized units, we make the transformation
\begin{equation}
\frac{r}{GM/c^2} \rightarrow \frac{r}{M}, 
\end{equation}
where  $M$ carries the unit cm, so that $r/M$ is dimensionless.
Therefore, to convert mass expressed in cm to mass expressed in grams, use
\begin{equation}
M_{\rm (g)} = \frac{c^2}{G} \, M_{\rm (cm)} ~~~{\rm or}~~~
M_{\rm (g)}=1.347 \times 10^{28} \, M_{\rm (cm)}.
\end{equation}
For example, the mass of the black hole near the center of the Milky Way Galaxy 
($M \approx 3.7 \times 10^6 ~M_{\odot}$ or $7.4 \times 10^{39}$ g in conventional 
units) is $5.5 \times 10^{11}$ cm.

Angular momentum, symbolized here as $J$ or $L$, has dimensions of length 
squared in geometrized units. The conversion is
\begin{equation}
\frac{c^3}{G} \, J_{\rm (cm^2)}=J_{\rm(g \, cm^2 \, s^{-1})}.
\end{equation}
For convenience with mathematical manipulations,
the spin of a black hole is characterized by the quantity $a=J/M$, where $J$ is the angular momentum
of the hole.  The conversion between geometric units
and c.g.s. units is 
\begin{equation}
a=\frac{J_{\rm (cm^2)}}{M_{\rm (cm)}}=\frac{J_{\rm(g \, cm^2 \, s^{-1})}}{M_{\rm (g)}c},
\end{equation}
so that spin has dimensions of length. Since mass also has dimensions of
length, it is often useful to define a dimensionless {\it spin parameter},
$a_*=a/M$. If the mass and angular momentum of a spinning object are known
in conventional units, a dimensionless $a/M$ follows from
\begin{equation}
a_*=\frac{a}{M}=\frac{c}{G} \, \frac{J}{M^2} \big|_{\rm c.g.s.}
\end{equation}

\subsection{The Schwarzschild Metric}

A spacetime {\it metric} provides a ``rule'' by which to relate measurements of events in different frames.
The {\it Minkowski metric}, 
\begin{equation}
\label{eq:mink}
d\tau^2=dt^2-(dx^1)^2-(dx^2)^2-(dx^3)^2,
\end{equation}
can be expressed in a more concise form if we define a set of
metric coefficients $\eta_{\alpha \beta}$. Thus one can write
$d\tau^2 = -\eta_{\alpha \beta} \,  dx^{\alpha} dx^{\beta}$, where $dx^0=dt$, $\eta_{00}=-1$,
$\eta_{jj}=1$ for $j=1,2,3$, and where summation
is performed over all repeated indices.

More generally, a spacetime metric has the form
\begin{equation}
d\tau^2=-g_{\mu \nu} \, dx^{\mu} dx^{\nu},
\end{equation}
where the metric coefficients can be functions of the spacetime coordinates.
The Schwarzschild solution to the Einstein 
field equations yields the metric
\begin{equation}
\label{eq:schfull}
d\tau^2=\biggl(1-\frac{2M}{r} \biggr)~dt^2 - \biggl(1-\frac{2M}{r} \biggr)^{-1} ~dr^2
-r^2~(d\theta^2+\sin^2 \theta ~d\phi^2),
\end{equation}
which is the vacuum solution to the spacetime outside of a spherical mass $M$
(for a derivation, see, e.g., \cite{weinberg} \cite{schutz}).
The metric coefficients $g_{\mu \nu}$ can be read off directly.
The $t$-coordinate is the time as measured by a distant observer. The $r$-coordinate,
called the {\it reduced circumference},
is defined such that the circumference of a circle centered on the gravitating mass
is precisely $2\pi r$. The $\theta$- and $\phi$-coordinates are defined such that
$r \, d\theta$ and $r \, d\phi$, respectively, measure differential distances along tangents to 
a circle at $r$, again, centered on the gravitating mass. For $r \gg 2M$, the Schwarzschild 
metric reduces to the Minkowski metric (Eq.\ \ref{eq:mink}), after a transformation between 
spherical polar coordinates and Cartesian coordinates. There is, however, no transformation 
of coordinates that can globally reduce the Schwarzschild metric to the Minkowski metric. 
Thus the spacetime described by the former is said to be curved, whereas the Minkowski 
spacetime is said to be flat --- ``flat'' in the sense that it is quasi-Euclidean.

Since the metric is spherically symmetric, the motion of test particles and photons
is restricted to a plane. This plane, with no loss of generality, can be
chosen to be the equatorial plane ($\theta=\pi/2$). For events occurring
within the equatorial plane, we can also set $d\theta=0$ in Eq. (\ref{eq:schfull}), 
which gives the simplified proper time interval,
\begin{equation}
\label{eq:schplane}
d\tau^2=\biggl(1-\frac{2M}{r} \biggr)~dt^2 - \biggl(1-\frac{2M}{r} \biggr)^{-1} ~dr^2
-r^2 ~d\phi^2.
\end{equation}
The proper length interval $ds$ ($ds^2=-d\tau^2$) is given by
\begin{equation}
\label{eq:properlength}
ds=\biggl(1-\frac{2M}{r} \biggr)^{-1/2} \, dr.
\end{equation}

\subsection{Gravitational Time Dilation and Gravitational Red Shift}

Even before Einstein arrived at his final formulation of the field equations, he
argued the case for gravitational time dilation. The standard version of his derivation
appeared in 1911, although he was aware of the effect as early as 1907 \cite{pais}. We provide
here a derivation that follows the one presented in \cite{weinberg}, except that we will 
take advantage of the explicit form of the Schwarzschild solution, thereby deriving a 
result that will be useful later.

Imagine that two identical clocks are manufactured in a 
region of flat spacetime and calibrated so that every hour 
the clocks chime together, each clock thus marking proper time intervals $\Delta
\tau$ corresponding to one hour.  Two experimenters, $A$ and $B$, are enlisted. $A$ moves one
clock to a position at distance $r_A$ from a gravitating object. $B$ moves the other clock
to a position at distance $r_B>r_A$ from the mass (Fig. 3). At each chiming, $A$ records the
time, and will maintain that the passage of time from chime to chime
$\Delta t_A$ is precisely one hour. According to Eq.\ (\ref{eq:schplane}), $A$ measures
$\Delta t_A=\Delta \tau /\sqrt{1-2M/r_A}$. Light pulses are sent from $r_A$ to $r_B$ at 
each chiming of $A$'s clock, and $B$ records the arrival times. Light pulses require time 
-- according to observers in {\it any} frame -- to travel from $A$ to $B$. Nevertheless, 
$B$ records a pulse-to-pulse time separation of precisely $\Delta t_A$, provided that
the light paths are identical for each pulse. When he compares the chime-to-chime record 
of his own clock, however, he finds that $\Delta t_B < \Delta t_A$, where $B$ will insist 
that $\Delta t_B$ is one hour.
$B$ measures $\Delta t_B=\Delta \tau /\sqrt{1-2M/r_B}$. $B$ is forced to conclude that time
runs slower at $r_A$ than at $r_B$. If the experiment were altered such that $B$ sent pulses
to $A$, then $A$ would be forced to conclude that time runs faster at $r_B$ than at $r_A$. This
effect is known as gravitational time dilation. 

For our purposes, it is more appropriate to express time dilation in terms of light frequencies.
Referring to the experiment above, and replacing the clocks by radiating atoms,
label by $\nu_{\rm obs}$ the frequency of light originating at $A$ as observed at $B$, 
and label by $\nu_{\rm em}$ the frequency of an identical source at $B$ as observed by $B$. 
The ratio of the frequencies is, therefore,
\begin{equation}
\frac{\nu_{\rm rec}}{\nu_{\rm em}}=\frac{\Delta t_B}{\Delta t_A}=
\biggl(\frac{1-2M/r_A}{1-2M/r_B} \biggr)^{1/2}.
\end{equation}
The case of primary interest -- observing radiation from a distant source -- corresponds 
to $r_B \rightarrow \infty$. This gives us the expression for the {\it gravitational redshift} 
for a Schwarzschild black hole, according to the observer at infinity,
\begin{equation}
\label{eq:redshift}
\nu_{\infty}=\nu_{\rm em}~\biggl(1-\frac{2M}{r} \biggr)^{1/2}.
\end{equation}
As $r \rightarrow 2M$, the observer at infinity finds that $\nu_{\rm obs} \rightarrow 0$. 
For this reason, the radius $r=2M$ is sometimes referred to as the {\it surface of 
infinite redshift}. It is important to remember that this expression is valid only for the case of
a stationary source and a stationary receiver. Below, we will accommodate the case of a source in orbit
around a black hole.

\begin{figure*}[top]
\centering
\includegraphics[width=12cm,height=7cm]{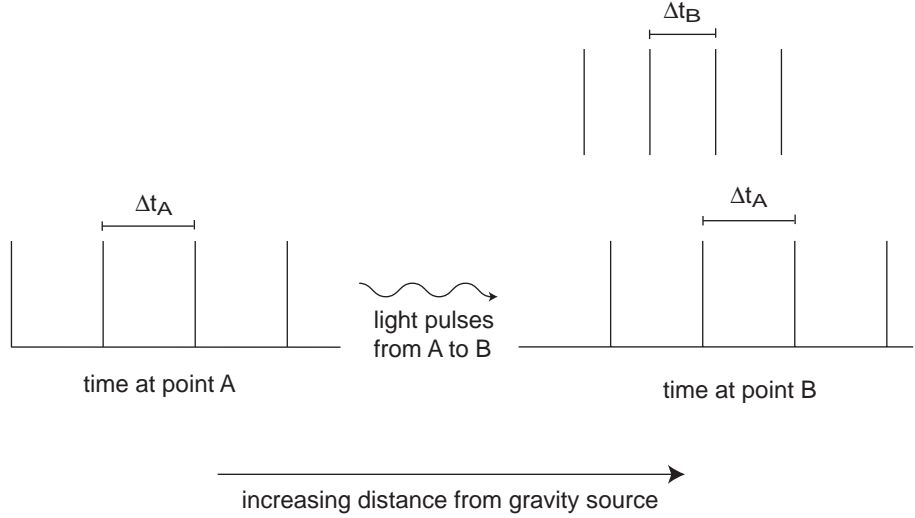}
%\label{fig:blackhole}
%\epsscale{0.8}
\caption{Gravitational time dilation. Two identically manufactured clocks are positioned
at points $A$ and $B$, with $r_B > r_A$ with respect to a gravitating mass. An observer at
$A$ sends pulses to observer at $B$ at each chiming of the clock at $A$. Observer at $B$
notes that chime frequency is higher at $B$ than at $A$ and concludes that time runs slow at $A$.
See text for futher discusion.}
\end{figure*}

\subsection{Conserved Quantities}

Before discussing the motion of particles and photons in the
Schwarzschild metric, we will need two conservation laws,
which follow from the equations of motion.
The equations of motion of free particles in curved spacetime can be derived from
the Euler-Lagrange equations, for an appropriate Lagrangian $\Lambda$, according to
\begin{equation}
\label{eq:lagrange}
\frac{d}{dp} \, \frac{\partial \Lambda}{\partial \dot{x}^{\alpha}}-\frac{\partial \Lambda}{\partial
x^{\alpha}}=0
\end{equation}
where the dot denotes differentiation with respect to a timelike parameter $p$. A useful
form for the Lagrangian is 
\begin{equation}
\Lambda=\frac{1}{2} \, g_{\alpha \beta} ~\frac{dx^{\alpha}}{dp} \, \frac{dx^{\beta}}{dp}.
\end{equation}
with $p=\tau/m$, as shown in \cite{shapiro}. For the Schwarzschild metric, again reducing 
the problem to motion in the $\theta=\pi/2$ plane, this is
\begin{equation}
2\Lambda=-\biggl(1-\frac{2M}{r}\biggr) \, \biggl(\frac{dt}{dp}\biggr)^2
+\biggl(1-\frac{2M}{r}\biggr)^{-1} \, \biggl(\frac{dr}{dp}\biggr)^2
+r^2  \, \biggl(\frac{d\phi}{dp}\biggr)^2
\end{equation}
Evaluating Eq. (\ref{eq:lagrange}) for the $t$-coordinate, we find
\begin{equation}
\label{eq:econserve}
\frac{E}{m}=\biggl(1-\frac{2M}{r} \biggr)~\frac{dt}{d\tau}
\end{equation}
where the constant of the motion, labeled $E/m$, is identified with the
energy per unit mass of the particle, since Eq. (\ref{eq:econserve}) reduces to the 
special relativistic expression $E=m~dt/d\tau$ for large $r$, i.e., the
time-component of the momentum four-vector. 
Similarly, evaluating Eq. (\ref{eq:lagrange}) for the $\phi$-coordinate, we find
a second constant of the motion
\begin{equation}
\label{eq:lconserve}
\frac{L}{m}=r^2 ~\frac{d\phi}{d\tau}.
\end{equation}
where $L/m$ can be identified as the angular momentum per unit mass.

\subsection{The Effective Potential}

The usual Newtonian approach to the problem of calculating the orbit
of a test particle under the influence of a spherically symmetric
gravitational field involves finding two constants of the motion --
energy and angular momentum -- and recasting the radial equation
in terms of an effective potential. The effective potential is the
sum of the gravitational potential and a centrifugal term. From an
analysis of the radial excursions in this effective potential, one
finds the two basic orbit classes -- elliptical and hyperbolic  --
corresponding to bound and unbound motion, respectively, with circular
orbits corresponding to a special case of elliptical orbits.
A similar, and quite fruitful, approach is adopted to analyze 
the orbit classes for motion in the Schwarzschild metric.

To derive the effective potential for the Schwarzschild metric,
notice that Equations (\ref{eq:schplane}), (\ref{eq:econserve}), and
(\ref{eq:lconserve}) can be treated as a set of three equations in the four unknowns
$d\tau$, $dt$, $dr$, and $d\phi$. Therefore, we can eliminate $dt$ and
$d\phi$ in Eq. (\ref{eq:schplane}) to obtain
\begin{equation}
\label{eq:orbit}
\biggl(\frac{dr}{d\tau} \biggr)^2 =
\biggl(\frac{E}{m} \biggr)^2 -
\biggl(1-\frac{2M}{r} \biggr) ~\biggl[1+\frac{(L/m)^2}{r^2} \biggr],
\end{equation}
and one defines the effective potential according to
\begin{equation}
\label{eq:potential}
\biggl(\frac{V}{m} \biggr)^2=\biggl(1-\frac{2M}{r} \biggr) ~\biggl[1+\frac{(L/m)^2}{r^2} \biggr] 
\end{equation}
Note that the effective potential cannot be associated with an actual potential energy,
as in Newtonian mechanics; in general relativity, it is impossible to separate energy
into a kinetic term and a potential term. Rather, the effective potential merely
allows us to characterize certain aspects of particle trajectories without
integrating the equations of motion.

One example of the effective potential is plotted in the
left panel of Fig. 4; as $r \rightarrow \infty$, $V(r) \rightarrow 1$ as can
be seen from Eq. (\ref{eq:potential}).
The curve is, in fact, qualitatively similar to
analogous curves derived from the Newtonian prescription, except for small radii, where the
two differ dramatically, i.e., the repulsive centrifugal barrier that extends to
arbitrarily small radii in the Newtonian potential is absent at small radii
in the Schwarzschild effective potential. Any inward bound particle that finds itself
within the region of this downturn of $V$, which has been referred to as ``the pit in 
the potential'' \cite{misner}, is destined to fall through the event horizon. Energies
corresponding to such capture orbits are indicated by the top two dashed lines in Fig. 4.
The top line, with $E/m >1$, is labeled ``capture/escape orbit'' because inward bound particles
are captured, whereas outward bound particles can escape to infinity. The energy level labeled
``capture orbit'' corresponds to a particle that, if inward bound, passes through the event
horizon, and, if outward bound, arrives at apastron, returns, since $E/m<1$, then passes through the
event horizon. Below that is the level for a particle with energy corresponding to an unstable
circular orbit at $r=r_{\rm us}$. Test particles with this energy, whether directed inward 
from $r>r_{\rm us}$ or directed outward from $r<r_{\rm us}$, will end up on
this orbit, and  remain there until perturbed. The next orbit is bound, since particles with
this energy transit between two turning points. The Newtonian analogy for this orbit is an
ellipse. In the Schwarzschild metric, the azimuthal oscillation frequency exceeds the radial
oscillation frequency, leading to an orbit with a precessing apastron, with prograde
precession. The final trajectory class, corresponding to the local minimum of $V$, is a stable
circular orbit, discussed in the next section.

\begin{figure}[t]
\centering
\includegraphics[width=14cm,height=5cm]{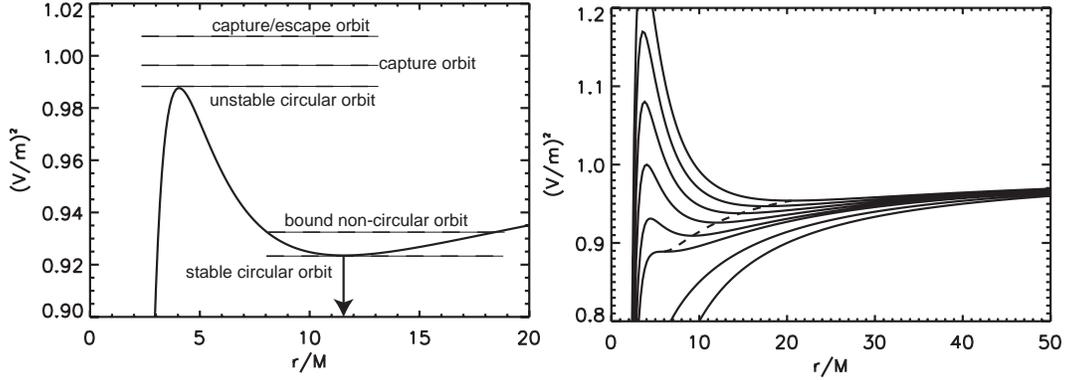}
\caption{{\it (Left panel)} Effective potential for a test particle trajectory in the
Schwarzschild metric (Eq. \ref{eq:potential}) with $L/mM=3.95$.
Dashed lines correspond to five different energies, increasing
from bottom to top. See text for a discussion.
{\it (Right panel)}  Effective potential for various values of the angular momentum:
from bottom to top, $L/mM$ = [0., 2.5, $2\sqrt{3}$, 3.7, 4.0, 4.3, 4.6, 4.9]. Intersections
of the dashed line with potential curves marks positions of stable circular
orbits. The innermost stable circular orbit occurs at $r/M=6$, corresponding
to $L/mM=2\sqrt{3}$.}
\end{figure}

\subsection{Circular Orbits in the Schwarzschild Metric}

The most important general relativistic effect for X-ray line shapes
is the gravitational redshift. The redshift formula is given in 
Eq. (\ref{eq:redshift}), and we intend to use that in our discussions of 
line profiles. However, as remarked there, the formula is valid only in the case of stationary
emitter and receiver, and the emitters in an accretion disk are not stationary. In fact,
we expect relativistic velocities. In order to preserve the use of the simple equation
for the redshift, we need first to know the photon energies as they would be measured
by {\it local stationary observers} ($dr/d\tau$=$d\phi/d\tau$=$d\theta/d\tau$=0), distributed 
in radius. We denote with the subscript {\it lso} measurements made by such observers. Once we 
know the local photon energy distribution, it is then a simple matter to calculate the distribution at
infinity by application of Eq. (\ref{eq:redshift}). Since disk material is, in the approximation that
disk annuli describe circular orbits, passing the stationary observer with azimuthal velocity $r \,
d\phi/dt_{\rm lso}$, we can simply apply a Lorentz transformation that relates the spectrum in the
disk frame to the spectrum measured by the stationary observer. The effective potential dictates a
relationship between the angular momentum and the radius of circular orbits. Once we find this
relationship, we proceed to the radial dependence of the velocity of circular orbits.

\subsubsection{Angular Momentum/Radius Relationship for Circular Orbits}

The condition for a circular orbit, $\partial V/\partial r =0$, 
yields a relationship between $r$ and $L$. Using Eq. (\ref{eq:potential}), this is
\begin{equation}
\label{eq:rL}
r^2-\biggl(\frac{L}{m} \biggr)^2 \frac{r}{M} +3 \biggl(\frac{L}{m} \biggr)^2=0,
\end{equation}
which has the two solutions
\begin{equation}
\label{eq:circle}
r_{\pm}=\frac{(L/m)^2}{2M}~\biggl[ 1 \pm \biggl( 1-\frac{12M^2}{(L/m)^2}\biggl)^{1/2} \biggr].
\end{equation}
The stability condition $\partial^2 V/\partial r^2 >0$ shows that a
circular orbit at $r_+$ is stable, while the one at $r_-$ is unstable.
Equation (\ref{eq:circle}) thus provides the radii of both the unstable circular orbit and the stable
circular orbit for a given value of $L/m$. We further see, that, owing to presence
of the square root, the domain of allowed $L/m$ yielding a circular orbit is restricted
to $L/m >2 \sqrt{3} M$. Setting $L/m$ to this minimum value in Eq.\ (\ref{eq:circle}), 
we find that the radius of the {\it innermost stable circular orbit} $r_{\rm isco}$ 
is equal to $6M$. 

The existence of the innermost stable circular orbit, a uniquely general 
relativistic feature, can be appreciated by inspection of the right panel 
of Fig. 4, which shows the effective potential for several values of the angular 
momentum. The intersection of the dashed line with each effective potential 
curve marks the radial position of each stable circular orbit. The trend in the behavior of $V$ with
$L/m$ is clear:  as $L/m$ decreases, the extent of the potential barrier decreases, as well. As 
$L/m \rightarrow 2 \sqrt{3} M$, the barrier flattens, and disappears for $L/m < 2 \sqrt{3} M$,
thereby eliminating the local minimum. All inward bound trajectories with $L/m < 2 \sqrt{3} M$
cross the event horizon.

\subsubsection{Energies of Circular Orbits}

From Eq. (\ref{eq:orbit}), setting $dr/d\tau=0$ for a circular orbit,
\begin{equation}
\label{eq:ecircular2}
\biggl(\frac{E}{m} \biggr)^2 =
\biggl(1-\frac{2M}{r} \biggr) ~\biggl[1+\frac{(L/m)^2}{r^2} \biggr]. 
\end{equation}
From Eq. \ref{eq:rL}, we have
\begin{equation}
\label{eq:rL2}
\biggl(\frac{L}{m}\biggr)^2=\frac{r^2}{r/M-3},
\end{equation}
which, when substituted into Eq. (\ref{eq:ecircular2}), gives
\begin{equation}
\label{eq:ecircular}
\frac{E}{m}=\frac{1-2M/r}{(1-3M/r)^{1/2}}
\end{equation}
This expression for energy is valid only
for $r \geq 6M$ because of the stability restriction.

At the ISCO, where $r/M=6$, we have $(E/m)_{\rm isco}=(8/9)^{1/2}$.
Therefore, the energy change per unit mass of a test particle as its circular orbit
degrades from a large distance to the ISCO is
\begin{equation}
\label{eq:extract1}
\Delta \biggl(\frac{E}{m} \biggr)=1-\biggl(\frac{8}{9} \biggr)^{1/2} \approx 0.057
\end{equation} 
i.e., the test particle loses an energy equivalent to about 6\% of its rest mass energy 
as it works its way down to the ISCO in a succession of circular orbits. This is often taken as the
maximum accretion  efficiency $\eta$ of a Schwarzschild black hole, in the sense that $L_{\rm acc}= 
\eta \dot{M}_{\rm acc}c^2$.

\subsubsection{Velocities of Circular Orbits}

We need to find the radial dependence of the velocities of circular
orbits $v_{\phi}=r \, d\phi/dt_{\rm lso}$ as they would be measured 
by locally stationary observers distributed
in $r$. Proper time is measured in the disk frame, so that we can write
\begin{equation}
\label{eq:vs1}
v_{\phi}=r \frac{d\phi}{d\tau} \frac{d\tau}{dt_{\rm lso}},
\end{equation}
and then relate $dt_{\rm lso}$ and $d\tau$ by a Lorentz transformation:
\begin{equation}
dt_{\rm lso}= (1-v_{\phi}^2)^{-1/2} \, d\tau.
\end{equation}
Using angular momentum conservation (Eq. \ref{eq:lconserve}), Eq. (\ref{eq:vs1}) becomes
\begin{equation}
\label{eq:vsL}
v_{\phi}^2=\frac{1}{r^2}~(L/m)^2 ~(1-v_{\phi}^2).
\end{equation}
Substituting for $L/m$ from Eq. (\ref{eq:rL}), and recalling that that equation is
valid only for stable circular orbits, some rearrangement gives
\begin{equation}
\label{eq:vshell}
v_{\phi}^2=\frac{M/r}{1-2M/r},
\end{equation}
which decreases monotonically with radius.
The domain of validity is the same as the domain for which stable circular
orbits can exist, i.e., $r \geq 6M$. The velocity of a particle in a circular
orbit at the ISCO can be found from Eq. (\ref{eq:vshell}) by substituting $r=6M$,
from which we find that $v_{\rm isco}=1/2$. 

For the sake of interest, let us find the linear velocity $v_{\infty}$ 
and angular velocity $\Omega_{\infty}$ of matter in a circular orbit as 
observed at infinity. To find the angular velocity, we start with the angular momentum
conservation law (Eq. \ref{eq:lconserve}) and write 
\begin{equation}
\label{eq:omega}
\Omega_{\infty}=\frac{L/m}{r^2} \, \frac{d\tau}{dt},
\end{equation}
so that we need to relate $dt$ and $d\tau$ for a circular orbit.
This is accomplished by manipulating a simplified form
of the metric (Eq. \ref{eq:schplane}), setting $dr=0$,
\begin{equation}
d\tau^2=\biggl(1-\frac{2M}{r} \biggr)~dt^2 -r^2 ~d\phi^2,
\end{equation}
which, after eliminating $d\phi/d\tau$, again using Eq. (\ref{eq:lconserve}), becomes
\begin{equation}
\biggl(1-\frac{2M}{r} \biggr) \biggl(\frac{dt}{d\tau}\biggr)^2
=1+\frac{(L/m)^2}{r^2}.
\end{equation}
Substituting for $L/m$ from Eq. (\ref{eq:rL2}) and simplifying gives
\begin{equation}
\frac{d\tau}{dt}=\biggl(1-\frac{3M}{r}\biggr)^{1/2} ~~~~~~(r \geq 6M).
\end{equation}
Substituting this back into Eq. (\ref{eq:omega}), again using Eq. (\ref{eq:lconserve})
gives
\begin{equation}
\Omega_{\infty}=\biggl(\frac{M}{r^3}\biggr)^{1/2}~~~~~~(r \geq 6M),
\end{equation}
which is identical to the Newtonian result. The linear velocity is simply
$r\Omega_{\infty}$, i.e.,
\begin{equation}
v_{\infty}=\biggl(\frac{M}{r}\biggr)^{1/2} ~~~~~~(r \geq 6M),
\end{equation}
which is also identical to the Newtonian result.

\subsection{The Motion of Light in the Schwarzschild Metric}

For light motion (technically, for the motion of massless particles),
we derive the governing equations by starting with the
equations of motion for massive particles, then take the limit $m \rightarrow 0$.
Starting with Eq. (\ref{eq:orbit}), we have
\begin{equation}
\biggl(\frac{dr}{dt} \biggr)^2 ~\biggl(\frac{dt}{d\tau} \biggr)^2=
\biggl(\frac{E}{m} \biggr)^2 -
\biggl(1-\frac{2M}{r} \biggr) ~\biggl[1+\frac{(L/m)^2}{r^2} \biggr]. 
\end{equation}
Using the equation for the conserved energy (Eq. \ref{eq:econserve}), the previous
expression gives
\begin{equation}
\biggl(\frac{dr}{dt} \biggr)^2 =\biggl(1-\frac{2M}{r} \biggr)^2 -
\biggl(1-\frac{2M}{r} \biggr)^3 ~\biggl[\frac{m^2}{E^2}+\frac{L^2}{r^2 E^2} \biggr]. 
\end{equation}
Taking the limit $m \rightarrow 0$, and defining the {\it impact parameter} 
$b=L/E$, we find
\begin{equation}
\label{eq:radiallight}
\biggl(\frac{dr}{dt} \biggr)^2 = \biggl(1-\frac{2M}{r} \biggr)^2
\biggl[1-\frac{b^2}{r^2} \, \biggl(1-\frac{2M}{r} \biggr) \biggr].
\end{equation}

For the azimuthal motion, we combine Eqs. (\ref{eq:econserve}) and (\ref{eq:lconserve}) to give
\begin{equation}
\frac{d\phi}{dt} \, \frac{E}{m} \, \biggl(1-\frac{2M}{r} \biggr)^{-1}=\frac{L/m}{r^2}, 
\end{equation}
which becomes
\begin{equation}
\label{eq:philight}
\frac{d\phi}{dt}= \frac{b}{r^2} \, \biggl(1-\frac{2M}{r} \biggr).
\end{equation}
Therefore, once we know the impact parameter (see below), the light path
can be calculated as parameterized by our measure of time $t$. 
By eliminating $dt$ between Eqs. (\ref{eq:radiallight}) and (\ref{eq:philight}),
we can plot the trajectory $r(\phi)$ according to
\begin{equation}
\label{eq:drdphilight}
\frac{dr}{d\phi}  = \frac{r^2}{b}~
\biggl[1-\frac{b^2}{r^2} \, \biggl(1-\frac{2M}{r} \biggr) \biggr]^{1/2}.
\end{equation}

\begin{figure}[top]
\centering
\includegraphics[width=10cm,height=7cm]{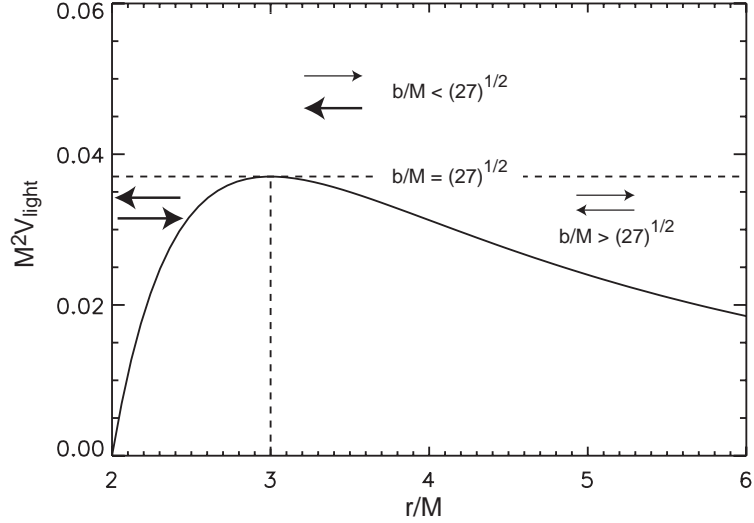}
\caption{Effective potential for light motion near a Schwarzschild black hole
(see Eq. \ref{eq:lightpotential}). Three sets of $(r,b)$ constitute the range of possibilties
for light motion in the Schwarzschild potential, where $b$ is the impact parameter.
Light arrows indicate escape orbits, heavy arrows are capture orbits. The vertical
dashed line shows the position of the unstable circular orbit.}
\end{figure}

\subsubsection{Effective Potential for Light Motion}

We wish to obtain an effective potential for light, analogous to that found
for particle motion, in order to gain a quick qualitative understanding of
light motion, without having to perform numerical integrations of the equations
of motion. Following \cite{taylor}, we need the relations between time
and radial displacements as measured by a local stationary observer and those 
measured by a distant observer, which results in
\begin{equation}
\frac{dr_{\rm lso}}{dt_{\rm lso}} =(1-2M/r)^{-1} ~ \frac{dr}{dt}.
\end{equation}
Then, Eq. (\ref{eq:radiallight}) can be written
\begin{equation}
\frac{1}{b^2} \, \biggl(\frac{dr_{\rm lso}}{dt_{\rm lso}}\biggr)^2 = \frac{1}{b^2}
-\frac{1}{r^2} \, \biggl(1-\frac{2M}{r} \biggr),
\end{equation}
and we identify as the effective potential for light
\begin{equation}
\label{eq:lightpotential}
V(r) =\frac{1}{r^2} \, \biggl(1-\frac{2M}{r} \biggr).
\end{equation}
Note the simplicity of the description of light trajectories
compared to particle trajectories, where, in the latter case, the
geodesics depend both on energy and angular momentum. 

The effective potential for light is plotted in Fig. 5.
The peak of $V$ occurs at $r=3M$, for which $V_{\rm peak}=1/27M^2$.
For the critical impact parameter, $b_{\rm crit}=3\sqrt{3}M$, 
$dr_{\rm lso}/dt_{\rm lso}=0$ at $r=3M$. This is the only radius for which 
a massless particle can move on a circular orbit. However, the orbit is clearly 
unstable; any perturbation to the orbit will lead to either to capture or escape. 

For the purposes of
determining the range of impact parameters that lead to capture by the black hole, the presence of
the  potential peak at $r=3M$ and the critical impact parameter divides
space into two. First, consider light rays emanating from radii $r_o>3M$. From
Fig. (5), we see that all outward directed photons escape to infinity. For inward
directed rays, those for which $b<3\sqrt{3}M$ are captured. For photons originating
inside $r=3M$, all inward directed photons are captured. Outward directed photons
are captured if $b>3\sqrt{3}M$, and escape otherwise.

\begin{figure}[top]
\centering
\includegraphics[width=12cm,height=6cm]{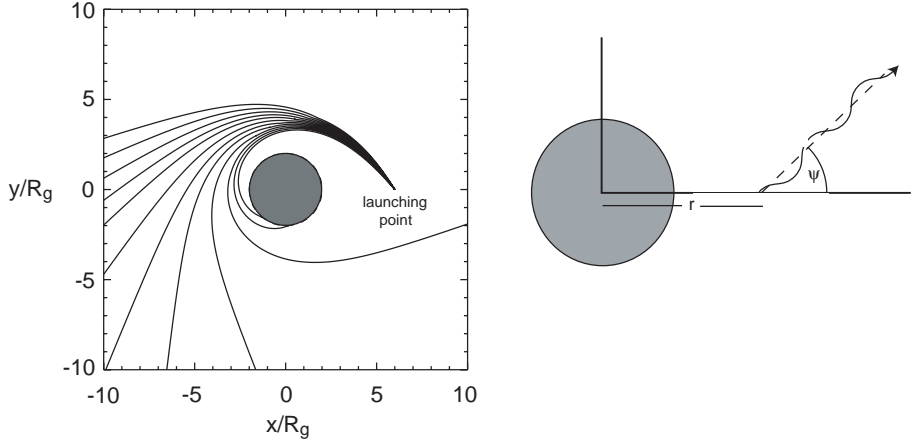}
\caption{({\it Left panel}) Light trajectories in the Schwarzschild metric (see Eq. \ref{eq:drdphilight})
for point of origin $r=6M$ for twelve impact parameters confined to the range 5.0$M$--5.9$M$.
({\it Right panel}) The angle $\psi$ between a light ray and a radius vector, as measured 
by a local stationary observer, used to define the impact parameter (see Eq. \ref{eq:bparameter}). }
\end{figure}

Using the effective potential, we can calculate the fraction of photons 
that will be captured by the black hole if emitted isotropically by a stationary source at $r$. 
First, we need to know how to calculate $b$ for various cases of interest. Suppose 
photons are being launched at various angles from a point $r$ near the black hole. 
We can find $b$ by referring to the azimuthal equation of motion. From Eq. 
(\ref{eq:philight}), reckoning time according to a local stationary observer, we have
\begin{equation}
r \, \frac{d\phi}{dt_{\rm lso}}=v_{\phi}=\frac{b}{r} \, \biggl(1-\frac{2M}{r} \biggr)^{1/2}
\end{equation}
Using the coordinate system illustrated in the right panel of Fig. (6), this becomes
\begin{equation}
\label{eq:bparameter}
b=r \, \biggl(1-\frac{2M}{r} \biggr)^{-1/2} \, \sin \psi_{\rm lso}
\end{equation}
As an aside, a distinction needs to be made between angles measured by a locally stationary
observer and those calculated by a distant observer. The relation between
$\psi_{\rm lso}$ and the same angle according to a distant observer
$\psi_{\infty}$ can be found as follows. Let the radial displacement
measured by the LSO be denoted by $ds$. Then
\begin{equation}
\tan \psi_{\infty}=\frac{r \, d\phi}{dr}=\frac{r \, d\phi}{ds} \,
\frac{ds}{dr}
=\biggl(1-\frac{2M}{r}\biggr)^{-1/2} \, \tan \psi_{\rm lso},
\end{equation}
where we have used Eq. (\ref{eq:properlength}) to relate $ds$ and $dr$.

\begin{figure}[top]
\centering
\includegraphics[width=12cm,height=8cm]{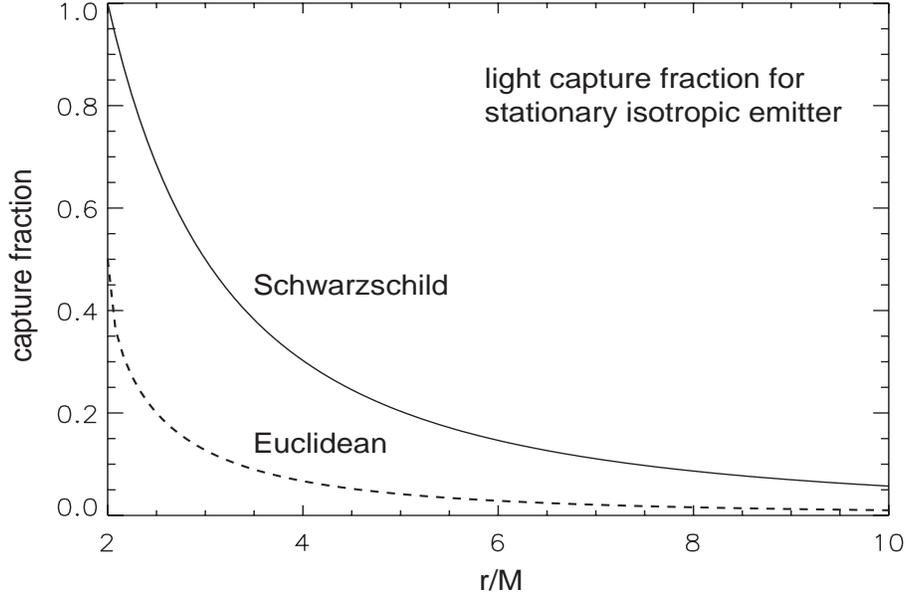}
\caption{Capture fraction for stationary isotropic photon emitters plotted against radius
for the Schwarzschild metric and for Euclidean space. In the latter case, the capture
fraction is simply the fractional solid angle subtended at $r$ by a sphere of radius $2M$. For
the Schwarzschild case, light bending is included.}
\end{figure}

Now, for $r>3M$, we already know that all outward directed photons escape. Inward
directed photons are captured if $\sin \psi_{\rm lso} < 3\sqrt{3}(M/r) \sqrt{1-2M/r}$. This 
results in
\begin{equation}
f_{\rm capt}=\frac{1}{2}~\biggl[ 1-\sqrt{1-27\biggl(\frac{M}{r}\biggr)^2
\biggl(1-\frac{2M}{r}\biggr)} \biggr]  ~~~~(r \geq 3M).
\end{equation}
For $r<3M$, we already know that inward directed photons are captured. Outward
directed photons are captured if $\sin \psi > 3\sqrt{3}(M/r) \sqrt{1-2M/r}$,
which results in
\begin{equation}
f_{\rm capt}=\frac{1}{2}~\biggl[ 1 + \sqrt{1-27\biggl(\frac{M}{r}\biggr)^2
\biggl(1-\frac{2M}{r}\biggr)} \biggr] ~~~~(r\leq 3M).
\end{equation}
The capture fraction is 1/2 for $r=3M$, and approaches unity as $r \rightarrow 2M$.
A plot of $f_{\rm capt}$ vs. $r$ is plotted in Fig. (7). Calculating the captured fraction
for material moving in a disk must also account for relativistic beaming. In other words,
a moving isotropic photon source will, to varying degrees, appear anisotropic to the 
stationary observer, with its emission concentrated into the direction of motion. 

\subsubsection{The Speed of Light}

One of the more peculiar aspects of light propagation near a black hole is that
those performing calculations in flat spacetime must account for the apparent reduction
of the speed of light. According to a distant observer, the speed of light 
for pure radial motion, from Eq. (\ref{eq:radiallight}), with $b=0$, is 
\begin{equation}
\label{eq:radiallight2}
\frac{dr}{dt} = \pm \biggl(1-\frac{2M}{r} \biggr),
\end{equation}
which approaches unity for large $r$, and approaches zero as $r \rightarrow 2M$.
For example, according to an observer at $r$, the time $\Delta t$ for a light
pulse to propagate radially from a position $r_o <r$ to $r$ is 
\begin{equation}
\frac{\Delta t}{M}=\frac{r-r_o}{M}+2 \ln \frac{r/2M-1}{r_o/2M-1},
\end{equation}
which shows the effect of the reduced speed of light (2nd term) compared
to the Euclidean value (1st term). Here we are measuring time in multiples
of $M$. The conversion between seconds and centimeters is $\Delta t_{\rm (s)}
=(GM/c^3) \Delta t_{\rm (cm)}$.

The apparent speed of light for pure azimuthal
motion is, from Eq. (\ref{eq:philight}),
\begin{equation}
\label{eq:philight2}
r\frac{d\phi}{dt}=\pm \frac{b}{r} \, \biggl(1-\frac{2M}{r} \biggr)
= \pm  \, \biggl(1-\frac{2M}{r} \biggr)^{1/2},
\end{equation}
where we have substituted for $b$ from Eq. (\ref{eq:bparameter}). Note that
this is not the same as the speed of light for pure radial motion (Eq.
\ref{eq:radiallight2}).

\begin{figure}[top]
\centering
\includegraphics[width=9cm,height=8cm]{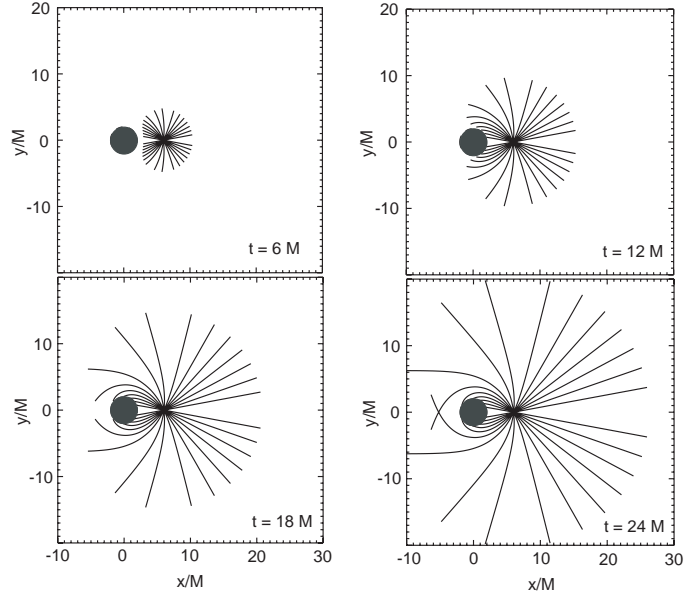}
\caption{Time dependence of an initially radial planar light pulse near a Schwarzschild
black hole. Time intervals are measured from initiation of pulse at $t=0$.
Deviation from perfectly circular wave fronts illustrates the combined effect of
a variable speed of light and light bending, as calculated by a distant observer. 
Plots based on integrations of Eq. \ref{eq:drdphilight}. }
\end{figure}

Combining Eqs. (\ref{eq:radiallight2}) and (\ref{eq:philight2}), we can evaluate
the speed of light for arbitrary trajectories according to a distant observer
\begin{equation}
c_{\infty}=\biggl[\biggl(\frac{dr}{dt}\biggr)^2+
\biggl(r\frac{d\phi}{dt}\biggr)^2 \biggr]^{1/2}
=\biggl(1-\frac{2M}{r}\biggr)\biggl(1+\frac{2b^2M}{r^3}\biggr)^{1/2}
\end{equation}
While this apparent non-constancy of the speed of light might seem somewhat disorienting, 
it is important to realize that we do not actually {\it measure} this speed.
Measurements must be made locally. We can calculate what a locally stationary
observer will measure by transforming Eqs. (\ref{eq:radiallight2}) and
(\ref{eq:philight2}) to the frame of such an observer.
\begin{equation}
\frac{dr_{\rm lso}}{dt_{\rm lso}}=\biggl(1-\frac{2M}{r}\biggr)^{-1} ~\frac{dr}{dt}
=\biggl[1-\frac{b^2}{r^2} \, \biggl(1-\frac{2M}{r} \biggr) \biggr]^{1/2}
\end{equation}
\begin{equation}
r\frac{d\phi}{dt_{\rm lso}}=\biggl(1-\frac{2M}{r}\biggr)^{-1/2} ~r\frac{d\phi}{dt}
=\frac{b}{r} \, \biggl(1-\frac{2M}{r} \biggr)^{1/2}
\end{equation}
These two equations give
\begin{equation}
c_{\rm lso}=\biggl[\biggl(\frac{dr_{\rm lso}}{dt_{\rm lso}}\biggr)^2
+\biggl(r\frac{d\phi}{dt_{\rm lso}} \biggr)\biggr]^{1/2} =1.
\end{equation}
Therefore, locally stationary observers always measure $c=1$. We, as distant observers,
must contend with the fact that light appears to propagate at varying speeds,
its speed depending on both the radial coordinate and the impact parameter.

An example that illustrates the effect of a variable speed of light is shown in
Fig. (8). A planar radial burst of light is emitted from $r=6M$ near a Schwarzschild
black hole. Shown in the figure are the trajectories of a set of rays with a distribution
of impact parameters, the endpoint of each trajectory indicating the progression of the
leading edge of the pulse at four times: $t=6M$, $12M$, $18M$, and $24M$, where
the time is measured as a multiple of $GM/c^3$. The reduced speed of light at small radii
compared to larger radii is evident in the plot for $t=6M$ by the distortion from a 
perfect circle of the endpoints of the light ray pattern. With a knowledge of 
the disk structure (or, more likely, a model), one can imagine exploiting the unique 
light propagation physics in a black hole metric to locate the
position of such a pulse as it irradiates different portions of the disk. Conversely, with a model
of the origin for a pulse, or X-ray flare, the disk structure can be mapped, since
the disk will respond to changes in the illumination pattern. Since
such an effect occurs on a timescale that is measured in multiples of $GM/c^3$, it
is practical to observe ``light echoes'' in AGN (hundreds to thousands of
seconds), but not in black hole X-ray binaries ($\sim1$ millisecond).  

This leads to the concept of {\it reverberation mapping}, which is among the programs planned 
for future X-ray observatories, such as {\it Constellation-X}. 
We do not discuss the details of this ambitious goal
here. However, a few comments are in order. Given that the model of X-ray fluorescence
from AGN and black hole X-ray binaries involves the spectral response of an accretion disk
to illumination by hard X rays, and that the hard X-ray continuum flux is observed to vary,
it makes sense to ask whether changes in the shapes and intensities of fluorescence
lines in response to continuum variability can elucidate both the nature of continuum production
and disk structure. The basics of such a program, for the case of a Schwarzschild black hole,
have been laid out in \cite{stella}. It was shown that a variable source of
localized hard  X-ray illumination placed at the center of a disk would cause a characteristic response 
in the profile of a fluorescence line. In this case, with a centrally-located illumination 
source, the inner disk would respond first, affecting the red and blue extremes of the profile.
Owing to light travel times, the rest of the profile would respond at later times. It is
shown that, in principle, the mass of the black hole can be determined. Measurements of
black hole spin (see \S6) are also plausibly within reach \cite{young}. In a more general
case, reverberation could involve a non-axial illumination source in the Kerr geometry, time-dependent
modifications to the disk structure, and multiple, overlapping X-ray flares, with a distribution
in both intensity and position. In such models,
the varying speed of light propagation figures prominently. For example, as discussed in
\cite{reynolds99}, non-axial variable illumination causes the redward portion of the 
line profile to lag the rest of the profile, indicative of an inward moving pulse
that sweeps across the disk, slowing to zero velocity at the horizon.

Interestingly, current studies of AGN in which the X-ray continuum is observed
to vary fail to show the expected variability in the iron K$\alpha$ line complex
\cite{lee2000} \cite{chiang} \cite{reynolds2000}, although efforts attempting to reconcile
this problem are underway \cite{miniutti}.

\subsection{Minimum and Maximum Frequency Shifts}

Assume that a locally stationary observer measures the frequency distribution of
a radiating particle that moves on a circular orbit at velocity $v_{\phi}$ and emits 
photons of frequency $\nu_o$ in its rest frame. The minimum and maximum frequencies 
measured by the LSO are determined by a Lorentz transformation:
\begin{equation}
\nu_{\rm lso}= \gamma\nu_o \, (1 \pm v_{\phi})
\end{equation}
where $v_{\phi}$ is given by Eq. (\ref{eq:vshell}), and $\gamma=
(1-v_{\phi}^2)^{-1/2}$. Then from Eq. (\ref{eq:redshift})
we can find the frequency observed at infinity according to
\begin{equation}
\nu_{\infty}=\biggl(1-\frac{2M}{r} \biggr)^{1/2} ~\nu_{\rm lso}.
\end{equation}
This is straightforward, and works out to
\begin{equation}
(\nu_{\infty})^{\pm}=\nu_o ~\biggl( \frac{1-2M/r}{1-3M/r} \biggr)^{1/2} ~
\biggl[ (1-2M/r )^{1/2} \pm ( M/r )^{1/2} \biggr].
\end{equation}
At the ISCO, we find $\nu_{\infty}^-=(\sqrt{2}/3)\nu_o$ and
$\nu_{\infty}^+ = \sqrt{2}\nu_o$ \cite{shapiro}.
This analysis also assumes that the observer at infinity sees the full red and blue
shifts of the emitted photons. In more realistic cases, the line profile is narrower,
since the disk is likely to be inclined.

Note that, although an observer at infinity marks the velocity of a circular orbit  
as being $v_{\infty}=\sqrt{M/r}$, we cannot properly calculate the photon 
energies (as above) unless we analyze the problem in the local frame of 
the orbiting matter. The lesson is that there is no Lorentz transformation 
that globally reduces curved spacetime to the Minkowski metric.

We revisit the minimum and maximum frequency shifts in a later section, after 
we derive the analogous quantities for Kerr black holes.

\section{Spinning Black Holes -- The Kerr Metric}

We have already remarked on the ubiquity of angular momentum in the cosmos.
Angular momentum is also likely to play a role in the behavior of spacetime near black holes.
Stellar precursors to Galactic black holes rotate, thus imbuing
these stars with angular momentum. Near the center of a galaxy, gas that supplies 
fuel to supermassive black holes in AGN is in motion, and thus carries angular momentum
with respect to the black hole. In both cases, then, we expect the black hole
to possess angular momentum. Therefore, improved treatments of the physics
in the black hole environment should include modifications to the spacetime
geometry induced by spin \cite{bardeen70}. 

In 1963, Roy Kerr \cite{kerr} presented his solution for the metric outside a spinning
object at the First Texas Symposium on Relativistic Astrophysics (see \cite{thorne}
for a description of these proceedings). The physical implications of this solution
were pursued over the course of the next few years. More recently, observable
properties of the emission and timing properties of accretion disks have led to
attempts to measure black hole spin.

The Kerr metric is given in Boyer-Lindquist coordinates \cite{boyer} as
\begin{equation}
\label{eq:kerr}
d\tau^2=\biggl(1-\frac{2Mr}{\Sigma} \biggr) \, dt^2 +
\frac{4Mar \, \sin^2 \theta}{\Sigma} \, dt \, d\phi 
-\frac{\Sigma}{\Delta} \, dr^2 -\Sigma \, d\theta^2
-R_{a,\theta}^2  \sin^2 \theta\, d\phi^2,
\end{equation}
where
\begin{equation}
\begin{array}{l}
\Sigma=r^2+a^2 \cos^2 \theta \\
\\
\Delta=r^2-2Mr+a^2 \\
\\
R_{a,\theta}^2=r^2+a^2 +2Mra^2 \Sigma^{-1} \sin^2 \theta. 
\end{array}
\end{equation}
For motion in the equatorial plane ($\theta=\pi/2$), the Kerr metric is
\begin{equation}
\label{eq:kerrplane}
d\tau^2=\biggl(1-\frac{2M}{r} \biggr) \, dt^2 +
\frac{4Ma}{r} \, dt \, d\phi - \frac{r^2}{\Delta} \, dr^2 - R_a^2  \, d\phi^2,
\end{equation}
where
\begin{equation}
\label{eq:ra}
R_a^2=r^2+a^2 +\frac{2Ma^2}{r}.
\end{equation}
We show below that $R_a$ is the reduced circumference in the equatorial plane
of a Kerr black hole.

\subsection{The Kerr Event Horizon}

The radial coordinate of the horizon $r_H$ is defined as that radius at
which the coefficient of the $dr^2$ term of the metric blows up.
For the Kerr metric this occurs when $\Delta =0$, which gives
\begin{equation}
\label{eq:kerrhorizon}
r_H=M + \sqrt{M^2-a^2}.
\end{equation}
This shows us that the maximum spin consistent with a real-valued radial coordinate for the
horizon is simply $M$. The radial coordinate of the horizon thus ranges from $2M$ ($a=0$) down
to $M$ ($a=M$). Detailed considerations concerning the spin-up of a black hole by accreting matter
shows that the maximum spin is not $M$, but rather $0.998M$ \cite{thorne73}. The case $a=M$ 
(or $a=0.998M$) is referred to as an {\it extreme Kerr black hole}, and the black hole is 
said to be {\it maximally spinning}. 

\subsection{The Static Limit and the Dragging of Inertial Frames}

The following derivation follows that in \cite{shapiro}.
The relation $g_{\mu \nu}p^{\mu}p^{\nu}=-m^2$ is equivalent to the condition
$g_{\mu \nu}u^{\mu}u^{\nu}=-1$, where 
$u^{\mu}$ is a component of the velocity four-vector. Consider an observer at fixed
$r$ and $\theta$, so that $dr=0$ and $d\theta=0$. Then
\begin{equation}
g_{tt} \, (u^t)^2 + 2 g_{t\phi} \, u^t u^\phi +g_{\phi \phi} \, (u^{\phi})^2=-1,
\end{equation}
where the metric coefficients can be read off from Eq. \ref{eq:kerr}.
Rearrangement of the previous equation gives
\begin{equation}
(u^t)^2 \biggl[ g_{tt} \,  + 2 g_{t\phi} \,  \frac{u^\phi}{u^t} +
g_{\phi \phi} \, \frac{(u^\phi)^2}{(u^t)^2} \biggr]=-1.
\end{equation}
If we let $u^\phi/u^t=d\phi/dt \equiv \omega$, then we require
\begin{equation}
\omega^2 +  2 \, \frac{g_{t\phi}}{g_{\phi \phi}} \, \omega +\frac{g_{tt}}{g_{\phi \phi}}<0.
\end{equation}
This restricts $\omega$ to the range
\begin{equation}
\frac{\vert g_{t\phi}\vert}{g_{\phi \phi}}\biggl[1-\sqrt{1-\frac{g_{tt} 
g_{\phi \phi}}{g_{t\phi}^2}} \biggr] \leq \omega 
\leq \frac{\vert g_{t\phi}\vert}{g_{\phi \phi}}\biggl[ 1+\sqrt{1-\frac{g_{tt} g_{\phi \phi}}{g_{t\phi}^2}} \biggr],
\end{equation}
or $\omega_{\rm min}<\omega<\omega_{\rm max}$.
Note that $g_{t\phi}$ is always negative, and that $g_{\phi \phi}$ is always positive.
The metric coefficient $g_{tt}$ can be positive or negative:
\begin{equation}
\label{eq:static1}
\begin{array}{l}
g_{tt} < 0 ~~{\rm if} ~~r> M+\sqrt{M^2-a^2 \cos ^2 \theta} \\
\\
g_{tt} >0 ~~ {\rm if} ~~M<r<M +\sqrt{M^2-a^2 \cos ^2 \theta}.
\end{array}
\end{equation}
Thus whenever $g_{tt} >0$ (small radii), we see that $\omega_{\rm min}>0$;
the observer is swept around the black hole, and no action on the part of the observer can
change that. Where $g_{tt} <0$ (large radii), $\omega_{\rm min}<0$, and retrograde motion 
is allowed. The critical boundary, the mathematical surface inside which matter is 
dragged irresistibly in the prograde direction is called the {\it static limit}, which, 
from Eq. \ref{eq:static1} is
\begin{equation}
\label{eq:static2}
r_{\rm stat}=M+\sqrt{M^2-a^2 \cos ^2 \theta},
\end{equation}
which corresponds to $g_{tt}=0$.

\begin{figure}[top]
\centering
\includegraphics[width=12cm,height=5cm]{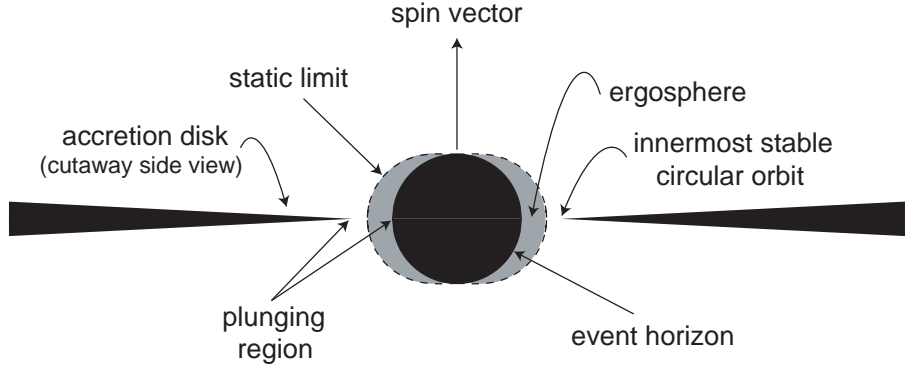}
\caption{Schematic of a disk-accreting Kerr black hole ($a=0.9M$) in cross-section, using $t$, $r$, 
$\theta$, $\phi$ coordinates, illustrating various geometric relations. The dark circle 
is the black hole, bounded by its event horizon. The gray region, delineated by the horizon 
and the static limit ({\it dashed lines}) is the ergosphere, cylidrically symmetric
with respect to the spin vector. The accretion disk is assumed to lie in the equatorial
plane of the hole. The disk is truncated near the hole according to the radius of the
innermost stable circular orbit appropriate to the spin. Inside the ISCO, matter free falls
through the plunging region and crosses the horizon. Tapering of the disk is schematic only,
suggesting a general trend of geometrical thickening with radius. Also, see Fig. (12).}
\end{figure}

The peculiar behavior of spacetime near the static limit can be illustrated \cite{taylor} if we
imagine launching light pulses in the azimuthal direction, first in the prograde direction,
then in the retrograde direction, at some radius $r$ near a spinning black hole. 
We wish to find $d\phi/dt$, the angular velocity of the light pulse
according to distant observers, for these two cases. 
For simplicity, we work in the equatorial plane. Then, with $\theta =\pi/2$, $dr=0$,
$d\theta=0$, and the condition $g_{\mu \nu}u^{\mu}u^{\nu}=0$, we find the following quadratic for
$d\phi/dt$,
\begin{equation}
R_a^2 \biggl(\frac{d\phi}{dt} \biggr)^2 -\frac{4Ma}{r} \frac{d\phi}{dt} 
-\biggl( 1-\frac{2M}{r}\biggr)=0,
\end{equation}
which has the solution
\begin{equation}
\label{eq:quadphi}
\frac{d\phi}{dt}=\frac{2Ma}{rR_a^2} ~\biggl[1 \pm 
\sqrt{1-\frac{r^2 R_a^2}{4M^2a^2} \biggl( 1-\frac{2M}{r} \biggr)} \biggr].
\end{equation}
From Eq. \ref{eq:static2}, the static limit in the equatorial plane is 
$r_{\rm stat}=2M$, which, when substituted into Eq. \ref{eq:quadphi}, gives the two
solutions $d\phi/dt=4Ma/rR_a^2=a/(a^2+2M^2)$ and $d\phi/dt=0$. The pulse emitted
in the retrograde direction appears to stand still. Inside the static limit, retrograde
pulses appear, in fact, to move initially in the prograde direction.
Although we do not pursue it here, it is worth noting that
the region between the static limit and the horizon is
called the {\it ergosphere}, after the Greek word {\it ergon}, for work, 
since energy can, in principle, be extracted from it \cite{penrose}. An illustration
of the size and shape of the ergosphere is shown in Fig. (9), along with other
components of an accreting black hole.

An expression for the rate at which spacetime is dragged around a spinning black
hole can be derived from the momentum conservation law, which, as in \S6 for
the Schwarzschild metric, we derive from the Euler-Lagrange equation for $\phi$.
A suitable Lagrangian for motion in the equatorial plane is given by
\begin{equation}
\label{eq:kerrlagrange}
\Lambda=-\frac{1}{2}\biggl(1-\frac{2M}{r} \biggr) \, \dot{t}^2 -\frac{2Ma}{r} \, \dot{t} \dot{\phi}
+\frac{r^2}{2\Delta} \,  \dot{r}^2 + \frac{R_a^2}{2} \, \dot{\phi}^2,
\end{equation}
where the dot denotes differentiation with respect to any parameter $p$
along the particle's world line \cite{shapiro}.
The Euler-Lagrange equation for the $\phi$-coordinate results in
\begin{equation}
\label{eq:kerrlconserve}
R_a^2 \, \frac{d\phi}{dp}-\frac{2aM}{r} \, \frac{dt}{dp}=L.
\end{equation}
If $dp=d\tau/m$, then we find an expression that can be compared to the
case for the Schwarzschild metric (Eq. \ref{eq:econserve}),
\begin{equation}
R_a^2 \, \frac{d\phi}{d\tau}=\frac{L}{m}+\frac{2aM}{r} \, \frac{dt}{d\tau}.
\end{equation}
to which it reduces for $a \rightarrow 0$. The second term on the right, which is
proportional to the black hole spin, shows that a particle with $d\phi/d\tau=0$
must have negative angular momentum. In fact,
for massive particles and massless particles alike, if $L=0$, then
\begin{equation}
\label{eq:drag}
\frac{d\phi}{dt}=\frac{2aM}{rR_a^2} \equiv \omega_{\rm drag}.
\end{equation}
According to a distant observer, a particle with zero angular momentum appears
to rotate in the direction of the black hole spin (i.e., following the right-hand rule).
This effect is known as the {\it dragging of inertial frames}.
The drag frequency increases with decreasing radius.
Frame dragging can also be described as follows: Imagine yourself in a rocket near a spinning
black hole. Looking at the fixed stars for reference, you find that you require
a thrust in the positive azimuthal direction (i.e., pointed toward 
the spin direction) to hold the rocket in a position such that the stars 
overhead appear not to move. It is said that spacetime itself is moving.

\begin{figure}[top]
\centering
\includegraphics[width=12cm,height=6cm]{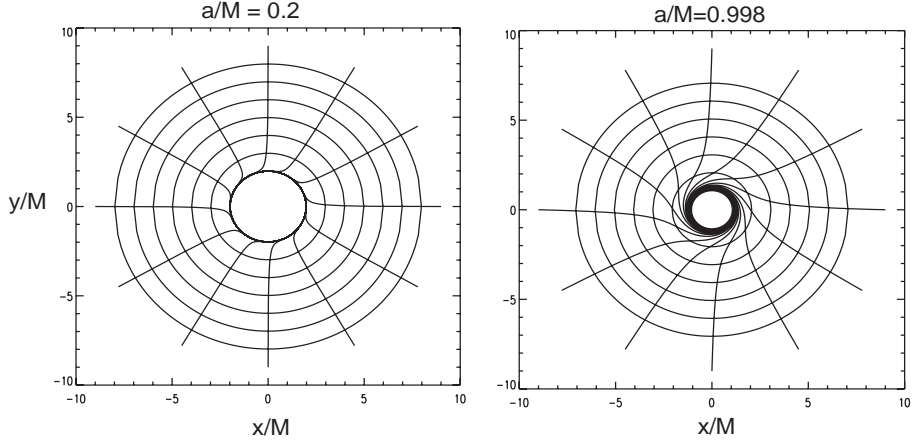}
\caption{Plots of photon trajectories with zero angular momentum near a Kerr black hole, for 
$a=0.2M$ {\it (left panel)} and $a=0.998M$ {\it (right panel)} illustrating the frame
dragging effect. Circles are lines of constant $r$. Plots are based on integrations of Eq.
\ref{eq:kerrspiral}. }
\end{figure}

The dragging of inertial frames occurs outside of any spinning body. As an example,
we use Eq. (\ref{eq:drag}) to estimate $\omega_{\rm drag}$ caused by the Earth's rotation.
In conventional units, the drag frequency at the Earth's surface is $\omega_{\rm drag} \approx 2 GI
\omega_E/(c^2 R_E^3)$, where $I$ is Earth's moment of inertia, $R_E$ is its
radius, and $\omega_E$ is its rotation frequency. The moment of inertia of the Earth is $0.33
M_ER_E^2$ \cite{allen}, where the deviation from 2/5 in the multiplier is due to
the non-uniform density distribution of Earth's mass. Taking $M_E=6.0 \times 10^{27}$ g,  
$R_E=6.4 \times10^8$ cm, and $\omega_E=7.3 \times 10^{-5}$ s$^{-1}$, we find
$\omega_{\rm drag} = 219$ mas yr$^{-1}$, large enough to be measured with current technology. 
Recently, the LAGEOS mission, a pair of laser-ranged satellites on orbits with aphelia
of approximately 12,000 km, measured $\omega_{\rm drag}$,  finding a value that is
$99 \pm 5$ percent of the value predicted by GR \cite{ciuf}. Gravity Probe-B
is expected to reduce the uncertainty in the measurement to 1$\%$.

Interestingly, we find for the Earth that $a=7.3 \times 10^2 ~M_E$. The restriction 
$a \leq M$ is clearly not a general constraint on material bodies; instead, it 
can be viewed as a necessary condition for matter lying inside its event horizon.

\subsection{Gravitational Redshift in the Kerr Metric}

In the Schwarzschild metric, we found the relation between proper time and coordinate time
by considering two stationary observers at different radii, and found that the $tt$-component of the metric
contained the necessary information. An analogous approach in the Kerr metric involves
an observer, called a {\it zero angular momentum observer} (ZAMO), moving along the 
$\phi$-direction with an angular velocity $\omega_{\rm drag}$ given by Eq. \ref{eq:drag}. 
By a simple transformation, the metric can be
diagonalized, i.e., we can decouple the $t$ and $\phi$ coordinates, then relate $dt$ to $d\tau$. Let
$d\phi=\omega_{\rm drag} \, dt + d\phi_r$, where $d\phi_r$ is an azimuthal displacement
measured in the ZAMO frame. The case of greatest interest is that of an emitter in the equatorial plane.
Therefore, for pure azimuthal motion, Eq. (\ref{eq:kerrplane}), with $dr=0$, becomes
\begin{equation}
d\tau^2=\biggl(1-\frac{2M}{r} \biggr) \, dt^2 +
\frac{4Ma}{r} \, dt \, d\phi -R_a^2  \, d\phi^2.
\end{equation}
Substituting the $d\phi$ transformation into this form of the metric yields a new
metric that uncouples the $t$- and $\phi$-coordinates:
\begin{equation}
\label{eq:phitrans}
d\tau^2 =\biggl(1-\frac{2M}{r} +\frac{4 M^2 a^2}{r^2 R_a^2} \biggr)~dt^2 -R_a^2 ~d\phi_r^2.
\end{equation}
This shows that the reduced circumference in the equatorial plane of a Kerr hole, 
the analog of $r$ used to describe the Schwarzschild geometry, is $R_a$. Equation (\ref{eq:phitrans})
gives us the relation between time for a ZAMO (set $d\phi_r=0$) and a distant observer,
which we write as a redshift formula for frequency:
\begin{equation}
\nu_{\infty}=\nu_o ~\biggl(1-\frac{2M}{r} +\frac{4 M^2 a^2}{r^2 R_a^2} \biggr)^{1/2},
\end{equation}
or, after expanding $R_a$, and rearranging,
\begin{equation}
\label{eq:kerrshift}
\nu_{\infty}=\nu_o ~\biggl(\frac{r^2+a^2 - 2Mr}{r^2+a^2+2Ma^2/r} \biggr)^{1/2},
\end{equation}
which reduces to Eq. (\ref{eq:redshift}) in the limit $a \rightarrow 0$, the
Schwarzschild case. As a consistency check, note that the surface of infinite redshift,
the outer solution of $r^2+a^2 - 2Mr=0$, is identical to our earlier result, 
Eq. (\ref{eq:kerrhorizon}).

The multiplier of $\nu_o$ in Eq. (\ref{eq:kerrshift}) is sometimes called the {\it lapse
function}. Since we use this function later in the calculation of orbital velocities, it is
defined here and given the label $\alpha$:
\begin{equation}
\label{eq:lapse}
\alpha=\frac{\sqrt{r^2+a^2 - 2Mr}}{R_a}.
\end{equation}

\subsection{Circular Orbits in the Kerr Potential}

Evaluation of the Euler-Lagrange equation for $\dot{t}$ provides
an energy constant of the motion,
\begin{equation}
\biggl(1-\frac{2M}{r}\biggr) \, \dot{t}+\frac{2Ma}{r} \, \dot{\phi}=E,
\end{equation}
using the Lagrangian in Eq. (\ref{eq:kerrplane}).
We have already evaluated the $\dot{\phi}$ component (Eq. \ref{eq:kerrlconserve}),
which we reproduce here,
\begin{equation}
-\frac{2Ma}{r} \, \dot{t} +R_a^2 \dot{\phi}=L.
\end{equation}
These two equations can be solved for $\dot{t}$ and $\dot{\phi}$ in terms
of the energy and angular momentum, yielding
\begin{equation}
\label{eq:tdotkerr}
\dot{t}=\frac{ErR_a^2-2LMa}{r\Delta}
\end{equation}
\begin{equation}
\label{eq:phidotkerr}
\dot{\phi}=\frac{L(r-2M)+2EMa}{r\Delta}.
\end{equation}
Substitiution of Eqs. (\ref{eq:tdotkerr}) and (\ref{eq:phidotkerr}) into the Lagrangian
(Eq. \ref{eq:kerrlagrange}), noting that $\Lambda=-m^2/2$, yields an equation for $\dot{r}$,
which we write in terms of $dr/d\tau$, substituting $p=\tau/m$,
\begin{equation}
\label{eq:Skerr}
r^3\biggl(\frac{dr}{d\tau}\biggr)^2 = S \equiv 
\biggl(\frac{E}{m}\biggr)^2  rR_a^2
-4Ma\biggl(\frac{L}{m} \biggr) \biggl( \frac{E}{m} \biggr)
-(r-2M) \, \biggl(\frac{L}{m} \biggr)^2-r\Delta.
\end{equation}

Stable circular orbits can be found by imposing three conditions on $S$:
(1) $S=0$ (turning point); (2) $\partial S/\partial r=0$ (circle); (3)
$\partial^2 S/\partial r^2 \leq 0$ (stability). The first two conditions
yield a system of two equations that can be solved for $E/m$ and $L/m$,
restricting those quantities for a given $r$ and $a$:
\begin{equation}
\label{eq:ekerr}
\frac{E}{m}=\frac{r-2M+a\sqrt{M/r}}{(r^2-3Mr+2a\sqrt{Mr})^{1/2}},
\end{equation}
\begin{equation}
\label{eq:lkerr}
\frac{L}{m}=\biggl(\frac{M}{r}\biggr)^{1/2} 
\frac{r^2-2a\sqrt{Mr}+a^2}{(r^2-3Mr+2a\sqrt{Mr})^{1/2}}.
\end{equation}

\begin{figure}[top]
\centering
\includegraphics[width=14cm,height=6cm]{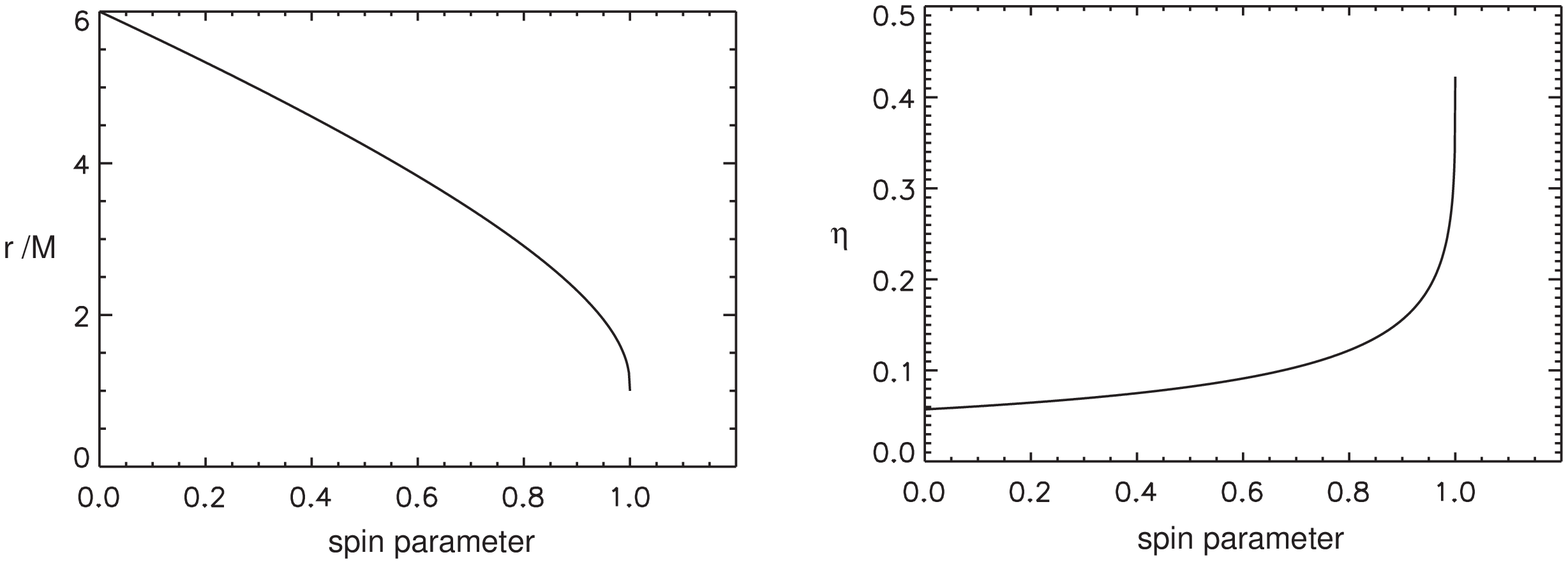}
\caption{({\it Left panel}) Radius of innermost stable circular orbit plotted
against the spin parameter $a_*$. Radii of the ISCOs range from $6M$ ($a_*=0$) to $M$ ($a_*=1$). 
For physically realizable Kerr black holes, the maximum spin is $a_*=0.998$,
corresponding to an ISCO radius of $1.24M$. ({\it Right panel}) Dimensionless
accretion efficiency $\eta$ plotted against $a_*$. Efficiency ranges from 0.057
($a_*=0$) to 0.42 ($a_*=1$). For $a_*=0.998$, $\eta=0.34$. }
\end{figure}

The third condition requires that
\begin{equation}
\biggl(\frac{E}{m}\biggr)^2-1+\frac{2M}{3r} \leq 0.
\end{equation}
Substituting from Eq. \ref{eq:ekerr}, and defining the dimensionless {\it spin parameter} 
$a_*=a/M$, the previous restriction becomes
\begin{equation}
\label{eq:quartic}
\biggl(\frac{r}{M} \biggr)^2 -6 \frac{r}{M}
+8a_*\biggl(\frac{r}{M} \biggr)^{1/2} -3a_*^2 ~\geq 0,
\end{equation}
where the equality can be treated as a quartic equation in $(r/M)^{1/2}$.
The quartic has been solved \cite{bardeen72} for the radius of the innermost 
stable circular orbit for a specified spin parameter:
\begin{equation}
\label{eq:riscokerr}
r_{\rm isco}=M \biggl[ 3+Z_2 - \sqrt{(3-Z_1)(3+Z_1+2Z_2)} \biggr],
\end{equation}
where $Z_1$ and $Z_2$ are defined by
\begin{equation}
\begin{array}{l}
Z_1=1+(1-a_*^2)^{1/3}~
[ ( 1+a_* )^{1/3} + (  1-a_* )^{1/3} ] \\
\\
Z_2=( Z_1^2 + 3 a_*^2 )^{1/2}.
\end{array}
\end{equation}
Equation (\ref{eq:riscokerr}) generalizes the result based on Eq. (\ref{eq:circle}) 
for the Schwarzschild metric. As shown in he left panel of Fig. (11), $r_{\rm isco}$ is a monotonically 
decreasing function of spin, ranging from a maximum of $6M$ for a Schwarzschild hole to 
a minimum of $M$ for a maximally-spinning Kerr hole. In Fig. (12) we show schematically
the variation of the position of the ISCO of an accretion disk compared to the size of
the event horizon and the shape of the ergosphere, where, for example, it is seen that
for large values of $a_*$ the ISCO lies inside the ergosphere, and approaches the event horizon.

If we assume that the accretion efficiency $\eta$ is determined by the energy of the 
innermost stable circular orbit, then the maximum efficiency
occurs for the case of maximal spin. With $r=M$ and $a=M$, Eq, (\ref{eq:ekerr}) gives
$E/m=1/\sqrt{3}$. Therefore, for a disk extending down to $r_{\rm isco}$
for a maximally spinning black hole, the accretion efficiency is
\begin{equation}
\label{eq:extract2}
\eta (a_*=1)=\Delta \biggl(\frac{E}{m} \biggr)=1-\biggl(\frac{1}{3} \biggr)^{1/2} \approx 0.423,
\end{equation}
which can be compared to Eq. (\ref{eq:extract1}) for the Schwarzschild metric.
The energy conversion efficiency $\eta$
is plotted as a function of the spin parameter $a_*$ in the right panel of Fig. (11). 

\begin{figure}[top]
\centering
\includegraphics[width=8cm,height=9cm]{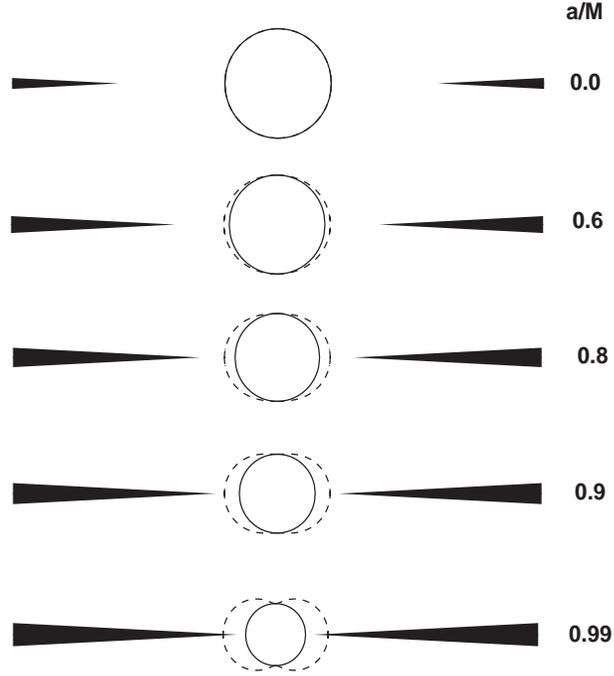}
\caption{Comparison of the radius of the event horizon ({\it solid circles}), the static limit
({\it dashed lines}) and ergosphere,  and position of the innermost stable circular orbit for 
black holes of various spin parameters.}
\end{figure}

The velocity of matter in a circular orbit around a Kerr black hole, as measured by a ZAMO, 
can be expressed in terms of the orbit's angular frequency as reckoned by a distant observer 
($\Omega$) as
\begin{equation}
\label{eq:vkerr1}
v_{\phi}=\frac{R_a}{\alpha} \, (\Omega-\omega_{\rm drag}),
\end{equation}
where $R_a$ (Eq. \ref{eq:ra}) is the reduced circumference of a spinning hole, 
$\omega_{\rm drag}$ (Eq. \ref{eq:drag}) corrects for the dragging of inertial frames, 
and the lapse function $\alpha$ (Eq. \ref{eq:lapse}) accounts for the transformation 
between measures of time.  An expression for $\Omega$ \cite{novikovthorne}
\cite{lightmanprice} is most easily found by starting with the equation of motion  
\begin{equation}
\label{eq:motion}
\frac{d^2 x^{\lambda}}{d\tau^2}+\Gamma_{\mu \nu}^{\lambda} \, \frac{dx^{\mu}}{d\tau}
\, \frac{dx^{\nu}}{d\tau} =0,
\end{equation}
where the {\it Christoffel symbols} are given in terms of the metric $g_{\mu \nu}$ and the
inverse metric $g^{\mu \nu}$ by
\begin{equation}
\Gamma_{\mu \nu}^{\lambda}=\frac{1}{2} \,g^{\eta \lambda} \, \biggl[
\frac{\partial g_{\nu \eta}}{\partial x^{\mu}} +  
\frac{\partial g_{\mu \eta}}{\partial x^{\nu}} -
\frac{\partial g_{\nu \mu}}{\partial x^{\eta}} \biggr],
\end{equation}
whose derivation can be found in \cite{weinberg}.
For a circular orbit, using Eqs. (\ref{eq:tdotkerr}) and (\ref{eq:phidotkerr}), the first
term on the left of Eq. (\ref{eq:motion}) vanishes for all $\lambda$. If we choose $\lambda=r$, then
\begin{equation}
\label{eq:motion2}
\Gamma_{\mu \nu}^{r} \, dx^{\mu} dx^{\nu} = \Gamma_{tt}^{r}~ dt^2 + 2 \, \Gamma_{\phi t}^{r} ~dt \, d\phi
+\Gamma_{\phi \phi}^{r} ~d\phi^2=0,
\end{equation}
or, letting $\Omega=d\phi/dt$,
\begin{equation}
\label{eq:motion3}
\Gamma_{\phi \phi}^{r} ~\Omega^2  + 2 \, \Gamma_{\phi t}^{r} ~\Omega + \Gamma_{tt}^{r}=0.
\end{equation}
An evaluation of the Christoffel symbols gives
\begin{equation}
\begin{array}{l}
\Gamma_{tt}^{r}= \frac{1}{2} g^{rr} (2M/r^2) \\
\\
\Gamma_{\phi t}^{r}=\frac{1}{2} g^{rr} (-2Ma/r^2)\\
\\
\Gamma_{\phi \phi}^{r} = \frac{1}{2} g^{rr} \, (2Ma^2/r^2 - 2r),\\
\end{array}
\end{equation}
which, when substituted into Eq. (\ref{eq:motion3}), leaves us with the simple quadratic equation
\begin{equation}
\biggl( \frac{Ma^2}{r^2}-r\biggr) \,\Omega^2 -\frac{2Ma}{r^2} \,\Omega+\frac{M}{r^2}=0.
\end{equation}
The solution corresponding to prograde motion  is 
\begin{equation}
\label{eq:kerromega}
\Omega=\frac{\sqrt{M/r}}{r+a\sqrt{M/r}}=\frac{1}{M} ~ \frac{1}{(r/M)^{3/2}+a_*}.
\end{equation}
This expression for $\Omega$ can also be derived by taking the ratio of
$\dot{\phi}$ and $\dot{t}$ from Eqs. (\ref{eq:phidotkerr}) and (\ref{eq:tdotkerr}),
respectively, then substituting for $E$ and $L$ from Eqs. (\ref{eq:ekerr}) and (\ref{eq:lkerr}),
respectively, but the algebra is tedious.

\begin{figure}[top]
\centering
\includegraphics[width=14cm,height=6.5cm]{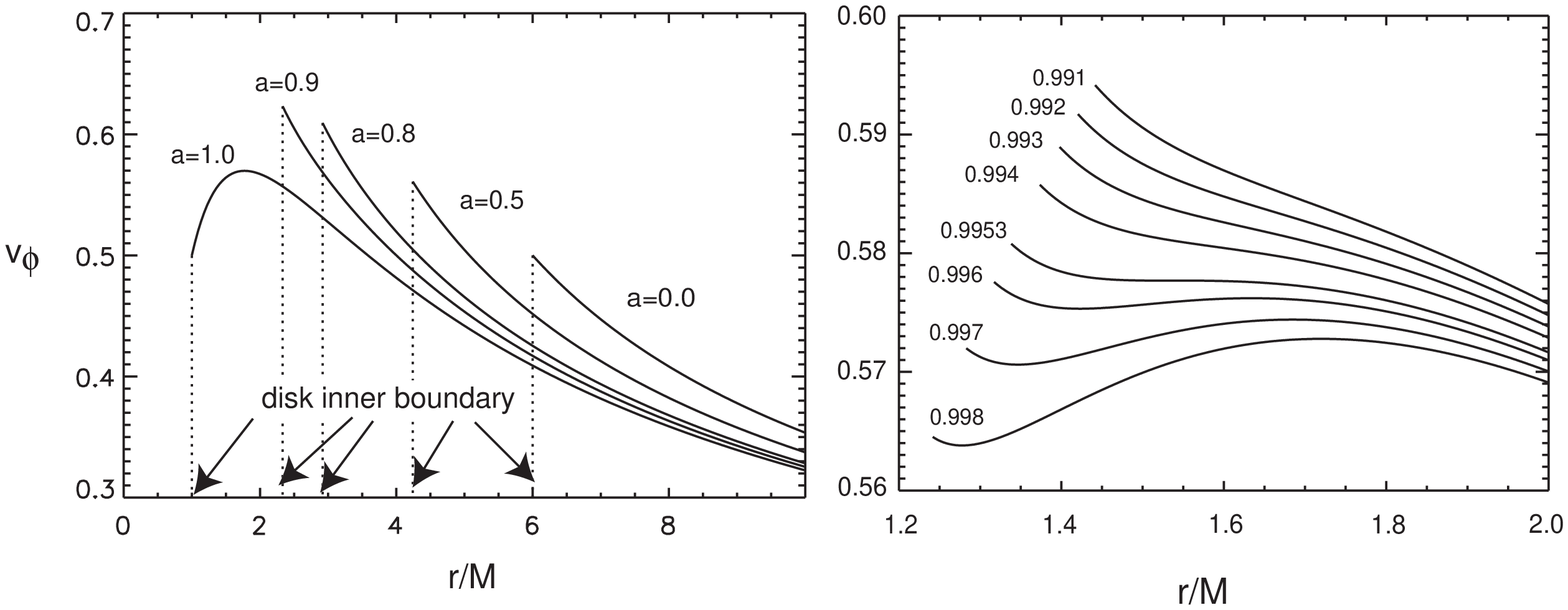}
\caption{{\it (Left panel)} Azimuthal velocities for equatorial circular 
orbits as a function of radius for
five values of black hole spin, as measured by locally stationary observers. 
Curves are labeled by the spin as fractions of
$M$. Vertical lines indicate radii corresponding to the innermost stable circular orbit.
{\it (Right panel)} Azimuthal velocities for a selected group of spin parameters (as labeled),
illustrating the development of a local minimum for $a_*>0.9953$.}
\end{figure}

Substituting Eqs. (\ref{eq:ra}), (\ref{eq:drag}), (\ref{eq:lapse}), and (\ref{eq:kerromega}) 
into Eq. (\ref{eq:vkerr1}) yields an expression for the velocity of circular orbits \cite{bardeen72},
\begin{equation}
\label{eq:vphikerr}
v_{\phi}=\biggl(\frac{M}{r}\biggr)^{1/2}~
\frac{r^2-2a\sqrt{Mr}+a^2}
{(r+a\sqrt{M/r})\sqrt{r^2-2Mr+a^2}},
\end{equation}
which reduces to the Schwarzschild case,
Eq. (\ref{eq:vshell}), for $a=0$.
For the other limit, $a \rightarrow M$, the numerator can be factored to give
\begin{equation}
v_{\phi}=\biggl(\frac{M}{r}\biggr)^{1/2}~
\frac{r^{3/2}+M^{1/2}r+Mr^{1/2}-M^{3/2}}
{(r+M^{3/2}r^{-1/2})(r^{1/2}+M^{1/2})},
\end{equation}
which shows that as $r \rightarrow M$, the velocity approaches $v_{\phi}=1/2$,
which is, perhaps surprisingly, identical to the velocity at the ISCO of a 
Schwarzschild black hole. The left panel of Fig. (13) shows the radial profile
of $v_{\phi}(r)$ for several different spin parameters. The maximum orbital velocity
attained is $v_{\phi}^{\rm max}=0.624$ for $a_*=0.9268$, which occurs at the ISCO with 
$r_{\rm isco}=2.14M$.

Another feature worth noting is that for the  extreme $a=M$ case, 
the maximum velocity is approached from above, i.e., $v_{\phi}(r)$ 
is not monotonic. For more realistic cases, where the maximum value 
of $a_*$ is 0.998, the velocity at the ISCO is always approached from 
below. But this velocity is not always the maximum velocity. For 
$a_*=0.9964$ and above, the radius of maximum velocity $r(v_{\rm max})$ 
is distinct from the ISCO, as can be seen in the right panel of Fig. (13) 
(lowest two curves), with the largest separation $r(v_{\rm max})-r_{\rm isco}$ 
occurring for the maximally spinning case, where $r(v_{\rm max})=1.72M$
and where $v_{\rm max}=0.573$. The non-monotonic behavior of  $v_{\phi}(r)$ 
sets in for $a_*>0.9953$.  
For $a_*$ in this range, $v_{\phi}(r)$ has a local minimum that moves 
inward as the spin parameter increases. Recently, it was proposed 
\cite{aschen} that this feature of $v_{\phi}$ may be related  to 
the fixed ratio of high frequency quasi-periodic oscillations (QPOs) 
observed in Galactic black hole X-ray binaries.

\begin{figure}[top]
\centering
\includegraphics[width=11cm,height=7cm]{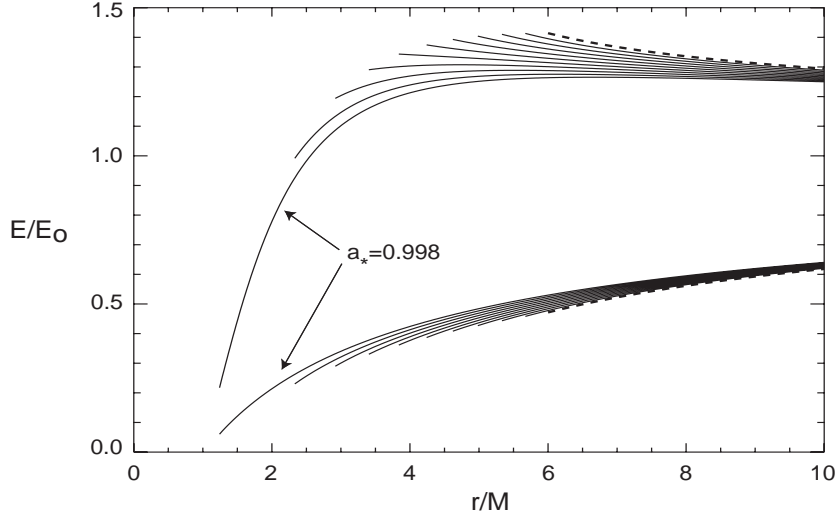}
\caption{Minimum and maximum normalized line energies observed at infinity vs. disk radius for several
black hole spin parameters, assuming circular orbits and 90 degree inclination. The upper 
set of curves shows the maximum blueshifts for the eleven spin parameters $a_*=$ [0., 0.1,
0.2, 0.3, 0.4, 0.5, 0.6, 0.7, 0.8, 0.9, 0.998], working downward from the dashed line 
(Schwarzschild case). The lower set of curves shows the corresponding maximum 
redshifts, working upward from the $a_*=0$ case. Pairs of curves delineate the 
maximum width of the line profile for a given radius, neglecting thermal Doppler 
widths, and any possible radiative transfer effects. Curves are truncated at the 
ISCO corresponding to each spin parameter.}
\end{figure}

Having obtained formulae for the gravitational redshift (Eq. \ref{eq:kerrshift}) and the velocities
of circular orbits (Eq. \ref{eq:vphikerr}), we now follow the same procedure used for Schwarzschild
black holes (\S5.8) to find the minimum and maximum line energies as observed at infinity as a function
of radius for several spin parameters. The result is shown in Fig. (14), where 
the extrema are plotted against radius for eleven values of the spin parameter.  From that figure one
sees that the differences between spinning black holes and a Schwarzschild black hole are fairly minor
for $r>6M$, viz., the maximum extent of the blue wing is the same to within about 10\%, for $r>6M$, and
the maximum redshift is even less distinct when comparing different spins. If, for example,
the disk of a maximally spinning hole did not, for some reason, produce line emission 
inside of about 6$M$, one would be hard-pressed to discern spin by measuring the line 
width alone. Figure (14) also shows that the line width is bounded by the Schwarzschild 
``envelope'' for $r>6M$ (dashed line).  The maximum blueshift is a monotonically decreasing 
function of spin; the maximum blueshift never exceeds the spin-zero case $E/E_o=1.41$ at $r=6M$, 
which may be surprising. For a given spin, and moving to smaller radii, the modest change 
in the maximum blueshift illustrates the trade-off between increasing orbital velocity and increasing
gravitational redshift. For large spin values, the maximum blueshift is actually redshifted from line center,
showing the dominant efffect of gravitational redshift. Thus the radius of maximum blueshift is not
always the innermost stable circular orbit. This is most obvious for maximally spinning holes,
where the energy of the blue side of the line approaches $\approx 0.2E_o$, since the ISCO approaches the
event horizon. The maximum redshift always occurs at the ISCO, with the minimum value
of $E/E_o$ being 0.060 for $a_*=0.998$. The maximum redshift increases monotonically with spin, since the
effects of increasing velocity and increasing gravitational redshift reinforce each other, rather
than offset each other, as they do when evaluating the maximum blueshift. One also sees from the figure
that the total line width changes with radius. For example, an emission profile from an
annulus near $r=2M$ in a maximally spinning hole would appear relatively narrow, with a highly 
redshifted centroid energy. The narrowing of the line follows from the overwhelming 
effect of gravitational redshift as the ISCO approaches the event horizon. 

In any realistic situation, one expects fluorescence from a large range of radii. 
A model line profile is built up annulus by annulus,
where account is taken of the intrinsic surface brightness variation, the fraction of emitted
photons that escape the system and find their way to the observer (given light bending), 
and the dynamical aspects discussed above. These model profiles can then be compared to 
spectroscopic data, which, in principle, allows one to extract several parameters of interest
(disk inclination, radial emissivity profile, disk inner edge, etc.).
Considered as a diagnostic of black hole spin, the distinguishing characteristic of line emission is,
from Fig. (14), the extent of the red wing. By contrast, the blue wing is more sensitive to the
inclination than to the spin. Given an observed line profile, the inclination, to a first approximation,  
can be determined by the position of the blue edge. The spin can then be estimated from the redward extent
of the line profile. It's not quite that easy, unfortunately, but computer programs allowing
fits to relativistic line profiles exist and are in wide use (e.g., \cite{fabian89}
\cite{laor}), made available, for example, as part of the XSPEC \cite{arnaud} or ISIS
\cite{houck} data analysis packages. 

The detection of highly redshifted iron K$\alpha$ emission has led to inferences of non-zero spin,
and has given rise to the idea that X-ray lines can be used to measure black hole spin. A
dramatic example of a red-shifted iron K$\alpha$ line profile has been documented for the Seyfert 1 galaxy
MCG--6-30-15 \cite{iwasawa96} \cite{iwasawa99}, in which the line profile was observed to vary
dramatically, from a profile for which the emission appeared to be dominated by large radii,
and then to a profile in which the bulk of the emission was observed to lie below the rest energy.
This latter result implies that the emission originated from $r<6M$, which, given the assumption
that radiation from within the ISCO is unobservable, implies black hole spin. In fact, a fit 
to the data requires $a_*>0.94$ \cite{dabrowski}. 

\subsection{The Plunging Region}

As orbiting matter in accretion disk reaches the ISCO, it is typically assumed to
lose all rotational support and free fall from there through the event horizon.
In terms of test particle trajectories, the gradually shrinking circular orbits
are replaced by a rapid spiraling infall. We can calculate such a trajectory
by eliminating $d\tau$ between Eqs. (\ref{eq:lconserve}) and (\ref{eq:orbit}), 
from which we find the following expression for the trajectory of a test particle in the 
Schwarzschild metric:
\begin{equation}
\label{eq:drdphimass}
\frac{dr}{d\phi}=\pm ~ \frac{r^2}{L/m}~\biggl[\biggl(\frac{E}{m}\biggr)^2- 
\biggl(1-\frac{2M}{r}\biggr) \biggl(1+\frac{(L/m)^2}{r^2}\biggr)  \biggr]^{1/2}.
\end{equation}
If we set $E/m=\sqrt{8/9}$
and $L/m=2\sqrt{3}M$, the values for a circular orbit at the ISCO, then add a slight radial
displacement inward, an orbit such as that shown in Fig. (15) results.

The sudden increase in radial velocity results in a reduced disk optical depth,
which may reduce the X-ray albedo.
The free falling matter is also conventionally assumed to be stress-free, so that a hot corona
cannot be supported within the ISCO. Therefore, irradiation inside the ISCO by hard X rays 
is reduced relative to matter at larger radii. Finally, the drop in density may imply
an increase in $\xi$, which reduces the overall efficiency of fluorescence line production. 
Taken together, these aspects of disks imply that any fluorescence must occur outside the 
ISCO. In other words, the radii over which observable fluorescence emission originates has a hard 
inner limit of $r=6M$ for a Schwarzschild black hole. Given such a constraint,
the assumption of a zero-spin black hole therefore limits the maximum redshift
of the line profile (see Fig. 14). Profiles violating this limit have thus been taken 
to imply spin (see \S6.4).

\begin{figure}[top]
\centering
\includegraphics[width=8cm,height=8cm]{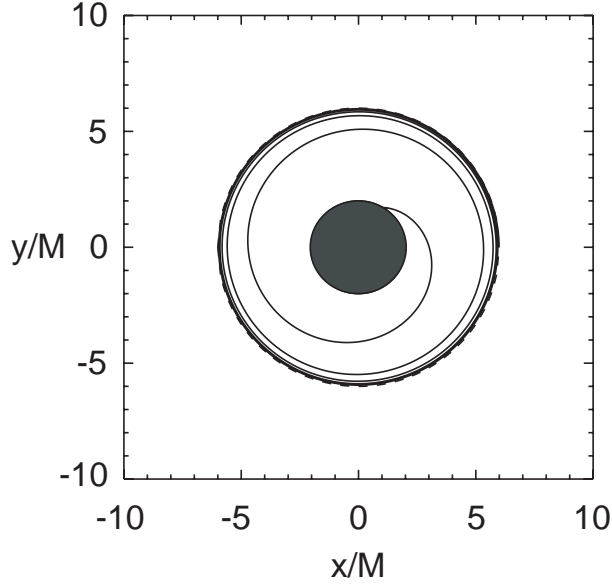}
\caption{Trajectory of free falling test particle inside the innermost stable circular orbit
--- the plunging region --- of a Schwarzschild black hole (see Eq. \ref{eq:drdphimass})
according to an observer at infinity.  The energy and angular momentum are set to the 
corresponding values at $r=6M$. In order to perturb the particle off the stable orbit at 
$r=6M$, the initial radius was set to $5.99M$.  }
\end{figure} 

The whole concept of a clearly demarcated inner disk edge has recently been called into question,
based both upon test particle trajectories \cite{reynolds97} and
upon three-dimensional MHD simulations of accretion disks \cite{krolikhawley}.
In the former paper, with a source of hard X-ray illumination located on the axis
above the disk plane, it was found that substantial fluoresecence line flux from
inside the ISCO may emerge from the system, and that this component of the flux
appears at redshifts exceeding the usual limit imposed by a hard cutoff at $r=6M$,
thus mimicking emission from a spinning black hole. The importance of emission from
inside the ISCO of a non-spinning hole has been contested \cite{young}, however, based
upon the absence of an iron K absorption edge in the well-studied source MCG--6-30-15
that is predicted by modeling calculations. In \cite{krolikhawley} 
stress is found to be continuous across the ISCO,  invalidating the usual assumption that the stress vanishes
there. Moreover, it was found that the matter density distribution need not conform to the simpler model
wherein the density decreases monotonically with decreasing radius, but that clumpiness
may permit the survival of matter with lower values of $\xi$. Thus there is some doubt
cast upon our ability to make clean measurements of black hole spin using the iron 
K$\alpha$ profile alone.
 
\subsection{The Motion of Light in the Kerr Metric}

A general treatment of light motion in the Kerr metric is considerably more complex
than the Schwarzschild case (see, e.g. \cite{rauch}. For example, except for the 
case of motion in the equatorial plane, trajectories are non-planar
(see Fig. 16). Discussions of the techniques used to calculate photon geodesics near Kerr black
holes may be found in \cite{rauch} \cite{beckwith} \cite{dovciak} \cite{li}.

We can, however, work out the relatively simple example of a massless particle with $L=0$  
moving in the equatorial plane of a spinning black hole. We find an expression for $dr/d\phi$ from the
following two equations:
\begin{equation}
\label{eq:spiralcouple}
\begin{array}{l}
R_a^2 \,d\phi-(2Ma/r) \,dt=0 \\
~ \\
(1-2M/r) \, dt^2 +(4Ma/r) \, dt \, d\phi - (r^2/\Delta) \, dr^2  
- R_a^2  \, d\phi^2 =0.
\end{array}
\end{equation}
The first of this set comes from Eq. (\ref{eq:kerrlconserve}) with $L=0$,
and the second is simply the metric (Eq. \ref{eq:kerrplane}) with $d\tau=0$.
Equation (\ref{eq:spiralcouple}) is a system of two equations in the variables $dt$, $dr$, and $d\phi$.
By eliminating $dt$, we arrive at an expression involving only $dr$ and $d\phi$
that describes the photon trajectory:
\begin{equation}
\label{eq:kerrspiral}
\frac{dr}{d\phi}=\pm ~\frac{R_a \Delta}{2Ma},
\end{equation}
for which a straightforward numerical integration produces the plots shown in Fig. (10),
where trajectories for $L=0$ photons follow the spacetime drag for two values of the
spin parameter.

\begin{figure}[top]
\centering
\includegraphics[width=9cm,height=6.3cm]{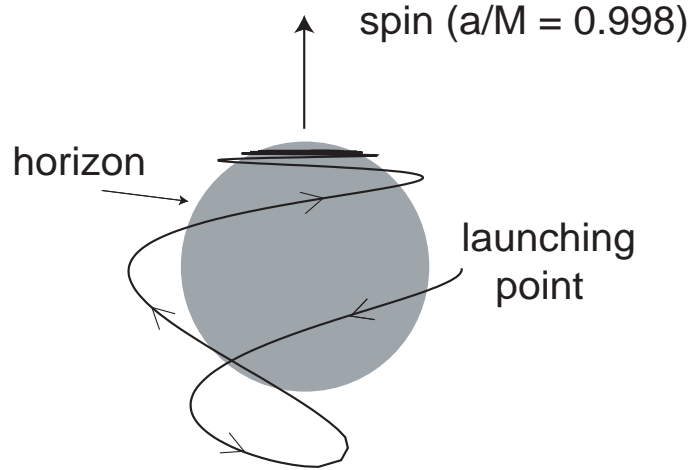}
\caption{Trajectory for a massless test particle near a maximally spinning black hole,
showing the generally non-planar character of photon trajectories in the Kerr spacetime.
The path is projected onto the $x$-$z$ plane, in which the angular momentum vector of
the black hole lies. Photon is launched from $r=1.27 R_g$, with an initial momentum vector
predominantly in the positive $y$-direction, with a small positive $x$ component and a small
negative $z$ component. The photon executes several ``circuits'' before impacting the horizon 
near the pole. The calculation was performed following the methods described in \cite{rauch}.}
\end{figure} 

\section{X-ray Fluorescence Spectroscopy of Accreting Black Holes}

Having introduced the basic aspects of GR and accretion disk theory,
we look now at the production of X-ray fluorescence line emission in black hole
accretion disks. We start with a brief discussion of photoionization codes, and introduce the
ionization parameter concept. Then we treat the phenomenon of ``reflection,'' the spectral response
of an optically thick medium to irradiation by a hard X-ray source. We close the section by
presenting a few aspects of fluorescence at the atomic scale. The treatment of the latter topic is primarily
from the point of view of atomic modeling, rather than from a quantum mechanical standpoint.

\subsection{X-Ray Photoionization Codes}

Photoionization codes, such as those described in references \cite{ferland} and 
\cite{xstar2}, are used to determine
the effect of a radiation field on a gas of specified chemical composition, and the self-consistent
effect that passage through the gas has on the radiation field. In other words,
the opacity determines the effect of the gas on the radiation field, but the
radiation field partly determines the opacity, primarily through its influence
on the charge state distribution (ionization balance) \cite{krolikkallman}. 
In addition to acting as the dominant agent of ionization, photoionization 
also heats the plasma, since suprathermal photoelectrons are thermalized after interacting
with the local population of Maxwellian-distributed electrons. Compton scattering
and the Auger effect (see below) also contribute to plasma heating. One may add a source of non-ionizing
heating to mimic the sum of various other processes \cite{kmt}. Plasma cooling
is primarily through recombination, collisionally-excited line emission, bremsstrahlung,
and inverse Compton scattering. Also, imposing the constraint of
charge neutrality controls the overall free electron to ion ratio.
The explicit calculation of heating and cooling thus couples
the energy equation ({\it radiative heating = radiative cooling}) to the ionization equations
and the neutrality equation, constituting a complex system of equations, whose 
solution requires an iteration scheme. 

Three approaches to calculating level populations can be used: (1) the nebular approximation,
in which only the ground state has a significant population, (2) the Saha-Boltzmann approach,
in which the populations of excited levels $k$ are given by $n_k=(g_k/g_1) \exp(-E_k/kT_e)$,
where
$g$ denotes the statistical weight factor, $E_k$ is the level energy with respect to ground,
and $kT_e$ is the local electron temperature, and (3) detailed level 
accounting, where level populations are calculated explicitly by diagonalizing the 
rate matrix, the elements of which include all rates into and out of each level, 
thus requiring many thousands of atomic rate coefficients. Once the level 
populations are specified, the local contribution to  the overall spectrum is 
determined. Finally, by more or less approximate methods, radiation transport of 
lines and continua from their sites of creation to the observer is accounted for 
\cite{hubeny} \cite{dumont} \cite{mauche} \cite{ross05}. Thorough discussions of 
photoionization codes and their applications can be found in \cite{ferland} 
\cite{davidson} \cite{halpern} \cite{km82} \cite{kallmanlex}.

\subsubsection{The Ionization Parameter}

The physical conditions in X-ray photoionized plasmas are,
for a given ionizing spectrum, often described in terms the
{\it ionization parameter} \cite{tarter}. 
This quantity arises naturally from the steady-state
equations of ionization equilibrium. Let $\beta_i$, $C_i$, and 
$\alpha_{i+1}$ denote the photoionization rate (s$^{-1}$) of charge
state $i$, the collisional ionization rate coefficent ($\rm{cm^3 ~s^{-1}}$)
of $i$, and the recombination rate coefficient ($\rm{cm^3 ~s^{-1}}$) 
of charge state $i+1$, respectively. The term $\alpha_{i+1}$ accounts
for all two-body recombination processes. Ignoring three-body recombination,
as well as coupling to charge states more than one charge away,
the steady state equations of ionization equilibrium can be written 
\begin{equation}
\label{eq:balance1}
[\beta_i +n_e C_i(T_e)]~ n_i  =n_e  \alpha_{i+1}(T_e)~ n_{i+1},
\end{equation}
where the rate coefficients for recombination and collisional ionization depend explicitly
on the electron temperature $T_e$.
In terms of the photoionization cross-section $\sigma_i$ and ionization
threshold energy $\chi_i$ of charge state $i$, the photoionization rate for a 
point source of ionizing continuum can be written as an integral over photon energy:
\begin{equation}
\label{eq:photorate}
\beta_i =\frac{L_x}{r^2}~ \int_{\chi_i}^{\infty}dE~\frac{S_E(E)}{4\pi E}~\sigma_i(E),
\end{equation} 
where $S_E$ is the spectral shape function, normalized on a suitable
energy interval. Denoting the integral in Eq. (\ref{eq:photorate}) by $\Phi_i$,
Eq. (\ref{eq:balance1}) becomes
\begin{equation}
\frac{L_x}{n_e r^2}~\Phi_i~ n_i + C_i (T_e)~n_i = \alpha_{i+1}(T_e) ~n_{i+1}.
\end{equation}
Let $\xi=L_x/n_e r^2$, which is called the {\it ionization parameter}. Then
\begin{equation}
\label{csd}
\frac{n_{i+1}}{n_i}=\frac{C_i(T_e)+\xi \Phi_i}{\alpha_{i+1}(T_e)}
\approx \frac{\xi \Phi_i}{\alpha_{i+1}(T_e)},
\end{equation}
where the second approximate equality applies for many cases of interest.
Such a plasma, where the ionization and energetics are dominated by
the influence of an X-ray field, is referred to as an {\it X-ray
photoionized plasma}.  More generally, when the source of X-ray illumination is not a point source, 
\begin{equation}
\xi=\frac{4\pi F}{n}
\end{equation}
where $F$ is the energy-integrated, angle-averaged, ionizing flux,
and $n$ is the particle number density. There are other, related forms
of the ionization parameter in use (e.g., \cite{kmt}), but we will use $\xi$ throughout.
 In terms of $\xi$, the photoionization rate corresponding to a single ionic cross-section is
\begin{equation}
\label{eq:photorate2}
\beta_i =n \xi~ \int_{\chi_i}^{\infty}dE~\frac{S_E(E)}{4\pi E}~\sigma_i(E).
\end{equation} 

\begin{figure*}[t]
\centering
\includegraphics[width=12cm,height=9cm]{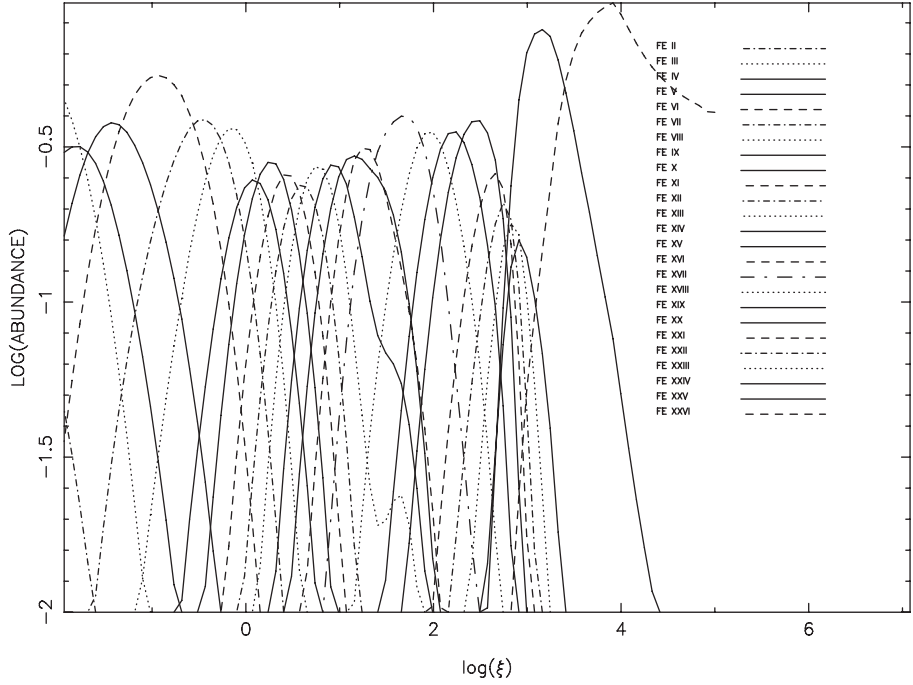}
\caption{Iron charge state distribution vs. ionization parameter.
Figure from Kallman et al. \cite{kallman04}.}
\end{figure*}

For the X-ray sources of interest, typical values of $\xi$ range from approximately 1--$10^4$ 
erg cm s$^{-1}$.  Since the energy equation and the equations of ionization balance are
coupled, the output from the calculation of the physical state of a gas using a photoionization
code gives both the temperature and the charge state distribution as a function of $\xi$.
A pertinent example is given in Fig. (17), which shows the distribution in $\xi$ of all 27 charge states
of iron \cite{kallman04}. 

\subsection{X-ray Fluorescence Lines}

As discussed above, in accretion-powered X-ray sources, such as AGN and X-ray binaries, 
reprocessing of a hard X-ray continuum in relatively cool matter
($10^5$ -- $10^6$ K) can generate intense iron K radiation from more neutral 
iron species of relatively low charge \cite{george91}. 
Iron fluorescence can be quite prominent, and, in fact, has long been known
to constitute an essential component of disk spectra \cite{nandra98} \cite{matsuoka}.

The reprocessing mechanism, for any species with more than two bound electrons, 
begins with photoionization of a $1s$ electron  by a photon with energy $\epsilon$ 
above the K edge, sending an element $A$ in charge state $i-1$ to charge state $i$, 
where $i$ is in a quasi-bound state that is coupled to the continuum, denoted by the 
double asterisk below: 
\begin{equation}
\epsilon +A_{i-1} \rightarrow A_i^{**}+e^- ~~~({\rm K~shell ~photoionization}).
\end{equation}
The intermediate state $A_i^{**}$, since it lies above the ionization threshold, 
has a non-vanishing $e^-$-- $e^-$ Coulomb repulsion matrix element with a final 
product state, part of which consists of a continuum electron. Therefore, there 
is a non-vanishing probability assigned to the reaction 
\begin{equation}
A_i^{**} \rightarrow A_{i+1} +e^- ~~~({\rm autoionization}).
\end{equation}
The ejection of an electron by this mechanism is called {\it autoionization}.
The configuration or energy level associated with $A_i^{**}$ is called an autoionizing
configuration or autoionizing level, respectively. Referred to as the {\it Auger effect} 
for $1s$ vacancy states, autoionization is often the dominant decay route for the 
$1s$-hole state. The reaction products are an {\it Auger  electron}, i.e., an electron 
with a kinetic energy that is characteristic of the atomic energy level structure, and 
an ion in charge state $i+1$, where, in this example, we assume for simplicity that 
$i+1$ is left in its ground state. Still, it is worth mentioning that autoionization
does not always leave a product ion in the ground state. It may be excited, and it
can even be autoionizing. In the latter case, several charge states are coupled by
autoionization; a single K-shell photoionization thus leads to a {\it vacancy cascade}, 
and must be accounted for when calculating the charge state distribution \cite{weisheit}.
An extreme case relevant to astrophysics occurs for K-shell photoionization
of neutral iron (Fe I), which initiates a vacancy cascade that can leave Fe X \cite{kaastra}.
However, even in the simplest case described above ($A_{i-1}$ $\rightarrow$
$A_i$ $\rightarrow$ $A_{i+1}$), three charge states are coupled, and we see that
the simple equations of ionization balance given in Eq. (\ref{eq:balance1}) need
modification for actual calculations of the charge state distribution. This does 
not affect the definition of $\xi$ given there, however.

\begin{figure}[top]
\centering
\includegraphics[width=10cm,height=7cm]{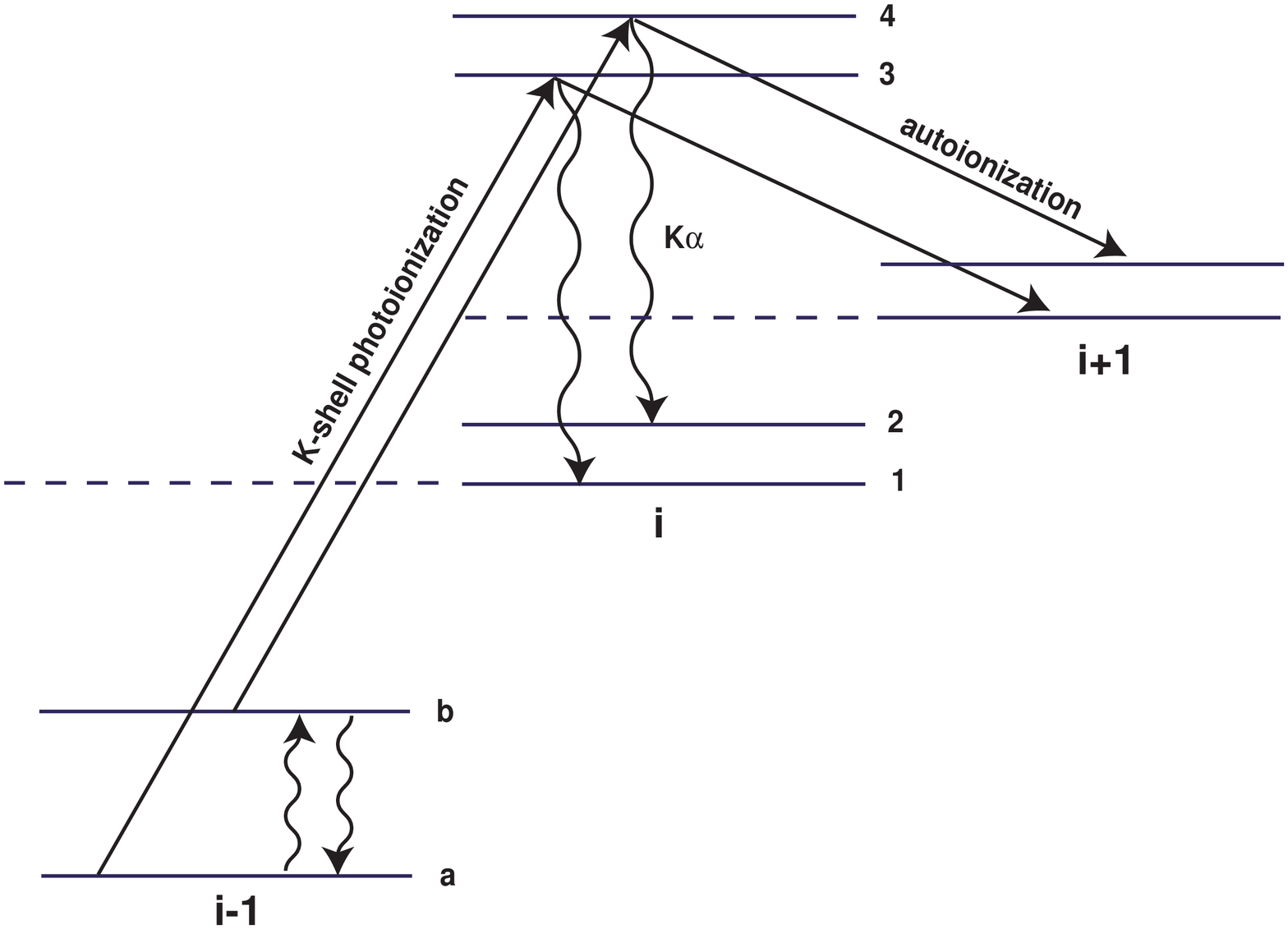}
\caption{Schematic of atomic processes involved in production of 
photoionization-driven K spectra. Three charge states, $i-1$, $i$, and $i+1$
are represented. Level 1 is the ground state of $i$, level 2 is an excited level, 
levels 3 and 4 are autoionizing, lying above the first ionization limit
of $i$ ({\it dashed line}).  The low-lying level $b$ in $i-1$ is populated
by photoexcitation from the ground level $a$, and decays radiatively back to $a$
({\it wavy lines}). (In the more general case, collisional excitation  and de-excitation
also affect the population distribution.) K-shell photoionization from levels
$a$ and $b$ populates levels 3 and 4, respectively, which can decay radiatively
by K$\alpha$ emission to levels 1 and 2, respectively.  Alternatively, levels
3 and 4 can autoionize, leaving as products an Auger electron and ion $i+1$. The relative
intensities of $3 \rightarrow 1$ and $4 \rightarrow 2$ thus depend on the level
population distribution in $1-i$. The escape probabilities of these two lines
depend on the level population distribution in $i$.}
\end{figure}

For iron, autoionization rates $\sim10^{12}$--$10^{14}$ s$^{-1}$ are typical.
Competing with autoionization of the state $A_i^{**}$ is spontaneous radiative decay.
Radiative transition rates are of the same order of magnitude as autoionization
rates in iron, but are typically smaller by factors of a few.
The case of highest probability involves a radiative
transition that fills the K-shell hole. The line energy $\epsilon_K$ is characteristic of
the atomic structure of the ion. We write the reaction as
\begin{equation}
A_i^{**} \rightarrow A_i^* +\epsilon_K ~~~({\rm spontaneous~radiative~decay}).
\end{equation}
The product ion $A_i^*$ is indicated as being an excited level, although there are
cases when it is a ground level. If the radiative transition
leaves an ion in a state such that its energy lies below the first ionization
potential, then that transition has led to {\it radiative stabilization} of the ion,
i.e., the ion is no longer subject to autoionization. Parity-changing transitions of
the type $np$-$1s$ are preferred over $nl$-$1s$, where $l \ne p$. Transitions filling
the K-shell hole, if $n=2$, are referred to as K$\alpha$ transitions, and if $n=3$,
K$\beta$ transitions. To make these concepts more concrete, we provide the following examples.

\vskip 10pt

\noindent
{\it Example 1: 11-electron ion -- autoionization}
\begin{equation}
\begin{array}{l}
1s^2 2s^2 2p^6 3s + \epsilon \rightarrow 1s 2s^2 2p^6 3s + e^- \\ \\
1s 2s^2 2p^6 3s \rightarrow 1s^2 2s^2 2p^5 + e^-
\end{array}
\end{equation}
The first reaction generates a photelectron, which heats the ambient plasma. The energy distribution
of photoelectrons is continuous, and depends on the shape of the ionizing spectrum.
The second reaction -- autoionization -- leaves the F-like ion in the ground level,
and ejects an Auger electron.

\vskip 10pt

\noindent
{\it Example 2: 11-electron ion -- radiative stabilization by K$\alpha$ emission}
\begin{equation}
\begin{array}{l}
1s^2 2s^2 2p^6 3s + \epsilon \rightarrow 1s 2s^2 2p^6 3s + e^- \\ \\
1s 2s^2 2p^6 3s \rightarrow 1s^2 2s^2 2p^5 3s + K\alpha \\ \\
1s^2 2s^2 2p^5 3s \rightarrow 1s^2 2s^2 2p^6 + \epsilon_L
\end{array}
\end{equation}
The second step is a radiative stabilization of the autoionizing level, which
produces a K$\alpha$ photon and leaves an excited Ne-like ion. The latter then
decays by emission of a $3s \rightarrow 2p$ soft X-ray line.
The end products are a Ne-like ground level, an Auger electron, one K$\alpha$ line,
and one Ne-like $3s \rightarrow 2p$ line.

Natural line widths are $\approx 0.4 ~A_{14}$ eV, where $A$ denotes the sum of radiative
and autoionization transition rates from the autoionizing level of interest,
and is given as a multiple of $10^{14}$ s$^{-1}$.
For a line with energy $\epsilon$, the thermal Doppler width is $\approx 0.4~ (\epsilon/6.5 ~{\rm
keV})(kT/{\rm 100 ~eV})^{1/2}$ eV. Note that the low temperature is 
characteristic of X-ray photoionized plasmas.  The commensurability of the natural 
line width and the thermal Doppler width for iron K  fluorescent lines implies large 
Voigt parameters (see, for example, \cite{rybicki}). Estimates of line-center 
optical depths based on a pure Doppler profile will tend to provide an overestimate 
by factors of a few, as well as underestimate the contribution of the line opacity 
in the damping wings. The curve of growth is affected rather dramatically; the large 
Voigt parameter results in the virtual elimination of the saturation part of the curve, 
leaving a transition from the linear part to the damping part. 

\subsubsection{K$\beta$ Emission}

While not as diagnostically useful as K$\alpha$ emission, the K$\beta$ complex
has the potential to provide corollary information.

\vskip 10pt

\noindent
{\it Example 3: 13-electron ion -- K$\beta$ emission}
\begin{equation}
\begin{array}{l}
1s^2 2s^2 2p^6 3s^2 3p + \epsilon \rightarrow 1s 2s^2 2p^6 3s^2 3p + e^- \\ \\
1s 2s^2 2p^6 3s^2 3p \rightarrow 1s^2 2s^2 2p^6 3s^2 + K\beta
\end{array}
\end{equation}
Competing with K$\beta$ emission in the second step are autoionization and K$\alpha$ emission.
There is a relatively small, but non-negligible, probability that a K$\beta$ 
photon is produced. Detailed calculations show that, for the near-neutral iron
charge states, the K$\beta$/K$\alpha$ intensity ratio varies from about 0.12 (Fe II)
to 0.15 (Fe IX) \cite{palmeri03}. 

For most cases of interest, photoionization-driven  K$\beta$ is important only
when the pre-ionization charge state has an occupied $3p$ subshell in its ground configuration. 
To see this, we take as a counterexample, a N-like ion, with ground configuration $1s^2 2s^2 2p^3$.
While, in principle, it is possible to create, say, C-like $1s 2s^2 2p^2 3p$ through
K-shell photoionization from the N-like ground configuration, 
this mechanism is of interest only if there is
a substantial population of the excited $1s^2 2s^2 2p^2 3p$ configuration, which is extremely
unlikely, since the radiative lifetime of such an excited level is usually less
than a nanosecond. Note also that it is not sufficient to have a populated $3s$
subshell in the ground configuration of the pre-ionization charge state. Decay by  K$\beta$
$3s \rightarrow 1s$, because of the parity selection rule, is extremely
improbable, since it competes with a much faster $2p \rightarrow 1s$ transition,
as indicated in Example 2 above. Therefore, as a rule of thumb, K$\beta$ is emitted
only by the iron ions Fe II -- Fe XIV (singly-ionized through Al-like) if 
photoionization dominates the excitation.

One possible mechanism leading to  K$\beta$ emission in charge states
more ionized then Fe XIV involves direct photoexcitation
by the same radiation field responsible for creating the charge state 
distribution \cite{liedahl03}. Thus for a continuum photon of energy $\epsilon$ 
reactions such as $1s^2 2s^2 2p^2 + \epsilon$ $\rightarrow$ $1s 2s^2
2p^2 3p$, a resonance absorption, followed by re-emission, can be efficient drivers of K$\beta$. Similarly, again
using a C-like ion as an example, the entire Rydberg series of lines $1s 2s^2 2p^2 np$ 
$\rightarrow$ $1s^2 2s^2 2p^2$ can be energized by resonance absorption 
of the continuum, so that K$\gamma$, K$\delta$, etc., may appear in a spectrum.
Gauging the overall importance of such a  mechanism involves radiation transport, 
thereby introducing macroscopic  properties of the gas into the calculation, which is case specific.
Generally, it is found that the efficacy with which photoexcitation 
competes with photoionization in driving line emission decreases with plasma column density
\cite{kink}. However, velocity gradients can complicate the analysis \cite{wojd}.

\subsubsection{Fluorescence Yield}

The two-step process described above --- inner-shell photoionization followed by
radiative stabilization accompanied by emission of a K photon --- is called {\it fluorescence}, 
possibly a misnomer, since the term applies also to other radiation processes.
The K {\it fluorescence yield}, which we denote by $Y_K$, is the quotient of the 
rate at which K lines are generated from an irradiated sample and the rate 
at which K-shell holes are produced in the sample \cite{bambynek}.
To motivate the concept of fluorescence yield, imagine a somewhat idealized
laboratory experiment in which a beam of ions of element $A$ interacts with a high-energy 
electron beam, say, in a crossed-beam setup. The electrons ionize the ions,
increasing their charge by one step, and the beams are magnetically separated downstream
of the interaction region. An experimenter counts the ionizations per unit time.
Suppose also that an X-ray detector records the X-ray line production, and the
ratio of X-ray count rate to ionization rate is recorded. A second
run with the beam current increased by a factor of two shows that both the ionizations
per second and the X-ray counts per second double. Next, an ion beam of a different
element $B$ is tested with the same setup. It is again found that the X-ray flux
doubles when the ionization rate doubles. However, the ratio of X rays to ionizations
is different for $B$. One draws the conclusion that the X-ray production is linearly
proportional to the ionization rate, but that the constant of proportionality differs
from element to element, and, possibly, from charge state to charge state for a given
element. One infers that the proportionality constant $Y$ is a characteristic of a given
species and, specializing to K line emission, defines 
\begin{equation}
\label{eq:ydefine}
Y_K = {\rm probability~of~K~line~emission~per~K~shell~ionization}. 
\end{equation}
The usual convention is to associate a fluorescence yield with the 
pre-ionization charge state. For example, Fe II K lines are associated with an Fe I yield. 
This makes good sense, since there are cases where K lines from a given charge state arise from
a photoionization of an ion several charge states removed. For example, Fe III line
production may add to the Fe I fluorescence yield.
We also distinguish emission from H-like and He-like ions from that of the lower
charge states. In X-ray photoionized plasmas, emission from H- and He-like ions
results predominantly from radiative recombination into excited levels. One {\it could} assign a
fluorescence yield to a H-like ion by taking the ratio of H-like K lines produced
per photoionization of the H-like ion, but that is actually somewhat at odds with the
convention used for the lower charge states. In this paper, we assume that the fluorescence
yield is defined for neutral atoms through Li-like ions.

In what follows, we show that the implication of Eq. (\ref{eq:ydefine}) --- that 
$Y$ is an intrinsic atomic property --- is false. For a set of energy levels $u$ 
that lie above the ionization threshold, and a set of stabilized levels $\ell$, 
all of which belong to charge state $i$ (see Fig. 18), one starting point for 
deriving an explicit expression for the fluorescence yield is the emissivity, 
summed over all K lines of charge state $i$,
\begin{equation}
\label{eq:emiss1}
j_K=\sum_u \sum_{\ell} n_{i,u} A^r_{u\ell},
\end{equation}
where we assume that transitions $u \rightarrow \ell$ are members of the
K$\alpha$ complex in $i$. Level population densities of level $u$ in charge state $i$
are denoted $n_{i,u}$. Radiative transition rates are denoted by $A^r$. The
summed K line emissivity has c.g.s. units cm$^{-3}$ s$^{-1}$.

The level population density $n_{i,u}$ is found by dividing the population
flux into level $u$ by the total decay rate of the level,
\begin{equation}
\label{eq:levelpop}
n_{i,u}=\frac{\sum_k n_{i-1,k} \beta_{ku}}
{\sum_j A^r_{uj} + \sum_m A^a_{um}},
\end{equation}
where $A^a_{um}$ is an autoionization rate connecting level $u$ in $i$
to level $m$ in $i+1$. Energy levels $k$ in the pre-ionization charge state
$i-1$ are represented as $n_{i-1,k}$. Photoionization connecting level $k$ in
$i-1$ to autoionizing levels $u$ in $i$ are denoted by $\beta_{ku}$, and are
calculated according to Eq. (\ref{eq:photorate}). 

Substituting Eq. (\ref{eq:levelpop}) into Eq. (\ref{eq:emiss1}) gives

\begin{equation}
j_K=\sum_k \sum_u \sum_{\ell} n_{i-1,k}  \beta_{ku}  \Gamma_{u\ell},
\end{equation}
where we define the {\it line fluorescence yield} $\Gamma_{u\ell}$ for each
transition $u \rightarrow \ell$ by
\begin{equation}
\Gamma_{u\ell}=\frac{A^r_{u\ell}}{\sum_j A^r_{uj} + \sum_m A^a_{um}}.
\end{equation}
The line fluorescence yield depends on purely atomic quantities, and can be
thought of as a radiative branching ratio where additional sinks are included,
viz., autoionization. 

Now we ``normalize'' the emissivity by dividing it by the total photoionization 
rate, which gives an expression for the fluorescence yield,
\begin{equation}
\label{eq:y1}
Y_K=\frac{\sum_k \sum_u \sum_{\ell} n_{i-1,k}  \beta_{ku}  \Gamma_{u\ell}}
{\sum_k \sum_u n_{i-1,k} \beta_{ku}}.
\end{equation}
The yield $Y_{K\alpha}$ is thus a weighted average
of the line fluorescence yields, where the weightings are determined by the level
population distribution of charge state $i-1$ and the level-to-level photoionization
rates, and should, in principle, be evaluated on a 
case-by-case basis, taking into account processes that affect the local
excitation conditions. While complicated enough, note that
$Y_K$ does {\it not} depend on the level population distributions 
of charge states $i$ and $i+1$, nor does it depend on the charge state 
distribution. We have added one further simplification, which is to ignore 
possible collisional redistribution of levels $u$, which is likely to be a 
valid assumption for most astrophysical densities. 

The above derivation shows that a fluorescence yield cannot be considered
as an intrinsic attribute of an atomic species, a fact that has been known for
some time \cite{jacobs}. Equation (\ref{eq:y1}) shows
that $Y_K$ depends on the level population distribution of $i-1$, which
depends in turn on the electron density, the electron temperature, and the local
radiation field. Nevertheless, tables of yields exist, and are are widely used
in astrophysics \cite{kaastra} \cite{house}. Part of the reason for this is
that atomic calculations covering the wide range of conditions expected in astrophysical
plasmas are scarce. Although this situation will change, in the meantime data
based upon laboratory experiments, where fluorescence lines are used for
the purposes of calibration, and ``theoretical'' data based upon approximation
schemes, such as $Z$-scaling, continue to be used. Notwithstanding
one's possible expectations that yields vary dramatically, depending on the plasma conditions,
detailed calculations of iron fluorescence \cite{kallman04} \cite{palmeri} show that numerical values
of $Y_K$ do not show substantial sensitivity to electron density between the low-density limit and the
Maxwell-Boltzmann limit for the ions Fe II -- Fe XVII. It is found that $Y_K$ hovers around the
``canonical'' value of 0.34 for these ions. A variation is observed for higher charge states, however.

Since electric dipole radiative rates scale as $Z^4$ for $\Delta n>0$ transitions, 
and since Auger rates are roughly constant with $Z$, $Y_K$ increases rapidly with $Z$ \cite{cowan}.
The trend of $Y_K$ with $Z$, based upon both experimental data and calculations
is shown in \cite{krause}. Among the elements currently relevant to cosmic 
X-ray spectroscopy, only nickel has a higher atomic
number than iron, and the nickel abundance is at least an order 
of magnitude smaller than that of iron \cite{anders}.
The relatively large iron fluorescence yields, coupled with its high abundance, 
conspire to make iron K fluorescence lines the most commonly
observed in astrophysics. Calculations of disk reflection models have shown that
fluorescence lines from the remaining cosmically abundant elements are relatively weak
\cite{matt97}, and this is validated by the lack of detection in black hole accretion
disk systems, although some (e.g., silicon K$\alpha$) are observed in other source 
classes (e.g., \cite{sako99}).

\subsubsection{Energy Distribution of Iron K Lines}

To date, spectrometers used to observe extrasolar X-ray sources 
have, at best, resolved the iron K$\alpha$ complex
into three features: the Fe XXVI Ly$\alpha$ doublet at 6.97 keV, the Fe XXV blend, near 
6.7 keV, and a blend of emission from a composite of lines
from Fe II-Fe XVIII near 6.4 keV.  For convenience, we refer 
to the 6.4 keV blend as arising in ``near-neutral'' material. 
The energy distribution of the iron K lines is shown in Fig. (19),
from \cite{kallman04}, where the potential for blending of the near-neutral 
lines is evident. Comparing adjacent charge states,
the small energy spacings of lines from near-neutral iron
arises from the fact that changes in the screening of the atomic potential 
experienced by $2p$ electrons are small compared to the K$\alpha$ line energies, i.e.,
the potential in the $n=2$ shell is dominated by the nuclear charge to such an extent
that screening by $n=3$ electrons is of small consequence. 
The blending persists into the L shell.  
For example, in assessing the 5 eV energy separation of the K$\alpha$ line
centroids\footnote{The centroid is defined as $\Sigma_i \epsilon_i j_i/\Sigma_i j_i$, where $\epsilon$ is
a line energy, and $j$ is an emissivity.} of Ne-like Fe XVII 
and F-like Fe XVIII, note that the upper configurations are $1s 2s^2 2p^6 3s$ and
$1s2s^2 2p^6$, respectively. The small difference in centroid energies results from the 
small screening of the nuclear potential by the $3s$ electron in the Fe
XVII $1s$-hole state. Starting with O-like Fe XIX, the ion-to-ion K$\alpha$ centroid separations
begin to increase. The removal of a $2p$ electron has a relatively large effect on the 
differential screening.  For example, the K$\alpha$ energy centroids
for Fe XVIII and Fe XIX are separated by 34 eV. 

\begin{figure}[t]
\centering
\includegraphics[width=12cm,height=8cm]{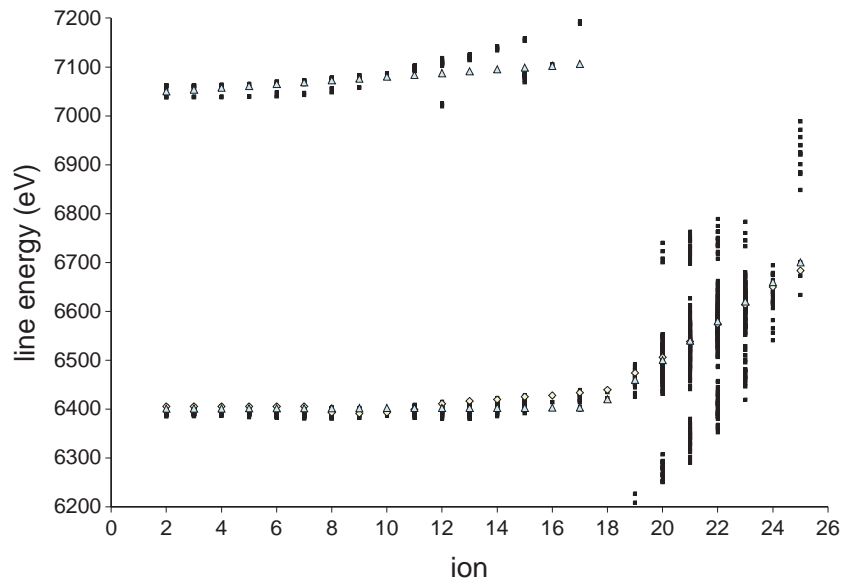}
\caption{Energy centroids for iron K$\alpha$ ({\it lower}) and K$\beta$ ({\it upper}.)
Ordinate denotes the number of bound electrons in the emitting ion: `2' for Fe II, etc.
Figure from Kallman et al. \cite{kallman04}.}
\end{figure}

The most commonly observed fluorescence lines are those from near-neutral ions.
Because of the blending and the small variation of line centroid energies,
the entire complex is often referred to as ``the iron line,'' with the understanding
the ``line'' is likely to be a composite structure. Reference to table values of line
energies often contain separate entries for K$\alpha_1$ and K$\alpha_2$, which partially
accounts for the intrinsic spectral structure. The distinction between these two features
results from the relative probabilities of a $1s$ vacancy being filled by a radiative
transition from an electron in the $2p_{3/2}$ subshell (K$\alpha_1$) or the 
$2p_{1/2}$ subshell (K$\alpha_2$). Thus $E(K\alpha_1)>E(K\alpha_2)$. Since there are
four $2p_{3/2}$ electrons in a filled L shell, and two $2p_{1/2}$ electrons, the
K$\alpha_1$ yield is  generally given as twice the K$\alpha_2$ yield. However, even this
greatly oversimplifies the K spectra from iron ions. In fact, the K$\alpha_1$--K$\alpha_2$
labels are primarily an observational convenience (when they are spectrally resolved).
When one accounts for the myriad possibilities introduced by angular momentum coupling,
and the variety of possible excitation conditions encountered in astrophysical plasmas,
the true spectrum may consist of hundreds of individual lines. In some cases
the K$\alpha_1$ and K$\alpha_2$ line complexes overlap in energy.

\subsubsection{Resonant Auger Destruction}

The blending of K$\alpha$ lines from near-neutral iron ions could be considered
as both a blessing and a curse. On the one hand, the blending allows a simple 
approach to spectral fitting: simply add a line at or near 6.4 keV to the trial
spectral model; no need to worry about a complex distribution of lines; no need
to worry about the charge state distribution. On the other hand, blending means
that while conditions in the disk plasma may vary considerably through the emitting
region, for a large range of physical parameter space,, these variations are not
conveyed by the spectrum. In this latter context, then, one might look to the L-shell
ions for potential diagnostics. The K$\alpha$ spectra of L-shell ions provide 
unique diagnostic information on plasmas that exist over intermediate ranges of ionization,
as can be seen in Fig. 17.  Interestingly, it appears that K$\alpha$ lines
from iron L-shell ions are not required 
in spectral fits to black hole accretion disk spectra, even though there is 
ample evidence for emission from the charge states that bracket them. The absence of 
L-shell emission has been attributed to a process known as {\it resonant Auger
destruction} \cite{ross96} \cite{band}, a radiation transfer effect that leads to severe
attenuation of K emission lines from L-shell ions.

To see how resonant Auger destruction (RAD) selectively attenuates K emission
from L-shell ions, as opposed to M-shell ions, consider the following example.

\vskip 10pt

\noindent
{\it Example 4: 10-electron ion -- resonant Auger destruction of K$\alpha$}
\begin{equation}
\begin{array}{l}
1s^2 2s^2 2p^6 + \epsilon \rightarrow 1s 2s^2 2p^6 + e^-
~~~~ ({\rm photoionization}) \\ \\
1s 2s^2 2p^6 \rightarrow 1s^2 2s^2 2p^5  + K\alpha 
~~~~~~~~~ ({\rm K}\alpha ~{\rm emission}) \\ \\
K\alpha + 1s^2 2s^2 2p^5 \rightarrow 1s 2s^2 2p^6
~~~~~~~~~ ({\rm resonant ~absorption})\\ \\
1s 2s^2 2p^6  \rightarrow 1s^2 2s^2 2p^4 +e^-
~~~~~~~~~~~ ({\rm Auger ~decay/photon ~destruction})
\end{array}
\end{equation} 
The third step is the inverse of the second, where it is understood that
the K$\alpha$ photon, somewhere along its line of flight, encounters an ion 
in the same state as the product ion that resulted from the original K$\alpha$ transition.
The fourth step, autoionization, is, on average, more probable than
radiative decay. Therefore, there is a high probability per scattering event that the
photon will be destroyed. If the autoionizing ion
in the fourth step decays radiatively instead, then, depending on the line
optical depth of the medium, another absorption can occur, with an equally large
destruction probability. The probability of the K$\alpha$ photon surviving more
than just a few scatters is very small, and very few photons can escape the medium.

To make a distinction between L-shell ions and M-shell ions and their response
to the RAD process, consider one more example.

\vskip 10pt

\noindent
{\it Example 4: 11-electron ion -- an improbable resonant Auger destruction of K$\alpha$}
\begin{equation}
\begin{array}{l}
1s^2 2s^2 2p^6 3s+ \epsilon \rightarrow 1s 2s^2 2p^6 3s + e^-
~~~~ ({\rm photoionization}) \\ \\
1s 2s^2 2p^6 3s\rightarrow 1s^2 2s^2 2p^5 3s + K\alpha 
~~~~~~~~~ ({\rm K}\alpha ~{\rm emission}) \\ \\
K\alpha + 1s^2 2s^2 2p^5  3s\rightarrow 1s 2s^2 2p^6 3s
~~~~~~~~~ ({\rm resonant ~absorption})\\ \\
1s 2s^2 2p^6 3s  \rightarrow 1s^2 2s^2 2p^5 +e^-
~~~~~~~~~~~~~~ ({\rm Auger ~decay/photon ~destruction})
\end{array}
\end{equation} 
While this set of reactions is possible, it is generally unimportant, 
since the third step is extremely unlikely. For this step to operate effectively,
the medium would require a high optical depth in the relevant excited level
of the $1s^2 2s^2 2p^5  3s$ configuration. For the densities of interest,
this will never be the case. By far, the dominant configuration of
10-electron ion is $1s^2 2s^2 2p^6$, and excitation channels
to the $2p$ subshell are closed. This is also the case for the RAD sequence
that begins with any ion that has $n=3$ electrons in its gound configuration.
Therefore, RAD has a negligible effect on M-shell ions, but a potentially
major effect on L-shell ions. 

\begin{figure}[t]
\centering
\includegraphics[width=12cm,height=9cm]{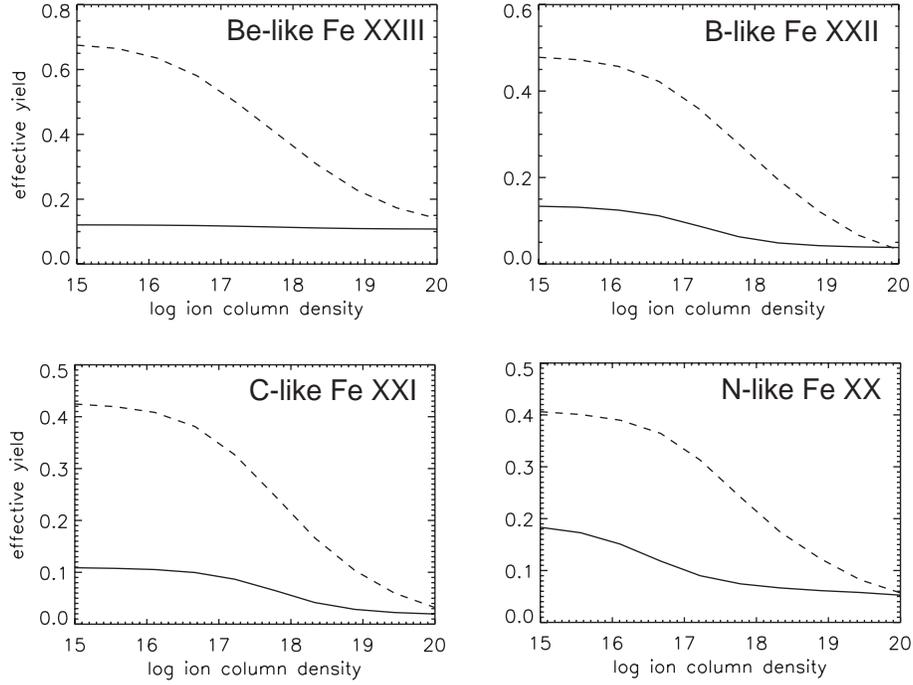}
\caption{Effective fluorescence yields as modified by resonant Auger destruction
plotted against ionic column density for Fe XX -- Fe XXIII.  Each plot shows two cases: 
({\it dotted line}) level populations of low-lying states driven by an 80 eV half-diluted blackbody 
and ({\it dashed line}) no perturbing radiation field. The fluorescence yield of N-like Fe XX
is determined by the production and escape probability of C-like Fe XXI lines, etc.}
\end{figure}

Consider a semi-infinite slab that is irradiated from above by a hard X-ray
continuum source. Only an upper layer of the slab
corresponding roughly to unity line optical depth will contribute to K$\alpha$ fluorescence 
from L-shell ions. Below that depth, escape is prohibited by
resonant Auger destruction. Compare this to the case of fluorescence from 
M-shell ions, where the fluorescing region for near-neutrals extends 
for roughly one continuum optical depth. This would suggest that
the emergent K$\alpha$ line flux from L-shell ions is dwarfed by that
from the near-neutrals. This provides a plausible explanation for the absence
of an L-shell component in iron K$\alpha$ spectra. Moreover, in the context of spectral
modeling of accretion disks, it appears to provide justification
for simply zeroing out the K$\alpha$ emission for L-shell ions.

This argument needs some modification, however. When we couch the argument only
in terms of absorption cross-sections and ion column densities, we ignore the fact
that the line optical depth depends on the level population of the lower level of
the relevant transition, which need not be the ground state. For a given 
transition between an upper level $u$ and a lower  level $\ell$, the line optical 
depth, in terms of the hydrogen column density $N_{\rm H}$, the elemental abundance 
$A_Z$, and the ionic fraction $ F_{\rm ion}$,  can be written
\begin{equation}
\tau_{\ell u}=N_{\rm H}A_Z F_{\rm ion} p_{\ell}~ \frac{\pi e^2}{mc} ~f_{\ell u} \phi(\nu)
\end{equation}
where $p_{\ell}$ is the fractional population density of level $\ell$, 
$f_{\ell u}$ is the absorption oscillator strength, and $\phi(\nu)$ is the line profile  
function in the rest frame of the absorber. Therefore, we must solve for the level populations
$p_{\ell}$  for the lower levels of each K$\alpha$ transition. In fact, many K$\alpha$ transitions
terminate on excited states, and such lines may be optically thin under certain conditions relevant to
accretion disk atmospheres
\cite{liedahl05}.

Given that the level population distribution in a charge state $i$ partly determines
the optical depth of K$\alpha$ lines from $i$, self-consistency demands that the
set $p_{\ell}$ in charge state $i-1$ also be calculated. The level populations 
of $i-1$ determine the relative
rates at which the upper levels of the K$\alpha$ transitions in $i$ are populated
by K-shell photoionization \cite{kallman04}. Referring back to the discussion
in \S7.2.2 concerning the effects of local excitation conditions on $Y_K$, this
constitutes a second reason why assessments of RAD cannot be decoupled
from microscopic considerations. It is shown in \cite{liedahl05} that the
level population distribution in $i-1$, responding to an underlying UV/soft X-ray
continuum, can significantly increase $Y_K$ for N-like, C-like, B-like, and
Be-like iron ions. If the definiton of $Y_K$ is modified so as to account for the 
escape probability of K$\alpha$ lines from a medium of specified column density
in charge state $i$, then a plot of $Y_K^{\rm eff}$ vs. $N_i$ allows us to evaluate
the overall effect of RAD, accounting for both microscopic and macroscopic influences.
This is illustrated in Fig. (20), which shows that RAD is not decisively effective
in quenching K$\alpha$ emission from L-shell ions until ionic column densities
$\sim10^{19}$ cm$^{-2}$ are reached. Calculations of vertical structure in irradiated
accretion disks suggest that, for these ions, this value is most likely on the high side of what
is expected in typical disk atmospheres \cite{nayakshin01}.

\subsection{X-ray Reflection}

The simplest assumption concerning vertical structure is to invoke a constant density, i.e.,
to forego a detailed consideration of the effect of the gravitational field entirely. 
Calculations of this type, while seemingly neglecting important physics, have nevertheless 
proven to be quite powerful in elucidating the physics of spectral formation in accretion 
disks \cite{rossfabian} \cite{matt93} \cite{zycki}. With a choice for the incident flux, 
the local density dictates the ionization parameter. Transport of the incident 
radiation down through the atmosphere leads to its attenuation, which results in a stratification of
$\xi$ ($\partial \xi/\partial z$ is positive). Photoionization codes are then used to generate local spectra,
which, accounting for opacity in the overlying layers, are propagated
through the atmosphere to the computational boundary, the result being the spectral
distribution of the radiation field at the ``surface.'' If desired, the  effects
of disk rotation, disk inclination, and relativistic effects are applied, thus generating
a model spectrum as observed at infinity.

The first applications of constant-density models were focused on an examination of the 
``reflection'' of the incident continuum from cold matter, ``cold'' in this context meaning
that H and He were assumed to be fully stripped, while the remaining elements retained all
of their electrons. A fraction of the radiation impinging on any gas will be re-radiated,
or reflected, back into the general direction of the radiation source. This fraction is 
known as the {\it albedo}. More generally, one is interested in the energy dependence of the albedo.
High-energy photons interacting with the cold material described above will either be absorbed
via photoionization or Compton scattered. 

A photoelectric cross-section falls roughly as $E^{-3}$ above
the photoionization threshold, or edge. Neutral iron has as its ground configuration
${\rm [Ar] \, 3d^6 4s^2}$, where the notation [Ar] symbolizes the configuration
set of ground state Ar. With electrons occupying four shells, Fe I has associated 
with it an N $(n=4)$ edge, an M edge $(n=3)$, an L edge $(n=2)$, and a K edge $(n=1)$. 
The rapid falloff in the cross-sections results in the K edge dominating the total 
cross-section for energies above the 7.1 keV K edge. Competing with iron opacity is 
Compton scattering, with a small contribution from photoabsorption by the remaining 
elements. Although the iron K edge cross-section at threshold is orders of magnitude 
larger than the Compton cross-section, the fact that iron is a trace element (the 
solar abundance relative to hydrogen is $4.7 \times 10^{-5}$ \cite{anders}) leads to 
a near equality in the opacities, so that Compton scattering is approximately as 
important as the iron photoelectric opacity at the iron K edge. A few keV above the 
iron K edge, Compton scattering becomes the dominant opacity source. 

Since the typical energy lost by a photon in Compton scattering off of cold electrons
is $\Delta E \approx E^2/mc^2$, not only do high-energy photons lose more energy than low-energy
photons, but they lose a larger fractional energy $\Delta E/E$, as well. Compton scattering
from a cold slab thus leads to a degradation in energy of the incident spectrum, 
high energy electrons migrating to lower energies, with a trend toward a steepening 
relative to the incident spectral shape.  At the same time, however, a fraction of 
the photons near and below $\sim10$ keV are absorbed, depleting the incident flux of 
soft X rays. The combined effect of these two processes produces the {\it Compton bump}, 
an apparent excess above an incident power-law spectrum in the approximate range 20--50 
keV \cite{guilbert} \cite{lightman} (see Fig. 21). Accompanying this signature of reflection are 
fluorescence lines \cite{george91} \cite{pounds90} (see below). The overall spectrum 
is a sum of the incident power-law and the reflection spectrum. The relative contribution 
of the reflection spectrum depends on the solid angle $\Delta \Omega$ that the reflector 
subtends at the hard X-ray source \cite{matt91} \cite{matt92}. For example, a disk 
geometry with illumination from above gives  a {\it covering fraction} $\Delta \Omega/4\pi$ 
of about 1/2. 

Some of the first serious attempts to model the vertical structure of hydrostatic accretion discs, 
including energy transport, opacity, and the  equation of state, were applied 
to disks in cataclysmic variables \cite{meyer} \cite{mineshige} \cite{canizzo}. 
These calculations made no provision for the influence of a hard X-ray source, 
however, and were not directly applicable to accreting neutron stars and black 
holes. Later developments saw the introduction of hydrostatic disk models for which the outer 
disk was assumed to be X-ray photoionized by a central source of illumination, 
appropriate to accreting neutron star X-ray binaries \cite{kokallman91} 
\cite{kokallman94} \cite{raymond} \cite{rozanska}.

\begin{figure}[t]
\centering
\includegraphics[width=12cm,height=8cm]{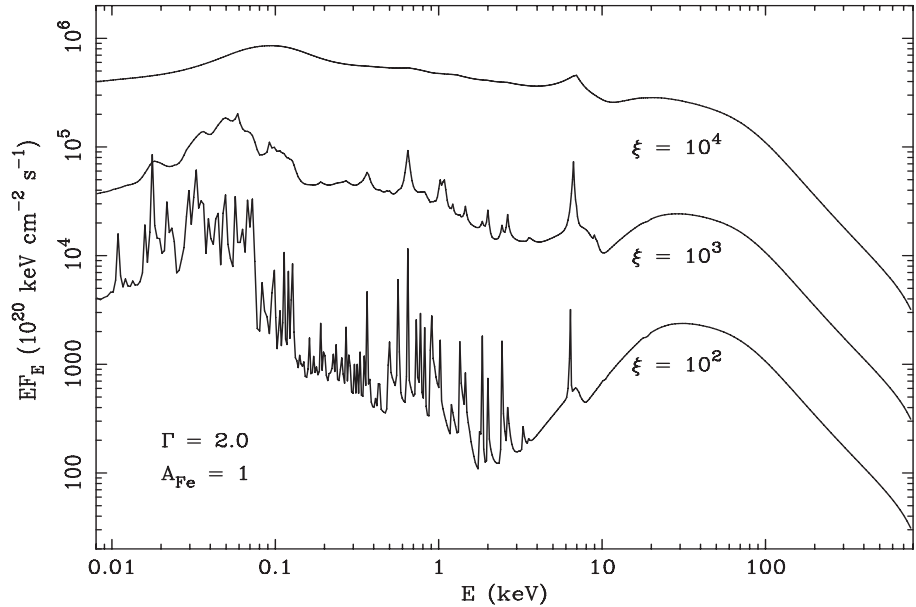}
\caption{Reflection spectra for constant-density slabs for three values of
$\xi$ (as labeled), assuming normal illumination
by a power law spectrum with photon index 2. Electron density is set
to $10^{15}$ cm$^{-3}$. The iron abundance is set to the solar value. From Ross and Fabian
\cite{ross05}. }
\end{figure}
 
A typical approach involves dividing the disk into a set of annuli, say, a few tens, then
using a disk photoionization code to calculate the atmospheric structure of each
annulus by integrating the equations of hydrostatic balance and 1-D radiation 
transfer for a slab geometry. A common alternative is to treat a single annulus.
A hard X-ray radiation field, incident from above at some specified angle, is assumed 
for each radius. The spectral shape is usually taken directly from observations, rather 
than calculated self-consistently. A typical choice for the relative normalizations of the 
X-ray field at each annulus is to scale it by a $r^{-3}$ power law. This 
is motivated by the picture described earlier wherein the hard X rays are 
supposed to originate from Compton scattering of the UV/soft X-ray  disk field, 
which scales as the local dissipation rate in the optically thick disk (see Eq. \ref{eq:ldisk}). 
The overall normalization of the X-ray flux to the UV flux is an input parameter, 
and is typically of order unity. In particular, if a single annulus is modeled, this ratio
is left to the discretion of the modeler, with the assumption that the disk is irradiated
by a localized disk flare \cite{nayakshin01} \cite{ball02}. The atmosphere is
divided into a set of zones of  (possibly) variable geometrical thickness. The absorption of the incident
radiation  field leads to heating through photoelectron thermalization and Compton heating. 
Photoelectron mean free paths are sufficiently small that the energy is deposited 
``on the spot.'' A further constraint is that local thermal equilibrium must be satisfied, i.e.,
$\Lambda (P, \rho,  F_{\nu}) = 0$, where the energy function depends on the gas pressure $P$,
the density $\rho$ and the local radiation field $F_{\nu}$.
The difference between heating and cooling $\Lambda$ includes Compton scattering, 
bremsstrahlung cooling, photoionization heating, Auger heating, collisional line cooling, 
and recombination cooling from H, He, C, N, O, Ne, Mg, Si, S, Ar, Ca, Fe, and Ni ions.
Photoionization produces spectral by-products, as well, such as 
fluorescence lines, radiative recombination continua, and lines produced in recombination 
cascades. This reprocessed radiation contributes to the  ``diffuse'' flux, which is 
added to the modified incident field, and is propagated upwards and downwards to the 
adjacent computational cells. Boundary conditions at the top and bottom of the annulus 
are also prescribed. One choice \cite{mario01} is to set the density at the bottom to 
the value given by the $\alpha$-disk model, and to set the gas temperature at the top 
to the Compton temperature \cite{rybicki}, given by the following integral over the radiation
spectrum:
\begin{equation}
kT_c=(4U)^{-1} \, \int_0^{\infty} d\epsilon ~\epsilon U_{\epsilon},
\end{equation}
where $U_{\epsilon}$ is the specific radiation energy density and $U$ is
the total radiation energy density. The Compton temperature for most cases
of interest is $\sim10^7$-$10^8$ K.

For modeling X-ray line emission from the disk atmosphere, the commonly used assumptions
of LTE and the diffusion approximation may not be valid.  The assumption of constant 
density may be invalid, as well, since the hydrostatic equilibration time scale is 
small or comparable to other relevant timescales  (cf., \cite{ross92}). For calculating the discrete emission 
line content of a disk spectral model, particularly recombination emission, the vertical 
ionization structure and the density stratification are crucial. By contrast, the output 
iron fluorescence energy centroid varies little until the ionization reaches the L shell (see Fig. 18). 
Still, even fluorescence emission can be affected by a Compton thick, 
fully ionized gas overlying the emission sites deeper in the atmosphere \cite{nayakshin00}.
Also, over certain ranges of $\xi$, the gas is subject to a thermal instability
that can lead to a steep transition between cold near-neutral gas and hot highly-ionized
gas \cite{nayakshin00} \cite{mario02}.

Examples of output from a set of reflection slab models are shown in Fig. (21) \cite{ross05}.
In these examples, the gas density is fixed at a constant value, and the slab is irradiated
from above by a continuum flux with spectral shape $AE^{-1} \exp(-E/300$ keV). The value
of $\xi$ is varied by varying the normalization $A$. As described in \cite{ross05},
the reflection spectrum for the largest value of $\xi$ ($10^4$) is nearly line-free,
since the gas is very highly ionized. Compton broadened emission from Fe XXXV and Fe
XXVI make the only conspicuous discrete contribution to the spectrum. At the two lower
values of $\xi$ the increasingly complex line spectra, with contributions from
$n=2 \rightarrow n=1$ transitions of H-like and He-like ions of C, N, O, Ne, Mg, 
Si, S, and a few $n=3 \rightarrow n=2$ Fe L-shell lines, are attributed to the 
progressively lower level of ionization.

\subsubsection{Thermal Instability}

X-ray irradiated gas is subject to a thermal instability in the $10^5$--$10^6$ K 
temperature range \cite{buff} \cite{habing}, suppressing X-ray line emission in that regime.
Application of the Field stability criterion \cite{field} indicates that
a photoionized gas may become unstable when recombination cooling of H-like 
and He-like ions is important. The $T$ and $\xi$ ranges where the instability occurs
depend on the metal abundances and the local radiation spectrum \cite{hess}, and,
from the theoretical point of view, depends on the atomic rates used \cite{savin}.
Within a range of {\it pressure ionization parameters} \cite{kmt}
$\Xi = P_{\rm rad}/P_{\rm gas}$, where $P_{\rm rad}$ is
the radiation pressure, thermal equilibrium is achieved by three distinct temperatures, only
two of which are stable to perturbations in temperature. 
Physically, the instability causes a steep temperature gradient 
as the gas is forced to ``move'' between thermally stable regimes, requiring
the formation of a transition region joining the hot and cold phases
whose size could be determined by thermal conduction.
The method by which the thermal instability is treated can significantly 
affect the predicted spectrum \cite{mario01}. One issue relevant to iron
K fluorescence emission is that the steep transition region may comprise
the iron L-shell ions. Therefore, even if the effects of RAD were nil, iron
L-shell ions may have insufficient optical depth to contribute substantially to the 
overall iron K spectrum. 

\subsubsection{Alternative X-ray Signatures of Relativistic Effects}

Prior to the launch of {\it Chandra} and {\it XMM-Newton}, the soft X-ray
spectrum ($\sim1$ keV) was generally modeled as a power-law attenuated
by an ionized absorber \cite{halpern84} -- the {\it warm absorber} model. 
A radical departure from this model was suggested in \cite{branduardi}, where it was proposed that
spectral features observed in the 5--35 \AA\ band of the Seyfert 1 galaxies MCG--6-30-15 and Mrk 766
could be explained as relativistically broadened emission lines from H-like carbon, 
nitrogen, and oxygen. Fitting parameters -- disk inclination, disk emissivity
profile -- were found to be roughly consistent with those found for the iron
K$\alpha$ line \cite{nandra97}. An analysis of a 130 ks grating observation with 
{\it XMM-Newton} of Mrk 766 \cite{mason03} led to the conclusion that a
relativistic line interpretation provides a better fit to the data than the warm absorber model.
A similar conclusion was found for the Seyfert 1 galaxy NGC 4051 \cite{ogle}.
If correct, these spectral
features, formed by radiative recombinaton onto bare  nuclei, could provide additional ``handles''
that would allow tighter constraints  to be placed on models of the accretion disk structure in the
relativistic regime. Since the sites of formation H-like lines are
spatially distinct from the formation sites of fluorescence lines a broader
range of parameter space would be accessible.

Still, numerous absorption lines from ionized species in the {\it Chandra} grating 
spectrum of MCG--6-30-15, also noted in \cite{branduardi}, indicated at least 
some influence from a warm absorber. In fact, the relativistic disk interpretation 
in its entirety was disputed in \cite{lee2001}. In the model proposed there, absorption in addition
to two warm absorber components is provided by a dust component containing FeO$_2$,
which is embedded in the warm absorber. The warm absorber model found additional
support from theoretical disk models, which showed that the line equivalent widths
required by the relativistic disk interpretation far exceeded theoretical predictions
\cite{ballantyne02}. An additional problem is the high nitrogen to oxygen ratio
required to account for the N VII feature, although a Bowen-like mechanism has been
suggested as an explanation \cite{sakobowen}. A further criticism of the relativistic
disk line model is that reflection models predict emission from iron L-shell
ions \cite{ballantyne02} that is not observed in the data. Radiation transport
calculations using a Monte Carlo approach indicate no strong iron L 
$n \rightarrow 2$ emission, however \cite{mauche}.

In subsequent work \cite{sako03} it was argued that the warm absorber model used 
in the {\it Chandra} interpretation of MCG--6-30-15 does not adequately describe the 
{\it XMM} Reflection Grating Spectrometer data, which has a better response at longer wavelengths.
Further analysis \cite{aturner} of the {\it XMM} data of MCG--6-30-15 showed that fits obtained with a
composite model,  involving both relativistic  disk lines and a warm absorber, yield ambiguous
results;  if a warm absorber is added to the best-fitting disk line model, then relatively 
little absorption is required; if relativistic lines are added to the best-fitting 
warm absorber model, then only weak emission lines are required. Given the potential
importance of relativistic recombination line emission, there is hope
that these issues will be resolved decisively in the near future.

Another class of relativistic signatures has been proposed recently.
The magnetic flare model \cite{galeev} of the hot corona and hard X-ray continuum observed in 
accretion disks posits the existence of
many small-scale, hot ($T>10^8$ K) magnetic loops. Following loop emergence,
Compton cooling of loop material, through the interaction with softer disk photons,
results in hard X-ray flaring events. Disk irradiation by these relatively localized 
flares produce fluorescence emission that is concentrated in the region of the disk near
the flare \cite{nayakshin01b}. In other words, regions of the disk that are small
in both radial and azimuthal extent, and which radiate intense transient
fluorescence emission are predicted by this model. The transiently fluorescing
region will continue to orbit the black hole, and will emit iron lines that
are narrow compared to the case of full disk irradiation. The centroid energy
is determined by the gravitational redshift associated with a small range of
annuli, and by the Doppler effect appropriate to the instantaneous projected
velocities over a narrow range of azimuth. If the lifetime of such a region is
sufficiently long, the centroid energy could be observed to change over the
time scale of an observation. It has been shown how the line flux received
from such a ``hot spot'' changes with position along the orbit, owing to the
effects of relativistic beaming and light bending \cite{dovciak04}.
Narrow spectral features lying at energies below 6.4 keV have indeed been reported,
e.g., \cite{iwasawa99} \cite{turner02} \cite{yaqoob03}. As pointed out in
\cite{iwasawa04}, however, inferring disk parameters from these kinds of data
push the limits of current observational capabilities. 

\section{Concluding Remarks}

The theory of accretion onto black holes has been in development for at least thirty
years. Owing to the complex physics governing accretion, it is probably safe 
to predict that decades of additional research will be required before the theory is
considered complete. Answers to outstanding questions concerning 
accretion onto black holes will not come without access to better data, and 
lie beyond the capabilities of current instrumentation. The current generation 
of X-ray observatories, {\it Chandra} and {\it XMM-Newton} are helping to 
identify the various avenues of inquiry \cite{paerels} that will motivate 
future observations with, for example, NASA's planned {\it Constellation-X} mission
and ESA's planned {\it XEUS} mission.

In the meantime, on the theoretical side, referring back to Shapiro and Teukolsky's
quote in \S1, an attack on the full inhomogeneous, non-axisymmetric, time-dependent, 
relativistic accretion disk problem with coupled radiative transfer is not far beyond the 
horizon. In any case, it seems evident that disk reflection models, with their
emphasis on radiation flow, and MHD simulations, with their emphasis on time-dependent 
gas flow, will need to be merged at some point in the not-too-distant future. 

Einstein developed the General Theory of Relativity at a time when it was not
demanded by astronomical observations. Only Einstein's physical intuition demanded it.
From Schwarzschild's solution, to the theory of stellar collapse, to the quasar model,
to the discovery of Cyg X-1, to the supermassive black hole at the Galactic center, to
the identification of relativistically modified X-ray emission lines,
we now find that General Relativity is fully acknowledged to be an indispensable 
component of the various theoretical formalisms used to describe black hole astrophysics.  
One wonders if Einstein's skepticism regarding the existence of black holes
would have remained intact in light of the current weight of evidence.

In addition to the special and general relativity theories,
Einstein's Nobel prize winning work on the photoelectric effect, his introduction of the 
A and B coefficients, his advances in statistical mechanics, his influence on de Broglie, 
Schrodinger, and Dirac, and many other of his contributions to theoretical physics, figure 
prominently in modern research on the topics discussed in this paper.

\section*{Acknowledgements}
We thank Gordon Drake and Peter Beiersdorfer for their advice and encouragement.
Our appreciation to Chris Mauche, whose careful reading of the text led to improvements 
in the overall presentation. Thanks also to Rob Fender, Mario Jimenez-Garate, Tim Kallman, 
Ben Mathiesen,  Paul Nandra, Martin Pessah, Randy Ross, and Masao Sako for contributions.
An anonymous referee contributed several helpful suggestions.
Work at LLNL was performed under the auspices of the U.S.\  Department 
of Energy by the University of California Lawrence Livermore National Laboratory under 
contract No.\ W-7405-Eng-48.

\end{document}